\documentclass[12pt]{amsart}
\usepackage{graphicx}
\usepackage[centertags]{amsmath}
\usepackage{amsfonts}
\usepackage{amssymb}
\usepackage{amsthm}

\newtheorem{thm}{Theorem}%[section]
\newtheorem{lem}{Lemma}[section]
\newtheorem{cor}[lem]{Corollary}
\newtheorem{conj}[thm]{Conjecture}
\newtheorem{prop}[lem]{Proposition}
\theoremstyle{definition}
\newtheorem{defn}[lem]{Definition}
\theoremstyle{remark}
\newtheorem{rem}{Remark}[section]
\numberwithin{equation}{section}

\newcommand{\ttl}[1]{\noindent \textbf{Part #1.}}

\newcommand{\norm}[1]{\left\Vert#1\right\Vert}
\newcommand{\abs}[1]{\left\vert#1\right\vert}
\newcommand{\set}[1]{\left\{#1\right\}}
\newcommand{\FF}[1]{\mathbb{F}_{\!#1}}
\newcommand{\T}[1]{\tilde{T}_{#1}}
\newcommand{\TT}[1]{T_{#1}}

\newcommand{\thet}{\vartheta}
\newcommand{\bthet}{{\bar\vartheta}}
\newcommand{\btheta}{{\bar\theta}}

\newcommand{\calC}{\mathcal{C}}
\newcommand{\calD}{\mathcal{D}}

\newcommand{\calI}{\mathcal{I}}

\newcommand{\calN}{\mathcal{N}}
\newcommand{\calO}{\mathcal{O}}

\newcommand{\calP}{\mathcal{P}}
\newcommand{\calS}{\mathcal{S}}

\newcommand{\calU}{\mathcal{U}}

\newcommand{\calW}{\mathcal{W}}

\newcommand{\calQ}{\mathcal{Q}}
\newcommand{\bbZ}{\mathbb{Z}}
\newcommand{\bbQ}{\mathbb{Q}}
\newcommand{\Z}{\mathbb{Z}}
\newcommand{\Q}{\mathbb{Q}}
\newcommand{\R}{\mathbb R}
\newcommand{\C}{\mathbb C}

\newcommand{\bbN}{\mathbb N}
\newcommand{\bbT}{\mathbb{T}}
\newcommand{\Tr}{ \mathrm{Tr}}
\newcommand{\Op}{ \mathrm{Op}}
\newcommand{\SL}{ \mathrm{SL}}
\newcommand{\GL}{ \mathrm{GL}}

\newcommand{\lcm}{ \mathrm{lcm}}

\newcommand{\Sp}{\mathrm{Sp}}

\newcommand{\Mat}{\mathrm{Mat}}
\newcommand{\Gal}{\mathrm{Gal}}
\newcommand{\Mor}{\mathrm{Mor}}

\newcommand{\ord}{\mathrm{ord}}

\newcommand{\im}{\mathrm{Im}}

\newcommand{\irr}{\mathrm{irr}}
\newcommand{\calH}{\mathcal{H}}

\begin{document}
\title[Arithmetic Quantum Unique Ergodicity ]
{Arithmetic Quantum Unique Ergodicity for Symplectic Linear Maps of the Multidimensional Torus}%
\author{Dubi Kelmer }%
\address{Raymond and Beverly Sackler School of Mathematical Sciences,
Tel Aviv University, Tel Aviv 69978, Israel}
\email{kelmerdu@post.tau.ac.il}

\thanks{}%

\date{\today}%
\dedicatory{}%
\commby{}%
\begin{abstract}
    We look at the expectation values for
    quantized linear symplectic maps on the multidimensional
    torus and their distribution in the semiclassical limit.
    We construct super-scars that are stable under the arithmetic symmetries
    of the system and localize on invariant manifolds.
    We show that these super-scars exist only when there
    are isotropic rational subspaces, invariant under the
    linear map. In the case where there are no such scars, we compute
    the variance of the fluctuations of the matrix elements for the desymmetrized system,
    and present a conjecture for their limiting distributions.
\end{abstract}
\maketitle

\section*{Introduction}\label{sINTRO}
    Quantization of discrete chaotic dynamics over a compact phase space, has proved to be an effective toy model
    for understanding phenomena in quantum chaos. The first such model was the quantization of the cat
    map, a symplectic linear map acting on the $2$-dimensional torus \cite{HB}.
    In this paper, we look at the multidimensional
    analog of this model, the quantization of symplectic
    linear maps on a multidimensional torus.
    We generalize some of the results obtained for the two
    dimensional case, and present some new phenomena occurring in
    higher dimensions.
\subsection*{Quantum Cat Map}
    In an attempt to gain better understanding of the correspondence between
    classical and quantum mechanics and in particular phenomena in quantum chaos,
    Hannay and Berry introduced a model for quantum mechanics on the
    torus \cite{HB}. The classical dynamics underlying this model is simply the iteration of a symplectic linear map, $A\in \Sp(2,\bbZ)$,
    acting on the $2$-torus, known colloquially as a cat map.
    For quantizing the torus, one takes a family of finite dimensional Hilbert spaces of states,
    $\calH_N=L^2(\bbZ/N\bbZ)$ (where $N$ stands for the inverse of Planck's constant).
    The quantization of smooth observables $f\in C^{\infty}(\bbT^2)$ are operators $\Op_N(f)$ acting on
    $\calH_N$, and the quantization of the classical dynamics, is a unitary operator
    $U_N(A)$,
    known as the quantum propagator. The connection with the classical system
    is achieved through an exact form of ``Egorov's theorem":
    \[U_N(A)^{-1}\Op_N(f)U_N(A)=\Op_N(f\circ A),\quad \forall f\in C^\infty(\bbT^2).\]
\subsection*{Quantum Ergodicity}
    When the matrix $A$ has no eigenvalues that are roots of unity, the
    induced classical dynamics is ergodic and mixing.
    The quantum analog of this, following the correspondence principle,
    is that the expectation values of an observable, $\langle
    \Op_N(f)\psi,\psi\rangle$
    (in an eigenfunction $\psi$ s.t $U_N(A)\psi=\lambda\psi$),
    should tend to the phase space average of the
    observable in the semiclassical limit.

    By an analog of ~Shnirelman's theorem, one can show that indeed
    almost all of these matrix elements converge to the phase space average ~\cite{BD}.
    This notion is usually referred to as ``Quantum Ergodicity"
    (QE), and was shown to hold for a large class of ergodic dynamical systems ~\cite{BD,DD,S,Z}.
    However, the stronger notion of ``Quantum Unique Ergodicity"
    (QUE), where there are no exceptional subsequences of eigenfunctions, doesn't hold for this model.
    Indeed, in \cite{FND} ~Faure, ~Nonnenmacher and De Bi\`{e}vre managed to construct a subsequence of eigenfunctions,
    for which the diagonal matrix elements do not converge to the phase space average but concentrate around a periodic orbit.
    Such exceptional subsequences are also referred to as scars.

    \subsection*{Arithmetic Quantum Unique Ergodicity}
    The existence of scars for the quantum cat map, is related to
    high degeneracies in the spectrum of the quantum propagator.
    If we denote by $\ord(A,N)$ the smallest integer such that
    $A^s\equiv I\pmod{N}$, then the quantum propagator satisfies that $U_N(A)^{\ord(A,N)}=I$, implying spectral
    degeneracies of order $\frac{N}{\ord(A,N)}$. In particular,
    since there are infinitely many values of $N$ for which $\ord(A,N)$ is of order $\log(N)$,
    there could be spectral degeneracies of order
    $\frac{N}{\log(N)}$. It is precisely for these values of $N$,
    that the scars in \cite{FND} were constructed.

    In \cite{KR2} Kurlberg and Rudnick introduced a group of symmetries of the system, i.e., commuting unitary operators that commute with
    $U_N(A)$, that remove most of the spectral degeneracies.
    These operators are called Hecke operators, in an analogy to a similar setup on the modular surface \cite{Iw,RS}.
    The space $\calH_N$ has an orthonormal basis consisting of
    joint eigenfunctions  called ``Hecke
    eigenfunctions''. For the desymmetrized system, Kurlberg and Rudnick
    showed that indeed
    $\langle\Op_N(f)\psi,\psi\rangle\stackrel{N\rightarrow\infty}{\longrightarrow}\int_{\bbT^2}f$,
    for any sequence, $\psi=\psi^{(N)}$, of ``Hecke eigenfunctions" \cite{KR2}.
    This notion is referred to as
    Arithmetic Quantum Unique Ergodicity, due to the arithmetic nature of these Hecke operators (both here and in the setting on the modular
    surface).

\subsection*{Higher dimensions}
    The Hannay-Berry model for the quantum cat map, can be
    naturally generalized for symplectic linear automorphisms of higher dimensional
    tori.
    For quantizing maps on the $2d$-dimensional torus, the Hilbert space of states,
    $\calH_N=L^2(\bbZ/N\bbZ)^d$, is of dimension $N^d$ (where,
    again, $N$ stands for the inverse of Planck's constant).
    The group of quantizable elements is the subgroup
    $\Sp_\theta(2d,\bbZ)$ defined
    \[\Sp_\theta(2d,\bbZ):=\set{\begin{pmatrix}
      E & F \\
      G & H \
    \end{pmatrix}\in\Sp(2d,\bbZ)\big|EF^t,GH^t \mbox{ are even matrices}}.\]
    The quantization of observables $f\in C^{\infty}(\bbT^{2d})$ and maps
    $A\in\Sp_\theta(2d,\bbZ)$, again satisfy ``exact Egorov":
    \[U_N(A)^{-1}\Op_N(f)U_N(A)=\Op_N(f\circ A),\quad \forall f\in C^\infty(\bbT^{2d}).\]

    Many of the results obtained on the two
    dimensional model (i.e, $d=1$), can be naturally generalized to higher
    dimensions. Nevertheless,
    there are still some new and surprising phenomena that occur in higher dimensions.

    \subsection*{Results}
    One new phenomenon that occur in high dimensions, is the
    existence of super-scars, that is, joint
    eigenfunctions of the propagator and all the Hecke
    operators localized on certain invariant manifolds
    \footnote{The name super-scars has been used before in a different context \cite{BoS}}.
    \begin{rem}
    The scars constructed in \cite{FND} (for $d=1$) are related to the large spectral degeneracies of the propagator.
    We note that for $d>1$, there are values of $N$ for which the order $\ord(A,N)$ could
    grow like $N$ (whenever the characteristic polynomial for $A$ splits modulo $N$) and possibly even slower
    (see \cite{RSO} for some numerical data on the order of $A$ modulo $N$).
    Consequently, for these values there are large spectral degeneracies
    of order $N^{d-1}$. However, the scarring described here
    is not related to these degeneracies. In fact, the action of the Hecke
    operators reduce almost all of the spectral degeneracies (see proposition \ref{pHES:1}).
    \end{rem}

    Let $A\in\Sp_\theta(2d,\bbZ)$ be a quantizable symplectic map.
    To any invariant rational isotropic
    subspace $E_0\subseteq \Q^{2d}$, we assign a manifold $X_0\subseteq \bbT^{2d}$
    of dimension $2d-\dim E_0$, invariant under the
    dynamics.
    \begin{thm}\label{tSCARS:1}
    Let $A\in\Sp_\theta(2d,\bbZ)$ with distinct eigenvalues.
    Let $E_0\subseteq \Q^{2d}$, be an invariant subspace that is isotropic with
    respect to the symplectic form. Then, there is a subsequence of
    Hecke eigenfunctions $\psi\in \calH_{N_i}$, such that the
    corresponding distributions
    \[f\mapsto \langle \Op_{N_i}(f)\psi,\psi\rangle,\]
    converge to Lebesgue measure on the manifold $X_0$.
    \end{thm}

    To illustrate this phenomenon, consider the following simple example
    (previously presented by Gurevich \cite{G} and by Nonnenmacher \cite{No}).
    Let $\tilde{A}\in\GL(d,\bbZ)$ and take $A=\begin{pmatrix} \tilde{A}^t & 0\\
    0 & \tilde{A}^{-1}\end{pmatrix}\in\Sp(2d,\bbZ)$.
    Then the space $E_0=\set{(\vec{n}_1,0)\in\Q^{2d}}$ is an invariant isotropic subspace,
    and the corresponding invariant manifold is $X_0=\set{\begin{pmatrix} 0\\ \vec{p}\end{pmatrix}\in\bbT^{2d}}$.
    The action of the quantum propagator
    corresponding to such a matrix is given by the formula
    $U_N(A)\psi(\vec{x})=\psi(\tilde{A}\vec{x})$
    (where the action of $\tilde{A}\in\GL(d,\bbZ)$ on $\vec{x}\in (\bbZ/N\bbZ)^d$ is the obvious one).
    One can then easily verify that the function $\psi_0(\vec{x})=\sqrt{N}\delta_0(\vec{x})$ is an eigenfunction
    of $U_N(A)$. On the other hand for any $f\in C^\infty(\bbT^{2d})$ a simple computation
    gives $\langle \Op(f)\psi_0,\psi_0\rangle=\int_{X_0}fdm_{X_0}$, that is the distribution $f\mapsto \langle
    \Op(f)\psi_0,\psi_0\rangle$ is Lebesgue measure on $X_0$.

    Theorem \ref{tSCARS:1} implies that any matrix $A\in\Sp_\theta(2d,\bbZ)$ that has a
    rational invariant isotropic subspace is \textbf{not} arithmetically QUE.
    We show that these are the only counter examples.

    \begin{thm}\label{tQUE:1}
    Let $A\in\Sp_\theta(2d,\bbZ)$ be a matrix with distinct eigenvalues.
    Then, a necessary and sufficient condition for the induced
    system to be Arithmetically QUE, is that there are no
    rational subspaces $E\subseteq \Q^{2d}$, that are invariant
    under the action of $A$, and are isotropic with respect to
    the symplectic form.
    \end{thm}

    \begin{rem}
    Note that the existence of a rational invariant isotropic subspace, is
    equivalent to the existence of an isotropic closed connected invariant subgroup of the torus.
    We can thus reformulate this theorem in
    these terms, i.e., the condition for Arithmetic QUE, is the
    absence of invariant isotropic sub-tori.
    \end{rem}
   \begin{rem}
    It is interesting to note, that the sufficient conditions to insure Arithmetic
    QUE, do not rule out matrices that
    have roots of unity for eigenvalues. So in a sense, Arithmetic
    QUE can hold also for matrices that are not classically ergodic.
    This phenomenon already occurs for matrices in
    $\SL(2,\bbZ)$, for example $A=\begin{pmatrix}
      1 & 1 \\
      -2 & -1 \
    \end{pmatrix}$ is not ergodic (because $A^4=I$), nevertheless it has two distinct eigenvalues
    and no rational invariant subspaces, hence arithmetic QUE
    does hold for this matrix.
    \end{rem}

    For systems that are arithmetically QUE, we can also give a bound on the rate of convergence.
    For $\vec{n}\in\bbZ^{2d}$ we denote by $2d_{\vec{n}}$
    the dimension of the smallest (symplectic) invariant subspace
    $E\subseteq\Q^{2d}$ such that $\vec{n}\in E$. For a
    smooth observable $f\in C^\infty(\bbT^{2d})$ define $d(f)=\min_{\hat{f}(\vec{n})\neq 0} d_{\vec{n}}$.
    \begin{thm}\label{tQUE:2}
    In the case where there are no rational isotropic subspaces, for any smooth $f\in C^\infty(\bbT^{2d})$ and any normalized Hecke eigenfunction
    $\psi\in\calH_N$, the expectation values of $\Op_N(f)$ satisfy:
    \[|\langle
    \Op_N(f)\psi,\psi\rangle-\int_{\bbT^{2d}}f|\ll_{f,\epsilon}
    N^{-\frac{d(f)}{4}+\epsilon}.\]
    \end{thm}
    \begin{rem}
    The exponent of $\frac{d(f)}{4}$ in this theorem is not
    optimal. The correct exponent is probably
    $\frac{d(f)}{2}$, in consistence with the fourth
    moments (proposition \ref{pFOURTH})
    and with the bounds for prime $N$ (corollary \ref{cBOUND}).
    For $N$ prime, in the case where there are no invariant rational subspaces the bound
    $O(N^{-d/2})$ was independently proved by Gurevich and Hadani \cite{GH2}.
    \end{rem}

    We note that the behavior of the matrix elements of an observable $\Op_N(f)$, is related to the decomposition of $N$ to its prime factors.
    Consequently, if we restrict ourselves to the case where $N$ is prime, we can obtain much sharper results
    (e.g., for the bounds on the number of Hecke operators and the dimension of the joint eigenspaces).

    We now consider only prime $N$, and restrict to the case where
    there are no isotropic invariant rational subspaces. In this
    case the matrix elements of a smooth observable $f\in C^\infty(\bbT^{2d})$ with respect to a Hecke basis $\{\psi_i\}$
    converge to their average $\int_{\bbT^{2d}}f$ and fluctuate around it. To study these fluctuations,
    we first give an asymptotic formula for their variance:
    \[S_2^{(N)}(f)=\frac{1}{N^d}\sum_{i} |\langle\Op_N(f)\psi_i,\psi_i\rangle-\int f dx|^2.\]

    Consider the decomposition $\Q^{2d}=\bigoplus E_\theta$, into
    symplectic irreducible invariant subspaces.
    To each space, we assign a quadratic form
    $Q_\theta:\bbZ^{2d}\rightarrow \bbZ[\lambda_\theta]$, where
    $\lambda_\theta$ is an eigenvalue of the restriction of $A$ to
    $E_\theta$, and define the product $Q=\prod Q_\theta$ (see section \ref{sVAR} for an explicit construction).
    For a smooth observable $f\in C^\infty(\bbT^{2d})$ and an element
    $\nu\in\prod\bbZ[\lambda_\theta]$, define modified Fourier
    coefficients
    \[f^\sharp(\nu)=\sum_{Q(\vec{n})=\nu}(-1)^{\vec{n}_1\vec{n}_2}\hat{f}(\vec{n}).\]
    Define $d_\nu=\frac{1}{2}\sum_{\nu_\theta\neq 0}\dim E_\theta$ and  $d_f=\min_{f^\sharp(\nu)\neq 0}
    d_\nu$.
    Note that if $\nu=Q(\vec{n})$, then $d_\nu=d_{\vec{n}}$ as
    defined in theorem \ref{tQUE:2}, hence for any smooth $f$ we have $d(f)\leq d_f$.
    For $f\in\C^\infty(\bbT^{2d})$ define $V(f)=\sum_{d_\nu=d_f}
    |f^\sharp(\nu)|^2$.

    \begin{thm}\label{tVAR:1}
    In the case where there are no rational isotropic subspaces, for a smooth observable $f\in
    C^\infty(\bbT^{2d})$, as $N\rightarrow \infty$ through primes,
    the quantum variance in the Hecke basis satisfies
    \[S_2^{(N)}(f)=\frac{V(f)}{N^{d_f}}+O(\frac{1}{N^{d_f+1}}).\]
    \end{thm}

    \begin{rem}
    We note that when there are symplectic invariant rational subspaces,
    one can construct observables for which $d_f<d$. We can
    thus produce a large family of examples (similar to the ones we described in \cite{K1}),
    for which the quantum variance is of a different
    order of magnitude from the one predicted for generic systems by the ~Feingold-~Peres formula \cite{EFK1,FP}.
    \end{rem}
    \begin{rem}
    In the case that there are isotropic invariant rational
    subspaces, the distribution can become degenerate (see remark
    \ref{rNONSYM}) and there is no definite behavior for the variance.
    \end{rem}

    After establishing the quantum variance,
    we renormalize to have finite variance $V(f)$ and give a conjecture for the
    limiting distribution, generalizing the Kurlberg-Rudnick conjecture for the two dimensional case \cite{KR1}.
    To simplify the discussion, we will restrict
    ourselves to elementary observables of the form
    $e_{\vec{n}}(\vec{x})=\exp(2\pi i\vec{n}\cdot\vec{x})$
    (see section \ref{sDIST} for treatment of any smooth observables).

    For an observable $\Op_N(e_{\vec{n}})$,
    the matrix elements in the Hecke basis can be expressed as a product of certain exponential sums.
    The sums in the product are of the form:
    \[E_q(\nu,\chi)=\frac{1}{|\calC|}\sum_{1\neq x\in\calC} e_{q}(\nu\kappa\frac{x+1}{x-1} )\chi(x)\chi_2(x),\]
    where $q$ is some power of $N$, $\calC$ is either the multiplicative group $\FF{q}^*$ or the
    group of norm one elements in the quadratic
    extension $\FF{q^2}/\FF{q}$, $\chi$ is a character of $\calC$ and $\chi_2$ is the quadratic character of $\calC$,
    $\nu\in\FF{q}$ and $\kappa\in\FF{q^2}$
    satisfies: $\forall x\in\calC,\;\kappa\frac{x+1}{x-1}\in\FF{q}$.

    The Kurlberg-Rudnick conjecture regarding the limit
    distribution \cite{KR1}, is naturally generalized to a conjecture regarding these
    exponential sums.
    \begin{conj}\label{cLIMIT:2}
    For each finite field $\FF{q}$, fix an element $0\neq\nu\in \FF{q}$ and consider the set of points on the line
    defined by the normalized exponential sums $\sqrt{q}E_q(\nu,\chi)$ for all characters $\chi:\calC\rightarrow\C^*$.
    Then, as
    $q\rightarrow\infty$ these points become equidistributed on the interval $[-2,2]$ with
    respect to the Sato-Tate measure. Furthermore, if for each field $\FF{q}$ we fix a number
    of distinct elements $\nu_1,\ldots,\nu_r\in \FF{q}$,
    then the limiting distributions corresponding to
    $\sqrt{q}E_q(\nu_1,\chi),\ldots,\sqrt{q}E_q(\nu_r,\chi)$
    are that of $r$ independent random variables.
    \end{conj}

    We now wish to deduce from this, a conjecture regarding the limiting distribution of the matrix elements.
    However, to do this we need to consider the decomposition of
    $\FF{N}^{2d}$ into invariant subspaces under the action of
    $A\pmod{N}$ (rather than the decomposition of $\bbQ^{2d}$ we used for the variance).
    For $\vec{n}\in\bbZ^{2d}$, let $E\subset \FF{N}^{2d}$ be the smallest (symplectic)
    invariant subspace containing
    $\vec{n}\pmod{N}$. Let $E=\bigoplus E_\bthet$ be the decomposition of $E$ into irreducible symplectic
    invariant subspaces, and let $2d_\bthet=\dim E_\bthet$.
    Then a matrix element for a Hecke eigenfunction $\langle\Op_N(e_{\vec{n}})\psi,\psi\rangle$,
    can be expressed as the product
    $\prod_{\bthet} E_{q_\bthet}(\nu_\bthet,\chi_\bthet)$,
    where $q_\bthet=N^{d_\bthet}$, the elements $\nu_\bthet$ are determined by the
    projections of $\vec{n}\pmod{N}$ to $E_\bthet$, and the
    characters $\chi_\bthet$ are determined by the eigenfunction.
    Consequently, if we denote by $\mathbf{P}_k$ the set of primes
    for which there are precisely $k$ invariant subspaces
    $E_\bthet$ in the decomposition we can deduce:
    \begin{conj}\label{cnLIMIT:1}
    As $N\rightarrow\infty$ through primes from $\mathbf{P}_k$,
    the limiting distribution of normalized matrix elements ${N}^{d_{\vec{n}}/2}\langle
    \Op_N(e_{\vec{n}})\psi_i,\psi_i\rangle$,
    is that of a product of $k$ independent random variables, each
    obeying the semi-circle law.
    \end{conj}

    It is interesting, that while the expression for the variance
    depends only on the rational properties of $A$, the limiting
    distribution already depends specifically on the action of $A$
    on $\FF{N}^{2d}$, and can vary for different values of (prime) $N$.
    Moreover, notice that at least one of the sets $\mathbf{P}_k$ is
    always infinite, so there is a sequence of primes for which there is a limiting distribution.
    However, there could be other values of $k$ for which the sets $\mathbf{P}_k$ are also infinite,
    resulting in different limiting distributions (see section \ref{sDIST} for some examples).

    \subsection*{Outline}
    This work is composed of three main parts. In the first part (section \ref{sQUANT}),
    we describe in detail the quantization procedure.  In the second part (sections \ref{sHECKE},
    \ref{sAQUE}), we develop Hecke theory and give the proof of theorem
    \ref{tQUE:2}. In the third part (sections \ref{sHECKP},\ref{sSCARS},\ref{sVAR},\ref{sDIST}), we restrict
    the discussion to the case where Planck's constant is an inverse of a prime
    number. For these values of Planck's constant the Hecke
    operators and eigenfunctions reveal structure closely related to the Weil representation over finite fields.
    We use this structure to construct scars proving theorem
    \ref{tSCARS:1}, and to calculate the quantum variance proving
    theorem \ref{tVAR:1}. We then generalize the Kurlberg-Rudnick
    conjecture, regarding the limiting distribution of (normalized) matrix
    elements, to deal with higher dimensional tori.

    \section*{Acknowledgments}
    I warmly thank my Ph.D. advisor Zeev Rudnick for introducing me to
    this subject and for his guidance throughout this project. I thank
    Par Kurlberg for long discussions of his work. I thank Shamgar
    Gurevich and Ronny Hadani for patiently explaining their work and
    for their stimulating suggestions. I also thank St\'{e}phane
    Nonnenmacher for his helpful comments. I thank my friends Lior
    Rosenzweig and Lior Bary-Soroker for our discussions. This work
    was supported in part by the Israel Science Foundation founded by
    the Israel Academy of Sciences and Humanities. This work was
    carried out as part of the author's Ph.D. thesis at Tel Aviv
    University, under the supervision of Prof. Zeev Rudnick.

\section{Quantized Linear Toral Automorphisms}\label{sQUANT}
    The quantization of the cat map on the $2$-torus, was originally introduced by Hannay and Berry \cite{HB},
    and is further described in \cite{DEG,Knabe,KR2}. For higher
    dimensions, the procedure is mostly analogous, and is described in
    \cite{BD1,RSO}. We take an approach towards the quantization procedure through
    representation theory, similar to the one taken in \cite{KR2}.
\subsection{Quantization procedure}\label{sQUANT:1}
    We start by giving the outline for the quantization of arbitrary symplectic maps.
    For a discrete time dynamical system, given by the iteration of a symplectic map $A$
    on a phase space $X$, the quantization procedure can be described as follows:
    The first step,  is constructing a one parameter family of Hilbert spaces $\calH_h$,
    parameterized by Planck's constant. For each space, there is
    a procedure that assigns to each smooth function $f\in C^\infty(X)$, an operator $\Op_h(f)$ acting on $\calH_h$.
    The connection with the classical system is fulfilled by the requirement that in the limit $h\rightarrow 0$,
    the commutator of the quantization of two observables $f,g$ reproduce the quantization of their Poisson bracket
    $\{f,g\}=\sum_j(\partial f/\partial p_j)(\partial g/\partial q_j)-(\partial f/\partial q_j)(\partial g/\partial p_j)$:
    \begin{equation}\label{eCP}
        \norm{\frac{1}{i\hbar}[\Op_h(f),\Op_h(g)]-\Op_h(\{f,g\})}\stackrel{h\rightarrow 0}{\longrightarrow} 0.
    \end{equation}
   The dynamical part of the quantization, is given by discrete time evolution of the algebra of operators.
   The evolution is through conjugation by a unitary map $U_h(A)$ of $\calH_h$ (referred to as the quantum propagator).
   We require that in the limit $h\rightarrow 0$ the classical dynamics is reproduced, in the sense that
   \begin{equation}\label{eEGOROV:1}
   \norm{U_h(A)^{-1}\Op_h(f)U_h(A)-\Op_h(f\circ A)}\stackrel{h\rightarrow 0}{\longrightarrow} 0.
   \end{equation}

   In our case, the classical phase space is the multidimensional torus
   and the classical observables are smooth function
   on the torus. For quantizing the torus, the admissible values of
   Planck's constant are inverses of integers $h=1/N,\;N\geq 1$.
   The space of states, is $\calH_N=L^2((\bbZ/N\bbZ)^d)$ of dimension $N^d$
   with inner product given by $\langle \psi,\phi\rangle=\frac{1}{N^{d}}\sum_{\vec{x}\pmod{N}}\psi(\vec{x})\overline{\phi(\vec{x})}$.
   To each observable $f\in C^\infty(\bbT^{2d})$, by an analog of ~Weyl quantization, we assign an
   operator $\Op_N(f)$ satisfying (\ref{eCP}). The classical
   dynamics is given by an iteration of a symplectic linear map
   $A\in \Sp(2d,\bbZ)$ acting on the torus, so that $\vec{x}=(\begin{array}{c}\vec{p}\\\vec{q}\end{array})\in\bbT^{2d}\mapsto A\vec{x}$ is a symplectic map of the torus. Given an observable
   $f\in C^{\infty}(\bbT^{2d})$, the classical evolution is
   defined by $f\mapsto f\circ A$. For a certain subset of matrices $A$, there is a
   unitary operator $U_N(A)$ acting on $\calH_N$ satisfying an
   exact form of (\ref{eEGOROV:1}), i.e.,
   \begin{equation}\label{eEGOROV:2}
   U_N(A)^{-1}\Op_N(f)U_N(A)=\Op_N(f\circ A).
   \end{equation}
   We now turn to describe these procedures in more detail.

 \subsubsection{Quantizing observables}
   In an analogous way to the quantization of observables on $\bbT^2$ \cite{HB,KR2}, introduce elementary
   operators $\TT{N}(\vec{n})$ (with $\vec{n}=(\vec{n}_1,\vec{n}_2)\in\bbZ^{2d}$), acting
   on $\psi\in \mathcal{H}_N$ via:
   \begin{equation}\label{eTN}
   \TT{N}(\vec{n})\psi(\vec{y})=e_{2N}(\vec{n}_1\cdot\vec{n}_2)e_N(\vec{n}_2\cdot\vec{y})\psi(\vec{y}+\vec{n}_1^t),
   \end{equation}
   where we use the notation $e_N(x)=e^{\frac{2\pi i x}{N}}$.
   For notational convenience we also define a twisted version of these
   operators:
   $$\T{N}(\vec{n}):=(-1)^{N\vec{n}_1\cdot\vec{n}_2}\TT{N}(\vec{n}).$$
   \begin{rem}
   The twisted operators were originally introduced in \cite{G}, and make some of the
   arguments simpler (e.g., the trace formula (\ref{eTR:1})).
   Moreover, these operators satisfy the intertwining equation
   (\ref{eEG}) for all of the symplectic group rather than for the subgroup $\Sp_\theta(2d,\bbZ)$.
   \end{rem}

   The main properties of the twisted elementary operators $\T{N}(\vec{n})$
   are summarized in the following proposition.
   \begin{prop}\label{pTN:1}
     For the operators $\T{N}(\vec{n})$ defined above:
     \begin{enumerate}
     \item $\T{N}(\vec{n})^*=\T{N}(-\vec{n})=\T{N}(\vec{n})^{-1}$ are unitary operators.
     \item The composition of two elementary operators is given by
       \[\T{N}(\vec{m}) \T{N}(\vec{n})=e_{2N}((1+N^2)\omega(\vec{m},\vec{n})) \T{N}(\vec{m}+\vec{n}),\]
       implying commutation relation
       \[\T{N}(\vec{m}) \T{N}(\vec{n})=e_N(\omega(\vec{m},\vec{n})) \T{N}(\vec{n})\T{N}(\vec{m}).\]
       where $\omega(\vec{m},\vec{n})=\vec{m}_1\cdot\vec{n}_2-\vec{m}_2\cdot\vec{n}_1$ is the symplectic
       inner product.
     \item For even $N$, $\T{N}(\vec{n})$ only depends on $\vec{n}$  modulo
     $2N$, while for odd $N$ it only depends on $\vec{n}$ modulo $N$.
     \end{enumerate}
   \end{prop}
   The proof is straightforward from (\ref{eTN}).

   For any smooth classical observable $f\in
   C^\infty(\mathbb{T}^{2d})$ with Fourier expansion
   $f(\vec{x})=\sum_{\vec{n}\in\bbZ^{2d}}\hat{f}(\vec{n})\exp(2\pi i\vec{n}\cdot\vec{x})$, where $\vec{n}\cdot\vec{x}=\vec{n}_1\cdot\vec{p}+\vec{n}_2\cdot\vec{q}$,
   define its quantization by
   \[\Op_N(f):=\sum_{\vec{n}\in\bbZ^{2d}}\hat{f}(\vec{n})\TT{N}(\vec{n}),\]
   or alternatively in terms of the twisted operators
   \[\Op_N(f)=\sum_{\vec{n}\in\bbZ^{2d}}\hat{f}(\vec{n})(-1)^{N\vec{n}_1\cdot\vec{n}_2}\T{N}(\vec{n}).\]
   Using the commutation relation given above, and the rapid decay of the fourier coefficients,
   relation (\ref{eCP}) can be verified.

   \subsubsection{The Heisenberg group}\label{sHEIS}
   The operators $\T{N}(\vec{n})$ defined above,
   are connected to a certain
   representation of a Heisenberg group $H_N$.

    For $N\geq 1$ the corresponding Heisenberg group is taken to be
    \[H_N=\set{(\vec{n},t)|\vec{n}\in (\bbZ/2N\bbZ)^{2d}\;,t\in
    \bbZ/2N\bbZ},\]
    with a multiplication law given by
    \[(\vec{n},t)\cdot(\vec{n}',t')=(\vec{n}+\vec{n}',t+t'+\omega(\vec{n},\vec{n}')).\]
    It is easily verified that the center of this group is given by
    \[Z(H_N)=\set{(\vec{n},t)\in H_N|\vec{n}\equiv 0\pmod{N}}.\]

    We now construct a unitary representation of $H_N$ on the
    space $\calH_N=L^2((\Z/N\Z)^d)$ by setting:
    \[\pi(\vec{n},t)=e_{2N}((N^2+1)t)\T{N}(\vec{n}).\]
    The relations given in proposition \ref{pTN:1}, insure that this is
    indeed a representation.
    Furthermore, the center of $H_N$ acts through the character
    $\xi(\vec{n},t)=e_{2N}((N^2+1)t)$.
    \begin{rem}
    This representation can be realized as an induced representation
    from the one dimensional representation of the normal subgroup
    $\set{(\vec{n},t)|\vec{n}_2=0\pmod{N}}$,
    given by $(\vec{n},t)\mapsto e_{2N}((N+1)t)$ for odd $N$ and $(\vec{n},t)\mapsto e_{2N}(t+\vec{n}_1\vec{n}_2)$ for even $N$.
    \end{rem}

    \begin{prop}\label{pHIZ:1}
      Let $\pi$ be a representation of the Heisenberg group, which is given by $\xi$ on
      the center (where $\xi$ is the character defined above), then:
      \begin{itemize}
      \item The characters of the representation $\pi$, are supported on the center.
      \item $\pi$ is irreducible if and only if the dimension of
    the representation is $N^d$.
    In this case, the class of the representation $\eta$ is
    determined by the character $\xi$.
      \end{itemize}
    \end{prop}
    \begin{proof}
    See \cite[lemma 1.2]{Ge}.
    \end{proof}
    In our case, the dimension $\dim(\pi)=\dim(\calH_N)=N^d$, and hence the representation $\pi$ is irreducible.
    Furthermore, from the condition on the characters of $\pi$, we deduce that
    the trace of the elementary operators $\T{N}(\vec{n})$ is given by
    \begin{equation}\label{eTR:1}
    \Tr(\T{N}(\vec{n}))=\left\lbrace
       \begin{array}{c l}
         N^d  & \vec{n}\equiv 0\pmod{N}\\
         0 & \mbox{otherwise}
       \end{array}
       \right.
    \end{equation}
    In particular, for fixed $\vec{n}\neq 0$ and sufficiently large $N$,
    the trace of $\T{N}(\vec{n})$ vanishes.
    \begin{cor}
    For any orthonormal basis for $\calH_N$, and any smooth observable $f\in C^\infty(\bbT^{2d})$, the average of the
    diagonal matrix elements of $\Op_N(f)$ converge to the phase
    space average as $N\rightarrow \infty$.
    \end{cor}

 \subsubsection{Quantizing maps}\label{sREP}
   In this section we show how to assign to a symplectic linear
   map $A\in \Sp_\theta(2d,\Z)$ acting on $\bbT^{2d}$, a unitary operator $U_N(A)$ acting on
   $L^2((\Z/N\Z)^d)$ s.t for all observables
   $f\in C^\infty(\bbT^{2d})$,
   \[U_N(A)^{-1}\Op_N(f)U_N(A)=\Op(f\circ A).\]

    Any symplectic matrix $A\in\Sp(2d,\bbZ)$, naturally acts on $H_N$ by automorphism via
    $(\vec{n},t)^A=(\vec{n}A,t)$. Composing
    the representation $\pi$ with the action of $A$, thus gives a new
    representation $\pi^A(\vec{n},t)=\pi(\vec{n}A,t)$, that is again
    irreducible and acts on the center through the same character
    $\xi(\vec{n},t)=e_{2N}((1+N^2)t)$.

    Therefore by proposition \ref{pHIZ:1}, for any $A\in\Sp(2d,\Z)$ the representations $\pi,\pi^A$ are unitarily
    equivalent, i.e., there is a unitary intertwining operator $U_N(A)$
    satisfying
    \[\pi^A(\vec{n},t)=U_N(A)^{-1}\pi(\vec{n},t)U_N(A),\quad\forall (\vec{n},t)\in H_N,\]
    and in particular $\forall \vec{n}\in \bbZ^{2d}$
    \[U_N(A)^{-1}\T{N}(\vec{n})U_N(A)=\T{N}(\vec{n}A).\]

    Assume now that in addition $A$ belongs to the subgroup
    \[\Sp_\theta(2d,\bbZ)=\set{\begin{pmatrix}
      E & F \\
      G & H \\
\end{pmatrix}\in\Sp(2d,\bbZ)\bigg|EF^t,GH^t, \mbox{ are even matrices}}.\]
    Then $\forall \vec{n}\in \bbZ^{2d}$, the image
    $\vec{m}=\vec{n}A$ satisfies
    $\vec{n}_1\cdot\vec{n}_2\equiv\vec{m}_1\cdot\vec{m}_2\pmod{2}$,
    hence for all observables $f\in
    C^\infty(\bbT^{2d})$,
    \[U_N(A)^{-1}\Op_N(f)U_N(A)=\Op(f\circ A).\]

    Because the operators $\T{N}(\vec{n})$, only
    depend on $\vec{n}$ modulo $2N$ (respectively modulo $N$ for odd $N$), the representation
    $\pi^A$ also depends only on $A\mod{2N}$ (respectively $\pmod{N}$). We can thus take the intertwining operator
    $U_N(A)$, to depend only on $A$ modulo $2N$ (respectively $N$).

    \begin{rem}
    Note that $U_N(A)$ is defined as an intertwining operator for any
    $A\in\Sp(2d,\bbZ)$. However, if $A\notin\Sp_\theta(2d,\bbZ)$
    then the operator $U_N(A)$
    no longer satisfies the Egorov identity. When restricting to
    the subgroup $\Sp_\theta(2d,\bbZ)$ the definition given here
    coincides with the standard definition given in \cite{KR2} (for $d=1$).
    \end{rem}

    \subsection{Formulas for the quantized cat map}\label{sQUANT:2}
    The irreducibility of $\pi$
    imply (through Schur's lemma), that the map $U_N(A)$ is unique up to multiplication by phase.
    In other words, if $U$ is a unitary map acting on $\calH_N$,
    satisfying the intertwining equation
    \begin{equation}\label{eEG}
    U\T{N}(\vec{n}A)=\T{N}(\vec{n})U,\quad \forall \vec{n}\in
    \bbZ^{2d},
    \end{equation}
    then after multiplying by some phase, $e^{i\alpha}U_N(A)=U$. On the other hand the
    contrary is also true, that is, if $U=e^{i\alpha}U_N(A)$,
    then it obviously satisfies (\ref{eEG}).

    In what follows, we give
    formulas for operators satisfying  (\ref{eEG}), thus obtaining formulas for the
    quantized maps.
    \subsubsection{Formulas through generators}\label{sFORM:1}
    The group $\Sp(2d,\bbZ)$ (and hence also $\Sp(2d,\bbZ/2N\bbZ)$) is generated by the family of matrices
    \[\begin{pmatrix} I & F \\ 0 & I \ \end{pmatrix},\quad
    \begin{pmatrix} E^t & 0 \\ 0 & E^{-1} \ \end{pmatrix},\quad
    \begin{pmatrix} 0 & I \\ -I & 0 \ \end{pmatrix},\]
    with $E\in \GL(d,\bbZ)$ and $F\in \Mat(d,\bbZ)$
    symmetric \cite[theorem 2]{Hua}.

    For these matrices the corresponding operators act by the following
    formulas (up to phase):
    \begin{equation}\label{eUN:1}
    U_N\begin{pmatrix} I & F \\ 0 & I \
    \end{pmatrix}\psi(\vec{x})=e_{2N}((1+N^2)\vec{x}\cdot F\vec{x})\psi(\vec{x})
    .\end{equation}

    \begin{equation}\label{eUN:2}
    U_N\begin{pmatrix} E^t & 0 \\ 0 & E^{-1} \ \end{pmatrix}\psi(\vec{x})=
    \psi(E\vec{x})
    .\end{equation}

    \begin{equation}\label{eUN:3}
    U_N\begin{pmatrix} 0 & I \\ -I & 0 \ \end{pmatrix}\psi(\vec{x})=
    \frac{1}{N^{d/2}}\sum_{\vec{y}\in
    (\Z/N\Z)^d}e_{N}(\vec{x}\cdot\vec{y})\psi(\vec{y})
    .\end{equation}
    One can verify directly that these formulas indeed satisfy
    (\ref{eEG}).
    Consequently, the action
    of any element $U_N(A),\;A\in\Sp(2d,\bbZ)$
    can be obtained, by composing the appropriate operators given above for the generators.

    \subsubsection{Formulas through averaging}\label{sFORM:2}
    A different approach to obtain formulas for the operators
    $U_N(A)$ is through averaging of the representation over the Heisenberg
    group (similar to the $p$-adic formula given in \cite[page 37]{Mo}).
    With this approach, for any $A\in\Sp(2d,\bbZ)$ satisfying $A\equiv\pm I
    \pmod{4}$, we obtain a formula for the propagator $U_N(A)$ in terms of the elementary operators
    $\T{N}(\vec{n})$. Moreover, if $N$ is odd the formula is valid with out the parity condition.

    Recall that we defined the operator $U_N(A)$ to be an
    intertwining operator of the representations $\pi$ and $\pi^A$. It is easily verified that an operator defined by
    averaging of the form
    \[F(\pi,\pi^A)=\sum_{h\in H_N/Z(H_N)}\pi(h)\pi^A(h^{-1}),\]
    is always an intertwining operator of these representations.
    Therefore, (by Schur's lemma) it will coincide
    with the original operator after multiplying by some constant (i.e.,
    $F(\pi,\pi^A)=c(A)U_N(A)$). Note that in general this constant might be zero.
    \begin{prop}\label{pUNA:1}
    Let $A\in Sp(2d,\bbZ)$ be a matrix satisfying $A\equiv -I\pmod{4}$. Denote by
    $\ker_N(A-I)$, the kernel of the map $(A-I):(\bbZ/N\bbZ)^{2d}\rightarrow (\bbZ/N\bbZ)^{2d}$.
    Then, the intertwining operator
    $F(\pi,\pi^A)=c(A)U_N(A)$ with
    $|c(A)|^2=N^{2d}|\ker_N(A-I)|$, and in particular $c(A)\neq 0$.
    \end{prop}
    \begin{proof}
    First, note that we can identify the quotient
    $H_N/Z(H_N)$ with $(\bbZ/N\bbZ)^{2d}$, so that
    \begin{equation}\label{eAVR}
    F(\pi,\pi^A)=\sum_{(\bbZ/N\bbZ)^{2d}}\T{N}(\vec{n})\T{N}(-\vec{n}A).
    \end{equation}
    Since the operator $U_N(A)$ is unitary,  $F(\pi,\pi^A)F(\pi,\pi^A)^*=|c(A)|^2I$.
    On the other hand, plugging in (\ref{eAVR}) gives,
    \[F(\pi,\pi^A)F(\pi,\pi^A)^*=\sum_{\vec{n},\vec{m}}\T{N}(\vec{n})\T{N}(-\vec{n}A)\T{N}(\vec{m}A)\T{N}(-\vec{m})=\]
    \[=\sum_{\vec{n},\vec{m}}e_N(\omega((\vec{n}-\vec{m})A,\vec{m}))\T{N}(\vec{n})\T{N}(-\vec{m})\T{N}(-\vec{n}A)\T{N}(\vec{m}A)=\]
    \[=\sum_{\vec{n},\vec{m}}e_N(\omega((\vec{n}-\vec{m})A,\vec{m})-\omega(\vec{n},\vec{m}))\T{N}(\vec{n}-\vec{m})\T{N}(-(\vec{n}-\vec{m})A).\]
    Now, change summation variable $\vec{k}=\vec{n}-\vec{m}$ to get
    \[F(\pi,\pi^A)F(\pi,\pi^A)^*=\sum_{\vec{k},\vec{m}}e_N(\omega(\vec{k}(A-I),\vec{m}))\T{N}(\vec{k})\T{N}(-\vec{k}A)=\]
    \[=\sum_{\vec{k}}\T{N}(\vec{k})\T{N}(-\vec{k}A)\sum_{\vec{m}}e_N(\omega(\vec{k}(A-I),\vec{m})).\]
    Since the second sum vanishes whenever $\vec{k}(A-I)\neq
    0\pmod{N}$, we get that
    \[F(\pi,\pi^A)F(\pi,\pi^A)^*=N^{2d}\sum_{\vec{k}\equiv\vec{k}A(N)}\T{N}(\vec{k})\T{N}(-\vec{k}A).\]
    Finally, when $A\equiv -I\pmod{4}$ the condition $\vec{k}\equiv\vec{k}A\pmod{N}$
    implies that $\T{N}(\vec{k})\T{N}(-\vec{k}A)=I$, which concludes the
    proof.
    \end{proof}
    When $A\equiv -I\pmod{4}$ the constant $c(A)$ does not vanish and we can divide by it to get a
    formula for $U_N(A)$:
    \begin{equation}\label{eUNA:1}
    U_N(A)=\frac{1}{c(A)}F(\pi,\pi^A),\quad (\forall
     A\equiv -I\pmod{4}).
    \end{equation}
    When $A\equiv I\pmod{4}$ the constant $c(A)$ might be zero.
    However, in this case $c(-A)\neq 0$ and since
    $U_N(A)=U_N(-A)U_N(-I)$ we get the formula:
    \begin{equation}\label{eUNA:2}
    U_N(A)=\frac{1}{c(-A)}F(\pi,\pi^{-A})U_N(-I),\quad (\forall
     A\equiv I\pmod{4}).
    \end{equation}

    \begin{rem}
    When $N$ is odd, the condition $\vec{k}\equiv\vec{k}A\pmod{N}$
    implies that $\T{N}(\vec{k})\T{N}(-\vec{k}A)=I$ for any
    $A\in\Sp(2d,\bbZ)$ (without the parity condition).
    Thus, for odd $N$ we can use both formulas for any symplectic matrix.
    \end{rem}

    From these formulas we get the following corollaries:
    \begin{cor} \label{cCOM:1}Let $A,B\in\Sp(2d,\bbZ)$ be matrices that commute modulo $N$.
    If $B\equiv \pm I\pmod{4}$ (or if $N$ is odd), then
    the corresponding operators $U_N(A)$,$U_N(B)$ commute as well.
    \end{cor}

    \begin{proof}
        If $B\equiv -I\pmod{4}$ (or if $N$ is odd), use formula (\ref{eUNA:1})
        for $U_N(B)$ and apply
        the intertwining equation (\ref{eEG}) for the action of $U_N(A)$.
        \[U_N(B)U_N(A)=U_N(A)\frac{1}{c(B)}\sum_{\vec{n}\in(\bbZ/N\bbZ)^{2d}}\T{N}(\vec{n}A)\T{N}(-\vec{n}BA).\]
        Now, change summation variable $\vec{n}\mapsto\vec{n}A$
        (using the fact that $A$ and $B$ commute), to get
        $U_N(B)U_N(A)=U_N(A)U_N(B)$.

        Otherwise, use formula (\ref{eUNA:2}) for $U_N(B)$.
        As above, the operators $F(\pi,\pi^{-B})$ and $U_N(-I)$ both commute with
        $U_N(A)$ and hence $U_N(B)$ commutes with $U_N(A)$ as well.
    \end{proof}
    \begin{cor}\label{cTR:1}
   The trace of $U_N(A)$ is given (up to phase) by:
    \begin{itemize}
        \item For $A\equiv -I\pmod{4}$ (or for odd $N$),
        \[|\Tr(U_N(A))|=\sqrt{|\ker_N(A-I)|}.\]
        \item For $N$ even, and $A\equiv I\pmod{4}$, either
        $\Tr(U_N(A))=0$ or
        \[|\Tr(U_N(A))|=
        \sqrt{\frac{|\ker_{2N}(A^2-I)|}{|\ker_N(A+I)|}}.\]
        \end{itemize}
        In particular $|\Tr(U_N(A))|\leq
        2^{d}\sqrt{|\ker_N(A-I)|}$.
    \end{cor}

    \begin{proof}
    In the first case, use formula (\ref{eUNA:1}) and take
    trace (note that $\Tr(\T{N}(\vec{n})\T{N}(-\vec{n}A))=0$ when $\vec{n}\neq
    \vec{n}A\pmod{N}$). Now, plug in $|c(A)|$ from proposition \ref{pUNA:1} to get the result.

    Otherwise, use formula (\ref{eUNA:2}).
    Using formula (\ref{eUNA:1}) for $U_N(-I)$ and taking trace we
    get that $\forall \vec{n}\in \bbZ^{2d}$,
    \[\Tr(\T{N}(\vec{n}(A+I))U_N(-I))=2^d.\]
     Therefore,
    \[|\Tr(U_N(A))|=\frac{2^d}{|c(-A)|}|\sum_{\vec{n}(N)}e_{2N}(\omega(\vec{n},\vec{n}A))|.\]
    Finally, similar to a Gauss sum, when the sum $\sum e_{2N}(\omega(\vec{n},\vec{n}A))$ does not vanish,
    its absolute value is given by
    \[|\sum_{\vec{n}(N)}e_{2N}(\omega(\vec{n},\vec{n}A))|=\frac{N^d\sqrt{|\ker_{2N}(A^2-I)|}}{2^d}.\]
    The bound $|\Tr(U_N(A))|\leq 2^{d}\sqrt{|\ker_N(A-I)|}$,
    is a consequence of the following observation,
    \[|\ker_{2N}(A^2-I)|\leq 2^{2d}|\ker_{N}(A^2-I)|\leq
    2^{2d}|\ker_{N}(A-I)||\ker_{N}(A+I)|.\]
    \end{proof}

    \subsection{Multiplicativity}\label{sQUANT:3}
    The quantum propagators, $U_N(A)$, are unique up to a phase
    factor and thus define a projective representation of
    $\Sp(2d,\bbZ/2N\bbZ)$, that
    is:
    \begin{equation}\label{eMUL:1}
     U_N(AB)=c(A,B)U_N(A)U_N(B).
    \end{equation}
    From corollary \ref{cCOM:1} we infer that: For odd $N$, if
    $AB=BA\pmod{N}$, then
    $c(A,B)=c(B,A)$ as well. For even $N$, this holds if $AB=BA\pmod{2N}$ and we restrict to the subgroup of matrices congruent
    to $\pm I$ modulo $4$. This property by itself already allows
    us to define the Hecke operators (see section \ref{sHECKE}). However, it is
    more convenient to work with a quantization such that the map $A\mapsto U_N(A)$ forms a
    representation of the symplectic group.
    In this section we show that such a quantization indeed exists:
    \begin{thm}\label{tMUL}
    For each $N>1$, there is a special choice of
    phases for the propagators, such that the map $A\mapsto U_N(A)$
    is a representation of  $\Sp(2d,\bbZ/N\bbZ)$ when $N$ is odd.
    Whereas for even integers, this map is a representation of the
    subgroup of $\Sp(2d,\bbZ/2N\bbZ)$ composed of all matrices congruent
    to $\pm I$ modulo $4$.
    \end{thm}

    In order to prove theorem \ref{tMUL} for all integers, it is sufficient to prove
    it separately for odd integers, and for integers of the form $N=2^k$ (see
    \cite[section 4.1]{KR2}).
    \subsubsection{Odd integers}
    When $N$ is an odd integer, we follow a proof of ~Neuhauser \cite{Ne}.
    As we apply this proof for the rings $\bbZ/N\bbZ$ (rather than finite fields as done in \cite{Ne}) we
    review the proof in some detail:

    Let $N\geq 1$ be an odd integer.
    Note that $-I$ is in the center of $\Sp(2d,\bbZ/N\bbZ)$, so by
    corollary \ref{cCOM:1}, $\forall A\in \Sp(2d,\bbZ/N\bbZ)$
    \[\quad U_N(-I)U_N(A)=U_N(A)U_N(-I).\]
    On the other hand, the operator $U_N(-I)$ acts by
    $U_N(-I)\psi(x)=\psi(-x)$ (formula \ref{eUN:2}). Hence, the space
    $\calH_N^+=\set{\psi\in \calH_N|\psi(-x)=\psi(x)}$,
    is an invariant subspace under the action of
    $\Sp(2d,\bbZ/N\bbZ)$.

    Denote by $U^+(A)$, the restriction of $U_N(A)$ to
    $\calH_N^+$, to get that
    \begin{equation}\label{eMUL:2}
    U^+(AB)=c(A,B)U^+(A)U^+(B).
    \end{equation}
    By taking determinants of equations (\ref{eMUL:1}) and (\ref{eMUL:2}) we get:
    \[\det(U_N(AB))=c(A,B)^{N^d}\det(U_N(A))\det(U_N(B)),\]
    \[\det(U^+(AB))=c(A,B)^{\frac{N^d+1}{2}}\det(U^+(A))\det(U^+(B)),\]
    (note that the dimension of $\calH^+$ is $\frac{N^d+1}{2}$).
    Define $\kappa(A)=\frac{\det(U_N(A))}{\det(U^+(A))^2}$, then
    $c(A,B)=\frac{\kappa(A)\kappa(B)}{\kappa(AB)}$, and
    $A\mapsto\kappa(A)U_N(A)$ is a representation of
    $\Sp(2d,\bbZ/N\bbZ)$.

    \subsubsection{Dyadic powers}
    For integers of the form $N=2^k$, we take a different
    approach by induction on the exponent $k$.

    We define a subspace $\calH_N^0\subset\calH_N$ of dimension
    $M^d=(N/2)^d$, invariant under the action of $\Sp_2(2d,2N)$ (i.e., the
    matrices congruent to $I$ modulo $2$) and under the action of certain elementary operators.
    We then construct a representation of the Heisenberg group
    $H_M$ on this space, and show that it is equivalent to the
    original representation on $L^2(\bbZ/M\bbZ)^d$. We can thus connect the
    restriction of the quantum propagators to the subspace $\calH_N^0$ with the quantum propagators on $\calH_M$,
    for which by induction we already have multiplicativity.

    Define the subspace
    \[\calH_N^0=\set{\psi\in \mathcal{H}_N| \psi(\vec{y})=0,\quad\forall \vec{y}\neq 0\pmod{2}},\]
    and the congruence subgroup
    $$\Sp_2(2d,2N)=\set{A\in\Sp(2d,\bbZ/2N\bbZ)|A\equiv I\pmod{2}}.$$
    \begin{lem}\label{lINV:1}
        For $N=2^k,\;k\geq 2$ and any $A\in \Sp_2(2d,2N)$,
        the space $\calH_N^0$ is invariant under the action of
        $U_N(A)$.
    \end{lem}
    \begin{proof}
        For any matrix
         $\left(%
        \begin{array}{cc}
        E & F \\
        G & H \\
        \end{array}\right)\in \Sp_2(2d,2N)$ we have a  Bruhat decomposition:
        \[\left(%
        \begin{array}{cc}
        E & F \\
        G & H \\
        \end{array}%
        \right)=\left(%
        \begin{array}{cc}
        {H^t}^{-1} & 0 \\
        0 & H \\
        \end{array}%
        \right)\left(%
        \begin{array}{cc}
        I & H^tF \\
        0 & I \\
        \end{array}%
        \right)\left(%
        \begin{array}{cc}
        I & 0 \\
        H^{-1}G & I \\
        \end{array}%
        \right).\]
        Consequently, the group
        $\Sp_2(2d,2N)$ is generated by the family of matrices
         \[u_+(X)=\left(%
        \begin{array}{cc}
        I & X \\
        0 & I \\
        \end{array}%
        \right),\;\;u_-(Y)=\left(%
        \begin{array}{cc}
        I & 0 \\
        Y & I \\
        \end{array}%
        \right),\;\;s(T)=\left(%
        \begin{array}{cc}
        T^t & 0 \\
        0 & T^{-1} \\
        \end{array}%
        \right)\]
        where $X,Y,T\in
        Mat(d,\Z/2N\Z)\;,X=X^t\;,Y=Y^t\;,\;X\equiv Y\equiv 0\;(\mod 2)\;,T\equiv I\;(\mod
        2)$.
        Therefore, it is sufficient to show that $\calH_N^0$ is
        invariant under the action of the corresponding operators.
        This can be done directly, using the formulas given in
        section \ref{sFORM:1}.
 \end{proof}

 \begin{lem} \label{lINV:2}
    For $N=2^k,\;k\geq 2$, the space $\calH_N^0$ is invariant under the action of
    $\T{N}(\vec{n})$ for all $\vec{n}=(\vec{n}_1,\vec{n}_2)$ such that $\vec{n}_1\equiv
    0\pmod{2}$.
    Furthermore, if $\vec{n}_1\equiv 0\pmod{N}$ and $\vec{n}_2\equiv 0\pmod{N/2}$ then the restriction
    $\T{N}(\vec{n})|_{\calH_N^0}=I$.
 \end{lem}
 \begin{proof}
 Direct computation using (\ref{eTN}).
 \end{proof}

    Define two subgroups of $\Sp_2(2d,2N)$,
    \[S_2(2N)=\set{\begin{pmatrix}
      E & F \\
      G & H \\
    \end{pmatrix}\in \Sp_2(2d,2N)\bigg|F\equiv0\pmod4}\]
    \[\hat{S}_2(2N)=\set{\begin{pmatrix}
      E & F \\
      G & H \\
    \end{pmatrix}\in \Sp_2(2d,2N)\bigg|G\equiv0\pmod4}\]
    Let $J=\begin{pmatrix}
      0 & I \\
      -I & 0 \\
    \end{pmatrix}$, then the map $A\mapsto -JAJ$ is an obvious isomorphism of these
    groups (in both directions). Another, less trivial isomorphism
    is given by the map $j:S_2\rightarrow\hat{S}_2$, defined by
 \begin{equation}\label{eS2}
   j(\begin{pmatrix}
      E & F \\
      G & H \\
    \end{pmatrix})= \begin{pmatrix}
      E & F/2 \\
      2G & H \\
    \end{pmatrix}.
 \end{equation}

 \begin{prop}
    For any $N=2^k$, there is a choice of phases so that for any $A,B\in S_2(2N)$,
    $U_N(AB)=U_N(A)U_N(B)$.\\
    There is another choice such that for any $A,B\in \hat{S}_2(2N)$,
    $U_N(AB)=U_N(A)U_N(B)$.
 \end{prop}
 \begin{proof}
    First note that it suffices to prove multiplicativity for
    $S_2(2N)$. Because, for any $B\in\hat{S}_2(2N)$
    there is $\tilde{B}\in S_2(2N)$ such that $B=-J\tilde{B}J$.
    Therefore, if we have multiplicativity for $S_2(2N)$, we can
    define for any $B\in \hat{S}_2(2N)$
    \[U_N(B)=U_N(J)^*U_N(\tilde{B})U_N(J),\]
    to get a multiplicativity for $\hat{S}_2(2N)$.

    We now show multiplicativity for $S_2(2N)$ by induction on k.
    For $k=1$, the group $S_2(4)$ includes only lower triangular matrices,
    for which the formulas given in \ref{sFORM:1} are multiplicative.

    For $k \geq 2$,
    by lemma \ref{lINV:1} the space $\calH_N^0$ is invariant under the action
    of $\Sp_2(2d,\bbZ)$ and hence also under the subgroup
    $S_2(2N)$. For $A\in S_2(2N)$ denote by
    $U_N^0(A)$ the restriction of $U_N(A)$ to $\calH_N^0$.

    Let $M=2^{k-1}=N/2$ and consider the Heisenberg group
    $H_{M}$ defined in section \ref{sHEIS}, together
    with the representation on $L^2(\bbZ/M\bbZ)$:
    \[\pi(\vec{n},t)=e_{2M}(t)\T{M}(\vec{n}).\]
    We now construct another representation on $\calH_N^0\subseteq
    L^2(\Z/N\Z)$:
    \[\tilde{\pi}(\vec{n},t)=e_N(t)\T{N}^0((2\vec{n}_1,\vec{n}_2)),\]
    where $\T{N}^0((2\vec{n}_1,\vec{n}_2))$ is the restriction
    of $\T{N}((2\vec{n}_1,\vec{n}_2))$ to $\calH_N^0$ (by lemma
    \ref{lINV:2} this is well defined).
    From the second part of lemma \ref{lINV:2} we see that the action on the center
    is given by
    $\tilde{\pi}(Mn,t)=e_N(t)I$.
    Consequently, by proposition \ref{pHIZ:1} there is a unitary
    operator $\calU:\calH_N^0\rightarrow L^2(\Z/M\Z)$ such that $\tilde{\pi}=\calU^{-1}\pi \calU$.

    The intertwining equation for $U_N(A)$, imply that the restricted operators satisfy
    \[U_N^0(A)^*\tilde{\pi}(n,t)U_N^0(A)=\tilde{\pi}(n j(A),t),\]
    where $j:S_2\rightarrow \hat{S}_2$ is the isomorphism defined in
    (\ref{eS2}). Consequently, $\calU U_N^0(A)\calU^{-1}$ is the intertwining
    operator between $\pi$ and $\pi^{j(A)}$,
    and by the uniqueness of the quantization we get:
    \[\calU U_N^0(A)\calU^{-1}=\kappa(A)U_M(j(A)).\]
    We can assume by induction that $A\mapsto U_M(A)$ restricted to $\hat{S}_2(2M)$ is
    multiplicative.
    Finally, for $A,B\in S_2(2N)$ we have
    $U_N(A)U_N(B)=c(A,B)U_N(AB)$, hence the restricted operators
    satisfy
    $U_N^0(A)U_N^0(A)=c(A,B)U_N^0(AB)$ as well.
    Conjugating by $\calU$ we get
    \[\kappa(A)U_M(j(A))\kappa(B)U_M(j(B))=c(A,B)\kappa(AB)U_M(j(AB)),\]
    implying $c(A,B)=\frac{\kappa(A)\kappa(B)}{\kappa(AB)}$.
    Therefore the map $A\mapsto \kappa(A)U_N(A)$ defined on
    $S_2(N)$, is multiplicative.

 \end{proof}
 Because the subgroup of matrices congruent to $\pm I$ modulo 4 is
 a subgroup of $S_2(N)$, this concludes the proof of theorem
 \ref{tMUL}.

\section{Hecke Theory}\label{sHECKE}
    In the following section we introduce Hecke theory for the
    multidimensional torus. For a given symplectic matrix
    $A\in\Sp_\theta(2d,\bbZ)$ with distinct eigenvalues, we follow the lines of \cite{KR2} and
    construct ``Hecke operators", a group of commuting operators that
    commute with the propagator $U_N(A)$. We show that this group of
    symmetries reduces almost all degeneracies in the spectrum.

    \begin{rem}
    The requirement that the matrix $A$ has distinct eigenvalues,
    is crucial for our construction. In fact when there are
    degenerate eigenvalues, the group of matrices commuting with
    $A$ modulo $N$ is not necessarily commutative. In such a case,
    it is not clear how one should define the Hecke group and Hecke operators.
    \end{rem}
    \begin{rem}
    In sections (\ref{sHECKE:2}) and (\ref{sHECKE:3}), in order to simplify the
    discussion, we will assume there are no rational isotropic subspaces
    invariant under the action of $A$.
    However, we note that the results presented
    in these sections (i.e., the bound on the number of Hecke
    operators in lemma \ref{lHECKE}
    and the dimensions of the joint eigenspaces in proposition \ref{pDIM}) are still valid with
    out this assumption, and the proofs are analogous.
    \end{rem}
\subsection{Hecke operators}\label{sHECKE:1}
    In \cite{KR2} Kurlberg and Rudnick constructed the Hecke
    operators (for $A\in\Sp(2,\bbZ)$) by identifying integral
    matrices with elements of the (commutative) integral ring of a certain
    quadratic extension of the rationals. We follow the same idea, except that for
    $A\in\Sp(2d,\bbZ)$ the correct ring to work with is the
    integral ring of a higher extension or rather a product of
    several such rings.

    Let $A\in\Sp_\theta(2d,\Z)$ with $2d$ distinct eigenvalues.
    Let $\{\lambda_i\}_{i=1}^{2d}$ be all of it's
    eigenvalues ordered so that $\lambda_{d+i}=\lambda_i^{-1}$.
    Denote by $\calD_i=\Z[\lambda_i]=\Z[\lambda_i^{-1}],\;i=1\ldots
    2d$, and
    define the ring
    \[\calD=\set{\beta=(\beta_1,\ldots,\beta_{2d})\in \prod_{i=1}^{2d}\calD_i|\exists f\in\Z[t],\;f(\lambda_i)=\beta_i}.\]

    This ring, is naturally isomorphic to the ring $\Z[t]/(P_A)$,
    where $P_A$ is the characteristic (and minimal) polynomial
    for $A$.
    Thus, there is an embedding $\iota:\calD\hookrightarrow \Mat(2d,\bbZ)$ (contained in the centralizer
    of $A$), given by
    \[\begin{array}{ccccc}
        \calD & \rightarrow & \Z[t]/(P_A) & \hookrightarrow & \Mat(2d,\Z) \\
        \beta & \mapsto & f & \mapsto & f(A) \
    \end{array}\]

    \begin{lem}\label{lSYMP:1}
    To any element $\beta=f(\lambda)\in\calD$ define an element
    $\beta^*\in\prod\calD_i$, such that
    $\beta^*_i=f(\lambda_i^{-1})$. Then, the map $\beta\mapsto \beta^*$ is an automorphism
    of $\calD$. Furthermore, to any $\vec{n},\vec{m}\in\bbZ^{2d}$ and any
    $\beta\in\calD$, the symplectic form $\omega$ satisfies:
    \[\omega(\vec{n}\iota(\beta),\vec{m})=\omega(\vec{n},\vec{m}\iota(\beta^*)).\]
    \end{lem}
    \begin{proof}
    The map $\beta\mapsto\beta^*$ is obviously injective, and it respects addition
    and multiplication. Therefore, to show that it is an automorphism it is
    sufficient to show that for any $\beta\in\calD$, $\beta^*\in\calD$ as well.

    Since $A$ is a symplectic
    map, the polynomial
    $h(t)=\frac{1-P_A(t)}{t}$ has integer coefficients. Therefore,
    for all $f\in\bbZ[t]$ the polynomial
    $g=f\circ h$ has integer coefficients as well.
    Notice that this polynomial satisfies $g(\lambda_i)=f(\lambda_i^{-1})$ for all eigenvalues.
    Hence, if $\beta\in\calD$
    such that $\beta=f(\lambda)$ then
    $\beta^*=g(\lambda)\in\calD$ as well.

    The second part is straightforward, indeed if $\beta=f(\lambda)\in\calD$, then
    \[\omega(\vec{n}\iota(\beta),\vec{m})=\omega(\vec{n}f(A),\vec{m})=\omega(\vec{n},\vec{m}f(A^{-1}))=\omega(\vec{n},\vec{m}\iota(\beta^*)).\]
    \end{proof}

    \begin{cor}
    For any $\beta\in\calD$, the matrix $\iota(\beta)$ is
    symplectic if and only if $\beta\beta^*=1$. Furthermore, for
    any integer $M>1$, if $\beta\beta^*\equiv 1\pmod{M\calD}$ then
    $\iota(\beta)$ is symplectic modulo $M$.
    \end{cor}

    Define a ``norm map"
    $\calN:\calD\rightarrow\calD$ sending $\beta \mapsto
    \beta\beta^*$.
    Given an integer $M>1$, the inclusion
    $\iota:\calD\hookrightarrow\Mat(2d,\Z)$ induces a map
    $\iota_M:\calD/M\calD\rightarrow \Mat(2d,\Z/MZ)$, and the norm map
    $\calN$ induces a well defined map
    $\calN_M:(\calD/M\calD)^*\rightarrow (\calD/M\calD)^*$.
    The norm map is multiplicative, hence the
    map $\calN_M$ is a group homomorphism and it's
    kernel correspond to symplectic matrices. Consequently
    \[\iota_M(\ker\calN_M)\subseteq \Sp(2d,\Z/M\Z),\]
    is a commutative subgroup of symplectic matrices, that commute with $A$ modulo $M$. We are now ready to define the Hecke group.

    \begin{defn}\label{dHECKE}
    Define the Hecke group
    \[C_A(N)=\left\lbrace\begin{array}{lc}
    \{\iota_N(\beta)|\beta\in\ker\calN_N\} & N \text{ odd}\\
    \{\iota_{2N}(\beta)|\beta\in\ker\calN_{2N},\;\beta\equiv\pm 1\pmod{4}\} & N \text{ even}\\
    \end{array}\right.\]
    Now take the Hecke operators to be $U_N(B),\;B\in C_A(N)$.
    \end{defn}

    \begin{rem}
    Note that if $A\not\equiv \pm I\pmod{4}$ and $N$ is even,
    then $U_N(A)$ is not one of the Hecke operators.
    Nevertheless, corollary \ref{cCOM:1} ensures that it still
    commutes with all of them.
    \end{rem}
\subsection{Galois orbits and invariant subspaces}\label{sHECKE:orb}
    The Structure of the Hecke group $C_A(N)$ is closely related to the
    decomposition of the rational vector space $\Q^{2d}=\bigoplus E_\theta$, into
    irreducible invariant subspaces under the left action of $A$. We
    now make a slight detour, and describe this decomposition in
    terms of Galois orbits of the eigenvalues of $A$.

    Let $\Lambda_\Q$ denote the set of eigenvalues of $A$, and
    $G_\Q$ the absolute Galois group. The group $G_\Q$ acts on
    $\Lambda_\Q$, and we denote by $\Lambda_\Q/G_\Q$ the set of
    Galois orbits. Since the matrix $A$ is symplectic, if $\lambda\in\Lambda_\Q$ is an
    eigenvalue, then $\lambda^{-1}\in\Lambda_\Q$ as well. To each orbit
    $\theta\in \Lambda_\Q/G_\Q$ there is a unique orbit $\theta^*$ such
    that $\lambda\in\theta\Leftrightarrow \lambda^{-1}\in\theta^*$.
    If $\theta=\theta^*$ we say that the orbit is symmetric, and
    otherwise nonsymmetric. For any orbit $\theta$ we define the
    symplectic orbit $\btheta=\theta\cup\theta^*$.
    \begin{prop}\label{pDECOMP}
    There is a unique decomposition into irreducible left invariant
    subspaces:
    $\Q^{2d}=\bigoplus_{\Lambda_\Q/G_\Q} E_\theta.$
    \begin{itemize}
    \item To each orbit $\theta\in\Lambda_\Q/G_\Q$, there is a
    corresponding subspace (denoted by $E_\theta$), such that the
    eigenvalues of the restriction $A_{|E_\theta}$ are the eigenvalues
    $\lambda\in\theta$.
    \item For any two orbits $\theta,\theta'$, unless
    $\theta'=\theta^*$, then $E_\theta$ and $E_{\theta'}$ are
    orthogonal with respect to the symplectic form.
    \item Let $\vec{v}_{\theta^*}$ be a left eigenvector for $A$ with
    eigenvalue in $\theta^*$. Then, the projection of $\vec{n}$ to
    $E_\theta$ with respect to the above decomposition vanishes, if and only if $\omega(\vec{n},\vec{v}_{\theta^*})=0$.
    \end{itemize}
    \end{prop}
    \begin{proof}
    Appendix \ref{sORBITS}, lemma \ref{lDECOMP} and corollary \ref{cPROJ:1}.
    \end{proof}
    \begin{rem}
    There is an alternative way to describe this decomposition, using the
    Characteristic polynomial $P_A$ of $A$. Any invariant
    irreducible subspace corresponds to an irreducible factor of
    $P_A$ (which is integral by Gauss's lemma). The roots of this
    irreducible factor are then precisely the eigenvalues in the
    Galois orbit. As a consequence we can deduce, that the product
    of all eigenvalues in one orbit is an integer that divides $1$
    and can thus be only $\pm 1$ (for a symmetric orbit by its definition the product is always $+1$).
    \end{rem}

    For each symplectic orbit $\btheta$ we define the space $E_\btheta=E_\theta+E_{\theta^*}$.
    Proposition \ref{pDECOMP} then implies that for $\theta$
    symmetric $E_\theta=E_\btheta$ is a symplectic space
    (i.e., the restriction of the symplectic form to this subspace is non-degenerate), while for
    $\theta$ nonsymmetric the spaces $E_{\theta},E_{\theta^*}$
    are both isotropic (i.e., the restriction of the symplectic form
    vanishes) and $E_\btheta=E_\theta\oplus E_{\theta^*}$ is again symplectic.

\subsection{Reduction to Galois orbits}\label{sHECKE:2}
    Consider the action of the absolute Galois group $G_\Q$, on the set
    of eigenvalues
    $\Lambda_\Q=\set{\lambda_1,\ldots,\lambda_{2d}}$.
    For each orbit $\theta\in\Lambda_\Q/G_\Q$ fix a representative $\lambda_\theta$
    (for nonsymmetric orbits we take $\lambda_{\theta^*}=\lambda_\theta^{-1}$).
    Let $K_\theta=\Q(\lambda_\theta)$
    be field extensions,
    and $\calO_{K_\theta}$ the corresponding
    integral rings.
    For any symmetric orbit, $\lambda_\theta$ and
    $\lambda_\theta^{-1}$ are Galois conjugates.
    Consequently, if we denote by $F_\theta=\Q(\lambda_\theta+\lambda_\theta^{-1})$, then
    $K_\theta/F_\theta$ are quadratic field extensions.

    Note that every element $\beta\in\calD$, is
    uniquely determined by its components on each orbit $\beta_\theta\in\bbZ[\lambda_\theta]\subseteq \calO_{K_\theta}$
    (because if $\lambda_i=\lambda_\theta^\sigma$ for some $\sigma\in G_\Q$, then
    $\beta_i=f(\lambda_i)=f(\lambda_\theta^\sigma)=\beta_\theta^\sigma$).
    We can thus identify the ring $\calD$ as a subring of
    $\prod_{\Lambda_\Q/G_\Q}\calO_{K_\theta}$.

    \begin{lem}\label{lNORM}
        The norm map $\calN$, acts on
        a component corresponding to a symmetric orbit $\theta$, through the corresponding field extension norm map,
        $\calN_{K_\theta/F_\theta}$, and on a component corresponding to a nonsymmetric orbit $\theta$
        by $\beta_\theta\mapsto \beta_\theta\beta_{\theta^*}$.
    \end{lem}
    \begin{proof}
    Let $\beta\in\calD$, then $\beta_\theta=f(\lambda_\theta)$ for
    some $f\in\bbZ[t]$. For any orbit $\theta$,
    $(\calN(\beta))_\theta=f(\lambda_\theta)f(\lambda_\theta^{-1})$.
    When the orbit $\theta$ is symmetric this is precisely
    $\calN_{K_\theta/F_\theta}(\beta_\theta))$, and when it is
    nonsymmetric it is $\beta_\theta\beta_{\theta^*}$.
    \end{proof}

    \begin{lem}\label{lSTRUC}
        There is $s\in\bbN$, such that
        \[s\prod_{\Lambda_\Q/G_\Q}\calO_{K_\theta}\subseteq \calD\subseteq
        \prod_{\Lambda_\Q/G_\Q}\calO_{K_\theta}.\]
    \end{lem}
    \begin{proof}
        The rings $\calO_{K_\theta}$ are isomorphic (as $\bbZ$ modules) to $\bbZ^{|\theta|}$,
        hence $\prod_{\Lambda_\Q/G_\Q}\calO_{K_\theta}\cong \bbZ^{2d}$.
        On the other hand $\calD\cong \bbZ[t]/P_A\cong \bbZ^{2d}$ as
        well (again as $\bbZ$ modules). The result is now immediate since any subgroup of
        $\bbZ^{2d}$ with the same rank satisfies this property.
    \end{proof}

    We can now estimate the number of Hecke operators.
    \begin{lem}\label{lHECKE}
        The number of elements in $C_A(N)$, satisfy
        \[N^{d-\epsilon}\ll_\epsilon |C_A(N)| \ll_\epsilon  N^{d+\epsilon}\]
    \end{lem}
    \begin{proof}
        To simplify the discussion,
        we will assume that there are no rational isotropic invariant rational subspaces
        (i.e., all orbits are symmetric).
        The Hecke group (for $N$ even) is a
        subgroup of $\iota_{2N}(\ker\calN_{2N})$ with index
        bounded by $2^{d^2}$, it is thus sufficient to show that for
        all $N$,
            \[N^{d-\epsilon}\ll |\ker\calN_N| \ll N^{d+\epsilon}.\]

        For each orbit $\theta\in \Lambda_\Q/G_\Q$, the norm map
        $\calN_{K_\theta/F_\theta}$, induces a map on the group of invertible elements
        $$\calN_{N\calO_{F_\theta}}:(\calO_{K_\theta}/N\calO_{K_\theta})^*\rightarrow(\calO_{K_\theta}/N\calO_{K_\theta})^*.$$ Let $\calC(N\calO_{F_\theta})$ be the
        kernel of this map.
        For any $\beta\in\calD$, denote by
        $\bar\beta\in\calD/N\calD$ its class modulo $N\calD$,
        by $\beta_\theta$ its component in $\calO_{K_\theta}$,
        and by $\bar\beta_\theta$ the class of $\beta_\theta$ modulo $N\calO_{K_\theta}$.
        Then the map $\bar\beta\mapsto\bar\beta_\theta$ is
        well defined ( because, if $\beta\in N\calD$ then obviously
        $\beta_\theta\in N\calO_{K_\theta}$),
        and by lemma \ref{lSTRUC}, the map
        \[\begin{array}{ccc}
        \calD/N\calD & \rightarrow & \prod_{\theta\in\Lambda}\calO_{K_\theta}/N\calO_{K_\theta}\\
        \bar\beta & \mapsto & (\bar\beta_\theta)_\theta\
        \end{array}\]
        has kernel and co-kernel of order bounded by $|\calD/s\calD|=s^{2d}$.
        Furthermore, the restriction of this map to the multiplicative group and to
        the subgroup of norm one elements
        also has bounded kernel
        and co-kernel. Thus, it is suffices to show that $\forall
        \theta\in \Lambda_\Q/G_\Q$
        \[N^{d_\theta-\epsilon}\ll |\calC(N\calO_{F_\theta})| \ll N^{d_\theta+\epsilon},\]
        where $d_\theta=\frac{|\theta|}{2}=[F_\theta:\Q]$.
        This is the estimate on the number of norm one elements
        in the ring $\calO_{K_\theta}/N\calO_{K_\theta}$ which
        is proved in appendix \ref{sCOUNT} (proposition \ref{pN1}).
    \end{proof}
    \begin{rem}
    If there are invariant rational isotropic subspaces the proof is
    analogous. For any symplectic orbit $\btheta=\theta\cup\theta^*$ corresponding to a nonsymmetric orbit,
    instead of evaluating the number of elements
    in $\calC(N\calO_{K_\theta})$ one needs to evaluate the size of
    $(\calO_{K_\theta}/N\calO_{K_\theta})^*$ and show
     $N^{d_\theta-\epsilon}\ll |(\calO_{K_\theta}/N\calO_{K_\theta})^*| \ll
     N^{d_\theta+\epsilon}$, where now $d_\theta=\frac{|\btheta|}{2}=|\theta|$.
    \end{rem}

    \subsection{Additional structure}\label{sSTRUC}
    So far we have identified a set of commuting integral matrices with the commutative ring $\calD$.
    We are now going to identify the action of these matrices
    on $\bbZ^{2d}$, with the action of $\calD$ on an appropriate ideal $\calI$.
    This identification allows us think of both the matrices and the lattice points on which they act as elements of the same space $\calD$,

    For every orbit $\theta\in \Lambda_\Q/G_\Q$,
    take a left eigenvector $\vec{v}_\theta$ with eigenvalue $\lambda_\theta^{-1}$ and coefficients in
    $s\calO_{K_\theta}$.
    Therefore $\vec{v}=(\vec{v}_\theta)_\theta$ is a (left) eigenvector with coefficients in
    $\prod s\calO_{K_\theta}\subseteq \calD$, such that $\vec{v}\iota(\beta^*)=\beta\vec{v}$.
    Define the map $\iota^*:\bbZ^{2d}\rightarrow \calD$, by
    $\iota^*(\vec{n})=\omega(\vec{n},\vec{v})$,
    and the ideal $\im(\iota^*)=\calI\subseteq \calD$.
    To see that $\calI$ is indeed an ideal notice that if $\nu=\iota^*(\vec{n})\in\calI$ and $\beta\in\calD$ with $B=\iota(\beta)$ then
    \[\beta\nu=\beta \iota^*(\vec{n})=\beta\omega(\vec{n},\vec{v})=\omega(\vec{n},\vec{v}\iota(\beta^*))=\omega(\vec{n}\iota(\beta),\vec{v})=\iota^*(\vec{n}B),\]
    so $\beta\nu\in\calI$ as well.
    Furthermore, by the third part of proposition \ref{pDECOMP}, we see that
    $(\iota^*(\vec{n}))_\theta=0$ if and only if the projection of
    $\vec{n}$ to $E_\theta$ vanishes.
    In particular $\iota^*(\vec{n})=0$ implies $\vec{n}=0$ and the map $\iota^*:\bbZ^{2d}\rightarrow \calI$
    is an isomorphism of $\bbZ$ modules.

    Now, for any integer $M\in\bbN$, the map $\iota^*$ induces a group isomorphism $\iota^*_M:(\bbZ/M\bbZ)^{2d}\to\calI/M\calI$.
    This map is compatible with the map $\iota_M:\calD/M\calD\to\Mat(2d,\bbZ/M\bbZ)$, in the sense that for any $B=\iota_M(\bar\beta)$, and $\vec{n}\in(\bbZ/N\bbZ)^{2d}$
    we have $\iota^*_M(\vec{n}B)=\bar\beta\iota_M^*(\vec{n})$ in $\calI/M\calI$.

    \subsection{Hecke eigenfunctions}\label{sHECKE:3}
        Since all the Hecke operators commute with $U_N(A)$, they
        act on it's eigenspaces, and since they commute with each
        other, there is a basis of joint eigenfunctions of
        $U_N(A)$ and the Hecke operators. Such a basis is called a
        Hecke basis. We now show that the Hecke symmetries
        cancel most of the degeneracies in the spectrum of $U_N(A)$, implying that the Hecke basis is essentially unique.

        The action of the Hecke group on the Hilbert space
        $\calH_N$, induces a decomposition into joint eigenspaces
        \[\calH_N=\bigoplus_\chi \calH_\chi,\]
        where $\chi$ runs over the characters of the Hecke group.
        \begin{prop}\label{pDIM}
        The dimension of any Hecke
        eigenspace satisfies
        \[\dim \calH_\chi\ll_\epsilon N^{\epsilon}.\]
        \end{prop}
        \begin{proof}
        Again, for simplicity we will assume all orbits are symmetric.
        The operator
        \[\calP_\chi=\frac{1}{|C_A(N)|}\sum_{
        C_A(N)}\chi(B)^{-1}U_N(B)),\]
        is a projection operator to the eigenspace $\calH_\chi$.
        Consequently, the dimension of $\calH_\chi$ is given by its trace, $\dim \calH_\chi=\Tr(\calP_\chi)$.

        By corollary \ref{cTR:1}, for any $B\in C_A(N)$,
        \[|\Tr(U_N(B))|\leq 2^d\sqrt{\ker_N(B-I)}.\]
        Note that while for even $N$ the operator $U_N(B)$ depends on
        $B$ modulo $2N$, this bound only depends on $B$ modulo $N$.
        Hence if $B=\iota_N(\beta)\pmod{N}$, then using the
        identification $\iota^*_N:\bbZ/N\bbZ\to\calI/N\calI$ we can
        write this bound as,
        \[|\Tr(U_N(B)|\leq 2^d\sqrt{\#\set{\nu\in\calI/N\calI|\nu(\beta-1)\equiv 0\pmod{N\calI}}}.\]
        Since both $\calD$ and the ideal $\calI$ are isomorphic (as $\bbZ$ modules) to $\bbZ^{2d}$, there is $s'\in\bbN$ such that
        $s'\calD\subseteq\calI\subseteq \calD$ and we can replace
        $\calI/N\calI$ by $\calD/N\calD$ to get
        \[|\Tr(U_N(B)|\leq (2s')^d\sqrt{\#\set{\nu\in \calD/N\calD|\nu(\beta-1)\equiv 0\pmod{N\calD}}}.\]
        We now replace the sum over $C_A(N)$ with a sum over $\ker(\calN_N)$ in the bound  $\dim(\calH_\chi)\leq
        \frac{1}{|C_A(N)|}\sum_{C_A(N)}|\Tr(U_N(A))|$ (losing at
        most a constant factor) to get the bound
        \[\dim(\calH_\chi)\ll \frac{1}{|C_A(N)|}\sum_{
        \beta\in \ker \calN_N}\sqrt{\#\set{\nu\in\calD/N\calD|\nu(\beta-1)\in N\calD}},\]
        Because the map $\ker(\calN_N)\rightarrow \prod_{\Lambda_\Q/G_\Q}\calC(N\calO_{F_\theta})$ has bounded kernel (as in the proof of \ref{lHECKE}),
        after multiplying by some bounded constant we can replace
        $\calD/N\calD$ with $\prod_{\Lambda_\Q/G_\Q}\calO_{K_\theta}/N\calO_{K_\theta}$, and
        the sum over the group $\ker \calN_N$, to a sum over
        $\prod_{\Lambda_\Q/G_\Q}\calC(N\calO_{F_\theta})$ to get
         \[\dim(\calH_\chi)\ll \frac{1}{|C_A(N)|}
         \prod_{\Lambda_\Q/G_\Q}S_1(N\calO_{K_\theta}),\]
         where,
         \[S_1(N\calO_{F_\theta})=\sum_{\beta\in
         \calC(N\calO_{F_\theta})}\sqrt{\#\set{\nu\in\calO_{K_\theta}/N\calO_{K_\theta}|\nu(\beta-1)=0}}.\]
         It now suffices to show that $\forall \theta\in
         \Lambda_\Q/G_\Q,\quad S_1(N\calO_{F_\theta})\ll_\epsilon
         N^{d_\theta+\epsilon}$
         (recall $\frac{1}{C_A(N)}=O_\epsilon (N^{-d+\epsilon}$).
         This is a counting argument on elements in the
         ring $\calO_{K_\theta}/N\calO_{K_\theta}$ that is proved
         in appendix \ref{sCOUNT} (proposition \ref{pS1}).
        \end{proof}
        \begin{rem}
        The proof in the nonsymmetric case is analogous. For any nonsymmetric orbit one
        needs to bound sums of the form
        \[\sum_{(\calO_{F_\theta}/N\calO_{F_\theta})^*}\sqrt{\#\set{\nu_1,\nu_2\in\calO_{K_\theta}/N\calO_{K_\theta}|\nu_2(\beta-1)=\nu_2(\beta^{-1}-1)=0}}.\]
        This can be done using the same methods.
        \end{rem}

\section{Arithmetic Quantum Unique Ergodicity}\label{sAQUE}
     This section is devoted to proving theorem \ref{tQUE:2}.
     We fix a matrix $A\in\Sp_\theta(2d,\bbZ)$ with distinct eigenvalues and no invariant isotropic rational subspaces,
     and show that for any smooth observable $f\in
     C^\infty(\bbT^{2d})$, the expectation values for $\Op_N(f)$ in any Hecke
     eigenfunction $\psi$, satisfy:
     \[|\langle
     \Op_N(f)\psi,\psi\rangle|\ll_{\epsilon,f} N^{-\frac{d(f)}{4}+\epsilon},\]
     where $d(f)=\min_{\hat{f}(\vec{n})\neq 0} d_{\vec{n}}$, and
     $2d_{\vec{n}}$ is the dimension of the smallest invariant
     subspace containing $\vec{n}$.

     Much of the proof goes along the lines of \cite{KR2}. The first step is to make a reduction to a theorem regarding elementary
     observables. Next, we show that it is sufficient to bound the
     fourth moment of the matrix elements (after restricting the
     elementary operator to an appropriate subspace). Finally, we use
     averaging over the Hecke group, to transform the moment
     calculation into a counting problem, which is then solved
     using the connection of the Hecke group with the groups $\calC(N\calO_{F_\theta})\subseteq(\calO_{K_\theta}/N\calO_{K_\theta})^*$.

    \subsection{Reduction to elementary observables}\label{sAQUE:1}
     In order to prove theorem \ref{tQUE:2} it is sufficient to
     prove it for elementary observables of the form
     $\Op_N(e_{\vec{n}}),\;0\neq\vec{n}\in\bbZ^{2d}$,
     that is, to show that the following theorem holds.
     \begin{thm}\label{tQUE:3}
    Let $0\neq\vec{n}\in\bbZ^{2d}$ and let $\psi$ be an
    eigenfunction of all the Hecke operators. Then, the diagonal
    matrix elements satisfy
    \[|\langle
     \T{N}(\vec{n})\psi,\psi\rangle|\ll_\epsilon
     \norm{\vec{n}}^{4d^2}
     N^{-d_{\vec{n}}/4+\epsilon}\]
     \end{thm}
     The proof of theorem \ref{tQUE:2} from theorem \ref{tQUE:3} is
     immediate, due to the rapid decay of the Fourier coefficients.
     \begin{rem}
     The estimate in theorem \ref{tQUE:3} is in fact valid also
     when there are invariant rational isotropic subspaces, as long as
     $\vec{n}$ is not contained in any of these subspaces.
     \end{rem}

     \subsection{Reduction to a moment calculation}\label{sAQUE:2}
     In order to prove theorem \ref{tQUE:3}, we estimate the fourth
     moment of the diagonal matrix elements in a Hecke basis,
     \[\sum_{\psi} |\langle
     \T{N}(\vec{n})\psi,\psi\rangle|^4.\]
     However, when summing over all the Hecke
     eigenfunctions, the fourth moment is of order $N^{d-2d_{\vec{n}}}$
     (where $2d_{\vec{n}}$ the dimension of the smallest (symplectic) invariant subspace containing
     $\vec{n}$). Thus, when $\vec{n}$ is contained in an invariant
     subspace of dimension $\leq d$, we can not use this method
     directly to bound the size of the individual matrix
     elements. Instead, we would like to make the sum only over a
     subset of the Hecke eigenfunctions.
     For that purpose, for each
     $\vec{n}\in\bbZ^{2d}$, and Hecke eigenfunction $\psi$, we introduce a
     subspace $\calH_{\vec{n},\psi}\subseteq \calH_N$, invariant under the action of $\T{N}(\vec{n})$ and the Hecke
     operators. Theorem \ref{tQUE:3} is then proved by estimating
     the fourth moment for the restriction of $\T{N}(\vec{n})$ to
     $\calH_{\vec{n},\psi}$.

     Let $\vec{n}\in\bbZ^{2d}$. Recall the decomposition into irreducible invariant
     subspaces $\Q^{2d}=\bigoplus E_\theta$ described in section
     \ref{sHECKE:orb}, and let $\Lambda_{\vec{n}}/G_\Q$ be the set of orbits
     $\theta\in\Lambda_\Q/G_\Q$ for which the projection of $\vec{n}$ to $E_\theta$
     vanishes. Denote by $E_{\vec{n}}$ the minimal invariant
     subspace containing $\vec{n}$. We can
     decompose
     $E_{\vec{n}}=\sum_{\theta\not\in\Lambda_{\vec{n}}/G}E_\theta$, and in particular
     \[2d_{\vec{n}}=\dim E_{\vec{n}}=\sum_{\theta\not\in\Lambda_{\vec{n}}/G_\Q}2d_\theta,\]
     where $2d_\theta=\dim E_\theta=|\theta|$ (recall that
     $E_\theta$ is of even dimension because it is symplectic).

     Define the lattice $Z_{\vec{n}}=E_{\vec{n}}\cap \bbZ^{2d}$, then by the third
     part of proposition \ref{pDECOMP} we have
     $Z_{\vec{n}}=\set{\vec{m}\in
     \bbZ^{2d}|\iota^*(\vec{m})_\theta=0,\;\forall\theta\in\Lambda_{\vec{n}}/G_\Q}$.
     \begin{defn}
     For $\psi\in\calH_N$ a Hecke eigenfunction, define the
     subspace $\calH_{\vec{n},\psi}\subseteq \calH_N$ to be the minimal subspace
     containing $\psi$ and invariant
     under the action of all $\T{N}(\vec{m}),\;\vec{m}\in Z_{\vec{n}}$.
     \end{defn}
     \begin{lem}
        The space $\calH_{\vec{n},\psi}$, is invariant under the action of the
        Hecke operators.
     \end{lem}
     \begin{proof}
        For $B=\iota_{2N}(\bar\beta)\in C_A(N)$ and $\vec{m}\in Z_{\vec{n}}$, let
        $\vec{m}'=\vec{m}\iota(\beta)$, where $\beta\in\calD$ is a representative of $\bar\beta$. Then $\vec{m}'\equiv \vec{m}B\pmod{2N}$
        and $\vec{m}'\in Z_{\vec{n}}$ (because
        $\forall \theta\in \Lambda_{\vec{n}}/G_\Q,\;\omega(\vec{m}',\vec{v}_\theta)=\beta_\theta\omega(\vec{m},\vec{v}_\theta)=0$).
        Now, if $\phi'=U_N(B)\phi\in U_N(B)\calH_{\vec{n},\psi}$ (for some $\phi\in \calH_{\vec{n},\psi}$), then
        \[\T{N}(\vec{m})\phi'=U_N(B)\T{N}(\vec{m}B)\phi=U_N(B)\T{N}(\vec{m}')\phi,\]
        hence $\T{N}(\vec{m})\phi'\in U_N(B)\calH_{\vec{n},\psi}$ as well
        (because $\calH_{\vec{n},\psi}$ is invariant under
        $\T{N}(\vec{m}')$).
        Therefore, the space $U_N(B)\calH_{\vec{n},\psi}$ contains $\psi$ and is
        invariant under the action of $\T{N}(\vec{m}),\;\forall \vec{m}\in
        Z_{\vec{n}}$. Thus, from the
        minimality condition $\calH_{\vec{n},\psi}\subseteq U_N(B)\calH_{\vec{n},\psi}$, and since $U_N(B)$
        is invertible we have $U_N(B)\calH_{\vec{n},\psi}=\calH_{\vec{n},\psi}$.
     \end{proof}

     When $\Lambda_{\vec{n}}=\emptyset$ then $Z_{\vec{n}}=\bbZ^{2d}$ and $\calH_{\vec{n},\psi}=\calH_N$, but otherwise it is a proper subspace
     and we can give an estimate for its dimension.
    \begin{prop} The dimension of the subspace
    $\calH_{\vec{n},\psi}$ satisfies
        \[\dim(\calH_{\vec{n},\psi})\ll_\epsilon N^{d_{\vec{n}}+\epsilon},\]
        where the implied constant does not depend on $\vec{n}$ or on $\psi$.
    \end{prop}
    \begin{proof}
     Consider a subgroup of the Hecke group
     \[C_0(N)=\set{\iota_N(\bar\beta)\in
     C_A(N)|\beta_\theta=1,\;\forall \theta\notin \Lambda_{\vec{n}}/G_\Q},\]
     in the sense that there is a representative $\beta\in\calD\subseteq \prod\calO_{K_\theta}$
     satisfying this condition. Notice that this group acts trivially on $Z_{\vec{n}}$ modulo $2N$ (i.e., $\forall B\in
     C_0(N)$ and $\forall \vec{m}\in Z_{\vec{n}},\;\vec{m}B\equiv
     \vec{m}\pmod{2N}$).

     Let $\chi(B),\;B\in C_A(N)$, be the eigenvalues corresponding to $\psi$,
     and consider the subspace
     \[\calH^0_\chi=\set{\phi\in\calH_N|U_N(B)\phi=\chi(B)\phi,\;\forall
     B\in C_0(N)}.\]
     Since $C_0(N)$ acts trivially on $Z_{\vec{n}}$, then
     $\forall \vec{m}\in Z_{\vec{n}}$ and $B\in C_0(N)$,
     $U_N(B)\T{N}(\vec{m})=\T{N}(\vec{m})U_N(B)$ commute, hence $\calH^0_\chi$
     is invariant under $\T{N}(\vec{m}),\;\forall\vec{m}\in Z_{\vec{n}}$.
     Obviously $\psi\in \calH^0_\chi$, hence from minimality
     $\calH_{\vec{n},\psi}\subseteq \calH^0_\chi$ and it suffices to bound the dimension of $\calH^0_\chi$.

     The eigenspace $\calH^0_\chi$ decomposes into joint eigenspaces of all
     the Hecke operators
     \[\calH^0_\chi=\bigoplus\calH_{\chi'},\]
     where the sum is only on characters  $\chi'$ that identify
     with $\chi$ on $C_0(N)$.
     Note that $\chi'_{|_{C_0(N)}}=\chi_{|_{C_0(N)}}$ imply that
     they differ by a character of $C_A(N)/C_0(N)$ (and
     viceversa).
     Therefore (by proposition \ref{pDIM}) the dimension
     \[\dim{\calH^0_\chi}=\sum_{\chi'|_{C_0}=\chi|_{C_0}}\dim{H_{\chi'}}\ll_\epsilon N^{\epsilon}[C_A(N):C_0(N)].\]
     Following the lines of the proof of lemma \ref{lHECKE}, one can show
     $[C_A(N):C_0(N)]\ll_\epsilon N^{d_{\vec{n}}+\epsilon}$, concluding the proof.
     Notice that the implied constants depend only on the set of
     orbits
     $\Lambda_{\vec{n}}/G_\Q$. But, as there are at most $2^d$
     possibilities for such subsets, we can take
     the same constant for all the spaces $\calH_{\vec{n},\psi}$.
     \end{proof}

     For $\vec{m}\in Z_{\vec{n}}$, denote by $\T{N}^{0}(\vec{m})$ the restriction of
     $\T{N}(\vec{m})$ to $\calH_{\vec{n},\psi}$. Then, similar to the original operators,
     the trace of the restricted operators vanishes for
     sufficiently large $N$.
     \begin{lem}\label{lTR}
        There is $r\in \bbN$ (depends only on $Z_{\vec{n}}$) such that for any $\vec{m}\in Z_{\vec{n}}$,
        \[|\Tr(\T{N}^{0}(\vec{m}))|\leq\left\lbrace\begin{array}{cc}
        \dim{\calH_{\vec{n},\psi}} & \vec{m}\equiv 0\pmod{N'}\\
        0 & \mbox{otherwise}\
        \end{array}\right.,\]
        where  $N'=\frac{N}{\gcd(N,r^2)}$.
     \end{lem}
     \begin{proof}
     Recall that the space $E_{\vec{n}}$ is a symplectic
     subspace. Let $\set{e_i,f_i}$ be a symplectic basis (i.e.,
     $\omega(e_i,f_j)=\delta_{i,j}$ and
     $\omega(e_i,e_j)=\omega(f_i,f_j)=0$), and let $r\in\bbZ$ such that
     $re_i,rf_i\in\bbZ^{2d}$. Fix $\vec{m}\in E_{\vec{n}}$, and consider the decomposition
     $\vec{m}=\sum_{i=1}^{d_{\vec{n}}}(a_ie_i+b_if_i)$.
     Then $ra_i=\omega(\vec{m},rf_i)$ and $rb_i=-\omega(\vec{m},re_i)$ are integers.

     Notice that for all $i=1,\ldots,d_{\vec{n}}$,
     \[\Tr(\T{N}^0(\vec{m}))=\Tr(\T{N}^0(-rf_i)\T{N}^0(\vec{m})\T{N}^0(rf_i))=
     e_N(ra_i)\Tr(\T{N}^0(\vec{m})),\]
     and by a similar argument
     $\Tr(\T{N}^0(\vec{m}))=e_N(rb_i)\Tr(T{N}^0(\vec{m}))$.
     Consequently, if $ra_i\not\equiv 0\pmod{N}$ or $rb_i\not\equiv 0\pmod{N}$ then
     $\Tr(\T{N}^0(\vec{m}))=0$. On the other hand if $\forall i,\;ra_i\equiv rb_i\equiv
     0\pmod{N}$, then $r^2\vec{m}=\sum_{i=1}^{d_{\vec{n}}}(ra_ire_i+rb_irf_i)\equiv
     0\pmod{N}$ and $\vec{m}\equiv 0\pmod{N'}$.

     \end{proof}
     \begin{rem}
     The integer $r$ in the above lemma, depends only on the lattice $Z_{\vec{n}}$, that is determined by
     the subset $\Lambda_{\vec{n}}/G_\Q$. We can thus take $r$ to be the same for
     all $\vec{n}$ (by taking the $\lcm$ for the $2^d$
     possibilities).
     \end{rem}

     The Hecke operators act on the space $\calH_{\vec{n},\psi}$ so
     there is a basis $\{\psi_i\}$ of joint eigenfunctions of all the
     Hecke operators (we can assume $\psi_1=\psi$). To prove theorem \ref{tQUE:3}, we will prove a stronger
     statement regarding the forth moment of matrix elements in this basis.
     \begin{prop}\label{pFOURTH}
     Let $\{\psi_i\}$ be a basis for $ \calH_{\vec{n},\psi}$
    composed of joint eigenfunctions of all the Hecke
    operators. Then, the fourth moment satisfies
     \[\sum_i|\langle
     \T{N}(\vec{n})\psi_i,\psi_i\rangle|^4\ll_\epsilon
     \norm{\vec{n}}^{16d_{\vec{n}}^2}
     N^{-d_{\vec{n}}+\epsilon}.\]
     \end{prop}
     The proof of theorem \ref{tQUE:3} from proposition \ref{pFOURTH}
     is now immediate. The first element in the sum is obviously
     bounded by the whole sum, so
     \[|\langle \T{N}(\vec{n})\psi,\psi\rangle|^4\ll_{\epsilon} \norm{\vec{n}}^{16d_{\vec{n}}^2}
     N^{-d_{\vec{n}}+\epsilon}\leq\norm{\vec{n}}^{16d^2}
     N^{-d_{\vec{n}}+\epsilon}.\]

    \subsection{Reduction to a counting problem}\label{sAQUE:3}
    We now reduce proposition \ref{pFOURTH} in
    to a counting problem, which is then solved in the following
    section.
    \begin{prop}\label{pRED:2}
    Let $\{\psi_i\}$ be a basis for $ \calH_{\vec{n},\psi}$
    composed of joint eigenfunctions of all the Hecke
    operators. Then, the fourth moment,
     $$\sum_i|\langle
     \T{N}(\vec{n})\psi_i,\psi_i\rangle|^4,$$
     is bounded by $ \frac{\dim
     \calH_{\vec{n},\psi}}{|C_A(N)|^4}$ times the number of
     solutions to
     \[\vec{n}(B_1-B_2+B_3-B_4)\equiv 0\pmod {N'},\;B_i\in C_A(N).\]
     where $N'$ is as in lemma \ref{lTR}.
    \end{prop}
    \begin{proof}
    Define an operator, $D=D(\vec{n})$,
    acting on $\calH_{\vec{n},\psi}$ through averaging over the Hecke group:
    \[D=\frac{1}{|C_A(N)|}\sum_{B\in C_A(N)}
    \T{N}^0(\vec{n}B).\]
    Recall that for $B\in C_A(N)$ there is $\vec{m}\in Z_{\vec{n}}$ such that $\vec{n}B\equiv\vec{m}\pmod{2N}$,
    so this is indeed well defined.
    The identity $\T{N}(\vec{n}B)=U_N(B)^*\T{N}(\vec{n})U_N(B)$ implies
    $\langle D(\vec{n})\psi_i,\psi_i\rangle=\langle
    \T{N}(\vec{n})\psi_i,\psi_i\rangle$,
    and since for any complex matrix $D=(d_{i,j}),\quad\sum_i|d_{i,i}|^4\leq
    \Tr((DD^*)^2)$, it is sufficient to bound $\Tr((DD^*)^2)$.
    Now, expand $(DD^*)^2$ as a product of 4 sums, and take trace (using lemma \ref{lTR}) to get the result.
    \end{proof}

     By proposition \ref{pDIM}, and lemma \ref{lHECKE}, we know
     \[\frac{\dim
     \calH_{\vec{n},\psi}}{|C_A(N)|^4}\ll_\epsilon \frac{1}{N^{4d-d_{\vec{n}}-\epsilon}}.\]
     Therefore, in order to prove proposition \ref{pFOURTH} from proposition
     \ref{pRED:2}, it remains show that the number of solution to
    \begin{equation}\label{eSOL:1}
    \vec{n}(B_1-B_2+ B_3-B_4)\equiv0\pmod{N'},\quad B_i\in C_A(N),
    \end{equation}
    is bounded by $O({\norm{\vec{n}}}^{16d_{\vec{n}}^2}N^{4d-2d_{\vec{n}}+\epsilon})$.

     \subsection{Counting solution}\label{sAQUE:4}
     We now bound the number of solutions to
     (\ref{eSOL:1}), thus completing the proof of theorem \ref{tQUE:2}.

     \begin{prop}\label{pSOL}
        The number of solution to
        (\ref{eSOL:1} ) is bounded by
        $O_{\epsilon}({\norm{\vec{n}}}^{16d_{\vec{n}}^2}N^{4d-2d_{\vec{n}}+\epsilon})$.
     \end{prop}
     \begin{proof}
     Let $\nu=\iota^*(\vec{n})\in \calI$, then the number of solutions
     to (\ref{eSOL:1} ) is the same as the number of solutions to
     \begin{equation}\label{eSOL:2}
     \nu(\beta_1-\beta_2+\beta_3-\beta_4)\equiv 0\pmod{N'\calI},\quad \beta_i\in\ker
     \calN_{2N}.
     \end{equation}
     In the same way as in the proof of proposition \ref{pDIM},
     it is sufficient to bound for each $\theta$ the number of
     solutions to
     \begin{equation}\label{eSOL:3}
     \nu_\theta(\beta_1-\beta_2+\beta_3-\beta_4)\equiv 0\pmod{N'\calO_{K_\theta}},\quad
     \beta_i\in \calC(N\calO_{F_\theta}),
     \end{equation}
     the product of which gives the number of solutions to (\ref{eSOL:2}) up to some bounded constant.

     If $\theta\in \Lambda_{\vec{n}}/G_\Q$, then $\nu_\theta=0$ and the best
     bound is the trivial bound of $|\calC(N\calO_{F_\theta})|^4=O_\epsilon(
     N^{4d_\theta+\epsilon})$. Otherwise, $0\neq\calN_{K_\theta/\Q}(\nu_\theta)\in\bbZ$ and the number of solutions to
     (\ref{eSOL:3}) is bounded by the number of solutions to
     \begin{equation}\label{eSOL:4}
     \beta_1-\beta_2+\beta_3-\beta_4\equiv 0\pmod{M\calO_{K_\theta}},\quad
     \beta_i\in \calC(N\calO_{F_\theta})
     \end{equation}
     where
     $M=\frac{N'}{\gcd(N',\calN_{K_\theta/\Q}(\nu_\theta))}$.
     The natural map $\calC(N\calO_{F_\theta})\rightarrow \calC(M\calO_{F_\theta})$
     has
     kernel of order at most $(\frac{N}{M})^{2d_\theta}\leq
     (r|\calN_{K_\theta/\Q}(\nu_\theta)|)^{2d_\theta}\ll
     \norm{\vec{n}}^{4d_\theta^2}$,
     hence the number of solutions to (\ref{eSOL:4}) is bounded
     by $\norm{\vec{n}}^{16d_\theta^2}$ times the number of
     solutions to
     \begin{equation}\label{eSOL:5}
     \beta_1-\beta_2+\beta_3-\beta_4\equiv 0\pmod{M\calO_{K_\theta}},\quad
     \beta_i\in \calC(M\calO_{F_\theta})
     \end{equation}

     Equation (\ref{eSOL:5}) is invariant under the action of the
     Galois group
     $\Gal(K_\theta/F_\theta)$. We thus get a second equation,
     \begin{equation}\label{eSOL:6}
     \beta_1^{-1}-\beta_2^{-1}+\beta_3^{-1}-\beta_4^{-1}\equiv 0\pmod{M\calO_{K_\theta}},\quad
     \beta_i\in \calC(M\calO_{F_\theta})
     \end{equation}

    The set of equation (\ref{eSOL:5},\ref{eSOL:6}) is equivalent
    to the following set of equations (see \cite[lemma 15]{KR2} ):
    \begin{equation}\label{eSOL:7}
    \begin{cases}
        (\beta_3-\beta_1)(\beta_3-\beta_2)(\beta_1+\beta_2)=0\pmod{M\calO_{K_\theta}} &  \\
        \beta_4=\beta_1-\beta_2+\beta_3=0\pmod{M\calO_{K_\theta}} &
    \end{cases}
    \end{equation}
     Since $\beta_4$ is determined by $\beta_1,\beta_2,\beta_3$,
     ignoring the second equation only increases the number of solutions.
     Finally the number of solutions to the first equation is bounded by $|\calC(M\calO_{F_\theta})|S_2(M\calO_{F_\theta})$
     where $S_2(M\calO_{F_\theta})$ is
     the number of solutions to
     \begin{equation}\label{eSOL:9}
     (1-\beta_1)(1-\beta_2)(\beta_1+\beta_2)=0\pmod{M\calO_{K_\theta}},\quad
     \beta_i\in \calC(M\calO_{F_\theta}),
     \end{equation}
     that satisfies $S_2(M\calO_{F_\theta})=O(M^{d_\theta+\epsilon})$ (proposition \ref{pS2}).

     To conclude, for $\theta\in \Lambda_{\vec{n}}/G_\Q$ the number of
     solutions to (\ref{eSOL:3}) is bounded by
     $O(N^{4d_\theta+\epsilon})$.
     Otherwise, it is bounded by
     $O(\norm{\vec{n}}^{16d_\theta^2}N^{2d_\theta+\epsilon})$. Therefore, since
     $\sum_{\theta\notin \Lambda_{\vec{n}}/G_\Q} d_\theta=d_{\vec{n}}$, the number of
     solutions to (\ref{eSOL:1}) is bounded by
     \[O(\prod_{\theta\in \Lambda_{\vec{n}}/G_\Q}N^{4d_\theta+\epsilon}\prod_{\theta\notin
     \Lambda_{\vec{n}}/G_\Q}\norm{\vec{n}}^{16d_\theta^2}N^{2d_\theta+\epsilon})=
     O(\norm{\vec{n}}^{16d_{\vec{n}}^2}N^{4d-2d_{\vec{n}}+\epsilon}).\]
 \end{proof}

\section{Hecke Theory For Prime $N$}\label{sHECKP}
    In the following section we restrict the discussion to the case
    where $N=p$ is a large prime. For this case, the structure of the
    Hecke group (hence also the behavior of Hecke eigenfunctions and matrix elements)
    is determined by the decomposition of the vector space
    $\FF{p}^{2d}$ (rather than $\Q^{2d}$) into irreducible invariant subspaces. This
    decomposition can be described using the Frobenius orbits of
    the eigenvalues of $A$. Analyzing the action of $A$ on $\FF{p}^{2d}$ enables us to obtain much sharper
    results from the ones presented above for composite $N$.

    The main difference between composite and prime $N$,
    is that instead of integral rings (we used in section \ref{sHECKE}) here
    we work with finite fields so that the counting arguments become sharp.
    For example, we can describe precisely the structure of the Hecke group (lemma \ref{lHECKE2}),
    and obtain sharp bounds for the dimension of the joint Hecke eigenspaces (proposition \ref{pHES:2}).
    Compare this to lemma \ref{lHECKE} and proposition \ref{pDIM} obtained for composite $N$.

\subsection{Hecke operators}
    Let $A\in\Sp_\theta(2d,\bbZ)$ be a matrix with distinct eigenvalues.
    Fix a large prime $N=p>\Delta(P_A)$ (the discriminant of the characteristic polynomial), then we can think of $A$, also as
    an element of $\Sp(2d,\FF{p})$ with distinct eigenvalues.
    In fact, in order to ensure that $A\pmod{p}$ has distinct
    eigenvalues it is sufficient to assume that $\Delta(P_A)\neq
    0\pmod{p}$.
    For $A\in\Sp(2d,\FF{p})$ with distinct eigenvalues, the centralizer of $A$ (in the symplectic group)
    is a commutative subgroup
    \[C_p(A)\subseteq \Sp(2d,\FF{p}),\]
    and we can take the Hecke operators to be $U_p(B),\;B\in C_p(A)$.
    Note that $p$ is odd, hence the operator $U_p(B)$ depends
    only on $B$ modulo $p$ and this definition of the Hecke
    operators makes sense.

    \begin{rem}
    when $N=p$ is a prime $\geq 5$, the map $B\mapsto U_p(B)$ (which is a
    representation of $\Sp(2d,\FF{p})$) identifies with the
    celebrated Weil representation of the symplectic group
    over the finite field $\FF{p}$. Consequently,
    the Hecke operators can be obtained by restricting the Weil representation
    to a maximal torus. These
    representations are described at length in \cite{Ge},
    and we follow the same lines in our analysis.
    \end{rem}

 \subsection{Reduction to irreducible orbits}
    Let $\Lambda_p/G_p$ denote the Frobenius orbits of the eigenvalues of
    $A$ modulo $p$. To each orbit $\thet\in\Lambda_p/G_p$ (with representative $\lambda_\thet$) denote
    by $\thet^*$ the orbit of $\lambda_\thet^{-1}$ and by
    $\bthet=\thet\cup\thet^*$ the symplectic orbit.
    We say that an orbit $\thet$ is symmetric if $\thet=\thet^*$ and nonsymmetric otherwise. Denote by
    $\Lambda_p/\pm G_p$ the set of symplectic orbits and let
    \[\FF{p}^{2d}=\bigoplus_{\Lambda_p/\pm G_p} E_\bthet,\]
    be the orthogonal decomposition into invariant irreducible symplectic subspaces
    (see appendix \ref{sORBITS} for more details).
    For each symplectic orbit $\bthet\in \Lambda_p/\pm G_p$ let $2d_\bthet=\dim(E_\bthet)=|\bthet|$
    denote the dimension of the corresponding subspace.
    \begin{rem}
    Note, that while this decomposition is similar to the decomposition of $\bbQ^{2d}$
    into invariant symplectic subspaces described in proposition \ref{pDECOMP}, they are not the same.
    The relation between the two decompositions is described in section \ref{sFROBRAT}.
    \end{rem}

    To each invariant subspace $E_\bthet$, take a symplectic basis.
    For any $\vec{n}\in\FF{p}^{2d}$,
    let $\vec{n}_\bthet\in \FF{p}^{2d_{\bthet}}$ be the projection of
    $\vec{n}$ to $E_{\bthet}$ in the symplectic basis.
    Since the decomposition is orthogonal then for any
    $\vec{n},\vec{m}\in \FF{p}^{2d}$
    \begin{equation}\label{eSYMD:1}
    \omega(\vec{m},\vec{n})= \sum_{\Lambda_p/\pm G_p}\omega(\vec{m}_{\bthet},\vec{n}_\bthet).
    \end{equation}
    We thus get an embedding,
    \begin{equation}\label{eEMBED}
    \prod \Sp(2d_{\bthet},\FF{p})\hookrightarrow\Sp(2d,\FF{p}),
    \end{equation}
    through the action of each factor on the corresponding subspace.
    Denote by $\calS\subseteq \Sp(2d,\FF{p})$ the image of
    $\prod\Sp(2d_{\bthet},\FF{p})$. For each $B\in \calS$ denote by
    $B_\bthet\in\Sp(2d_\bthet,\FF{p})$ the restriction of $B$ to
    $E_\bthet$ in the symplectic basis. In order to keep track of dimensions,
    we denote by $\T{p}^{(d)}(\cdot),U_p^{(d)}(\cdot)$, the quantized
    elementary operators and propagators for $\bbT^{2d}$.
    \begin{prop}\label{pTENZ}
    There is a unitary map
    \[\calU:L^2(\FF{p}^d)\rightarrow
    \bigotimes_{\Lambda_p/\pm G_p}L^2(\FF{p}^{d_{\bthet}}),\]
    such that
    \begin{enumerate}
    \item For any $\vec{n}\in\FF{p}^{2d}$,
    \[\calU \T{p}^{(d)}(\vec{n})\calU^{-1}=\bigotimes_{\Lambda_p/\pm G_p}
    \T{p}^{(d_{\bthet})}(\vec{n}_\bthet).\]
    \item For any $B\in \calS$,
    \[\calU U^{(d)}_p(B)\calU^{-1}=\bigotimes_{\Lambda_p/\pm
    G_p}U_p^{(d_{\bthet})}(B_{\bthet}).\]
    \end{enumerate}
    \end{prop}
    \begin{proof}
    Define $\TT{p}^\otimes(\vec{n})=\bigotimes_{\Lambda_p/\pm G_p}
    \T{p}^{(d_{\bthet})}(\vec{n}_\bthet)$.
    It is easily verified from (\ref{eSYMD:1}),
    that $\TT{p}^\otimes(\vec{n})$ obey the same commutation
    relation as in proposition \ref{pTN:1}. Therefore, there is a unitary map $\calU$
    such that $\calU \T{p}^{(d)}(\vec{n})\calU^{-1}=\TT{p}^\otimes(\vec{n})$ for all
    $\vec{n}\in\FF{p}^{2d}$.

    As for the second part, recall $U_p^{(d_{\bthet})}(B_{\bthet})$ all satisfy the intertwining equation, and from the first part $\calU \T{p}^{(d)}(\vec{n})\calU^{-1}=\TT{p}^\otimes(\vec{n})$.
    Consequently, if we define $\tilde{U}_p(B)=\calU^{-1}\bigotimes_{\Lambda_p/\pm
    G_p}U_p^{(d_{\bthet})}(B_{\bthet})\calU$, then $\tilde{U}_p(B)$
    is also an intertwining operator:
    \[\tilde{U}_p(B)^{-1}\T{p}^{(d)}(\vec{n})\tilde{U}_p(B)=\T{p}^{(d)}(\vec{n}B).\]
    Thus, from uniqueness of the quantization the operators
    $\tilde{U}_p(B)$ and $U^{(d)}_p(B)$ differ by a character of
    $\calS$ (recall that the quantization is multiplicative).
    Finally, since $\calS\cong \prod \Sp(2d_{\bthet},\FF{p})$ has no
    nontrivial multiplicative characters, indeed
    \[\calU^{-1}U^{(d)}_p(B)\calU=\bigotimes_{\Lambda_p/\pm
    G_p}U_p^{(d_{\bthet})}(B_{\bthet}).\]
    \end{proof}

    Notice that any element in $B\in C_p(A)$ leaves the spaces $E_\bthet$
    invariant, hence $C_p(A)\subseteq\calS$. Let $C_p(A_\bthet)\subset\Sp(2d_{\bthet},\FF{p})$ be the centralizer of  $A_{\bthet}$ in
    $\Sp(2d_\bthet,\FF{p})$, then the embedding (\ref{eEMBED}) induce an isomorphism
    \begin{equation}\label{eCENT}
    \prod_{\Lambda_p/\pm G_p}C_p(A_\bthet)\rightarrow C_p(A).
    \end{equation}
    We can thus recover the quantization of any element in $B\in C_p(A)$,
    from the tensor product of the quantization of corresponding elements $B_\bthet\in C_p(A_\bthet)$.

    We now want to look at the quantization of $A_\bthet$ together
    with its centralizer $C_p(A_\bthet)\subset \Sp(2d_\bthet,\FF{p})$, for one irreducible
    symplectic orbit $\bthet\in \Lambda_p/\pm G_p$. For the rest of this section, the orbit
    $\bthet$ will be fixed and for notational convenience the subscript will be omitted.
\subsection{Irreducible orbit}
    Let $A\in \Sp(2d,\FF{p})$ be a matrix with $2d$ distinct
    eigenvalues, such that there is only one irreducible symplectic orbit (symmetric or nonsymmetric).
    We now look at the quantization of $A$ together with its
    centralizer $C_p(A)$.
    \begin{rem}
    For a symplectic matrix $A\in\Sp(2d,\bbZ),\;d\geq 2$, the requirement
    that $A\pmod{p}$ has only one irreducible orbit can not hold
    for all primes. However, for a two dimensional matrix $A\in\SL(2,\bbZ)$
    this is indeed the case (since for any $A\in\SL(2,\FF{p})$ there could be only one orbit).
    The distinction between symmetric and nonsymmetric orbits in this case,
    correspond to inert and splitting primes respectively (c.f \cite{DGI,KR1}).
    \end{rem}
\subsubsection{Identification with finite fields}\label{sFIELDS}
    We now identify the action of the Hecke group $C_p(A)$ on the vector space $\FF{p}^{2d}$ with the action
    of (a multiplicative subgroup) of the finite field $\FF{p^{2d}}^*$ on itself by multiplication.
    Compare this to the identification
    $\iota^*_N:(\bbZ/N\bbZ)^{2d}\to\calI/N\calI$ that we defined in section \ref{sSTRUC} (note that the identification of the Hecke group here is
    more precise from the inclusions we used in the proof of lemma \ref{lHECKE}
    to estimate the number of Hecke operators).

    Take a pair of eigenvalues
    $\lambda,\lambda^{-1}$ in a field extension of $\FF{p}$.
    If we denote by $q=p^d$, then in the symmetric case
    $\FF{p}(\lambda)=\FF{q^2}$ and in the nonsymmetric $\FF{p}(\lambda)=\FF{q}$.
    Let $\vec{v},\vec{v}^*$ be eigenvectors for $\lambda,\lambda^{-1}$ respectively. In the
    symmetric case, where the eigenvalues are Galois conjugates, $\tau(\lambda)=\lambda^{-1}$,
    we take $\vec{v}^*=\tau(\vec{v})$ to be Galois conjugates as well.
    By lemma \ref{lFIELD:1}, in the nonsymmetric case (respectively
    symmetric), the map
    $$(\nu_1,\nu_2)\mapsto\Tr_{\FF{q}/\FF{p}}(\nu_1\vec{v})+\Tr_{\FF{q}/\FF{p}}(\nu_2\vec{v}^*)$$
     (respectively $\nu\mapsto\Tr_{\FF{q^2}/\FF{p}}(\nu\vec{v})$)
    is an isomorphism from $\FF{q}\oplus \FF{q}$ (respectively $\FF{q^2}$) to $\FF{p}^{2d}$.
    By lemma \ref{lFIELD:2}, under this identification,
    \begin{equation}\label{eSFORM}\omega(\vec{n},\vec{m})=\Tr_{\FF{q}/\FF{p}}(2\kappa(\mu\nu^*-\nu\mu^*)),\end{equation}
    where,
    $\nu=\omega(\vec{n},\vec{v}^*),\nu^*=\omega(\vec{n},\vec{v})$,
    $\mu=\omega(\vec{m},\vec{v}^*),\mu^*=\omega(\vec{m},\vec{v})$,
    and $\kappa=(2\omega(\vec{v},\vec{v}^*))^{-1}$

    This identification with finite fields, enables us to identify
    the centralizer as a subgroup of the multiplicative group
    $\FF{q^2}^*$, and to identify the orbits of elements in
    $\FF{p}^{2d}$ under the action of the centralizer.

\begin{lem}\label{lHECKE2}
    In the symmetric case,
    $C_p(A)\cong\ker(\calN_{\FF{q^2}/\FF{q}})$, while in the nonsymmetric case,
    $C_p(A)\cong \FF{q}^*$.
\end{lem}
\begin{proof}
    First for the symmetric case.
    For any $B\in C_p(A)$ the vectors $\vec{v},\vec{v}^*$ are eigenvectors with eigenvalues $\beta,\beta^{-1}\in\FF{q^2}$.
    Therefore the action of $B\in C_p(A)$ on $\FF{q^2}$ is given by
    \[\nu=\omega(\vec{n},\vec{v}^*)\mapsto \omega(\vec{n}B,\vec{v}^*)=\omega(\vec{n},\vec{v}^*B^{-1})=\beta\nu.\]
    On the other hand, any element
    $\beta\in\FF{q^2}$ defines (by multiplication) a linear
    transformation on $\FF{q^2}$ that commutes with the action
    of $A$. Given formula (\ref{eSFORM}) for the symplectic form,
    the condition for the action of $\beta\in\FF{q^2}$ to be
    symplectic, is precisely that $\beta\tau(\beta)=1$.
    We can thus identify $C_p(A)$ with the norm one elements in $\FF{q^2}/\FF{q}$.

    For the nonsymmetric case, the action
    of $C_p(A)$ on $\FF{q}\oplus\FF{q}$ is given by
    $(\nu_1,\nu_2)\mapsto (\beta\nu_1,\beta^{-1}\nu_2)$. Here any
    element $(\beta_1,\beta_2)\in \FF{q}\times\FF{q}$ defines
    a linear action that commutes with the action of $A$, and
    the elements that preserve the symplectic form, are precisely
    the elements $(\beta,\beta^{-1})$. We can thus identify these
    elements with $\FF{q}^*$.
\end{proof}

\begin{cor}\label{cORBITS}
For $\vec{n}\in \FF{p}^{2d}$ define
$\calQ(\vec{n})=\omega(\vec{n},\vec{v})\omega(\vec{n},\vec{v}^*)\in\FF{q}$.
Let $\vec{n},\vec{m}\in\FF{p}^{2d}$. If
$\calQ(\vec{n})=\calQ(\vec{m})\neq 0$ then there is $B\in C_p(A)$
s.t $\vec{n}B=\vec{m}$.
\end{cor}
\begin{proof}
We use the identification with finite fields. In the symmetric
case, let $\nu=\omega(\vec{n},\vec{v}^*)$ and
$\mu=\omega(\vec{m},\vec{v}^*)$. We thus need to find
$\beta\in\ker\calN_{\FF{q^2}/\FF{q}}$, such that $\beta\nu=\mu$.
The requirement that $\calQ(\vec{n})=\calQ(\vec{m})\neq 0$ implies
that $\calN_{\FF{q^2}/\FF{q}}(\mu\nu^{-1})=1$ and we can take
$\beta=\mu\nu^{-1}$.

In the nonsymmetric case, denote
$(\nu,\nu^*)=(\omega(\vec{n},\vec{v}^*),\omega(\vec{n},\vec{v}))$
and
$(\mu,\mu^*)=(\omega(\vec{m},\vec{v}^*),\omega(\vec{m},\vec{v}))$.
Then the requirement $\calQ(\vec{n})=\calQ(\vec{m})\neq 0$,
implies that $\frac{\nu}{\mu}=\frac{\mu^*}{\nu^*}$. Set
$\beta=\mu\nu^{-1}$, then
$(\beta\nu,\beta^{-1}\nu^*)=(\mu,\mu^*)$.
\end{proof}
\begin{rem}
Notice that the converse is obviously true, that is, if
$\vec{m}=\vec{n}B$ for some $B\in C_p(A)$ then
$\calQ(\vec{n})=\calQ(\vec{m})$.
\end{rem}

\subsubsection{Hecke eigenspaces}
    Consider the quantization of an irreducible
    element $A\in\Sp(2d,\FF{p})$, together with its centralizer $C_p(A)$.
    To any character $\chi$ of $C_p(A)$ let $\calH_\chi$ denote
    the corresponding eigenspace. Both in
    the symmetric and nonsymmetric cases, the centralizer
    is a cyclic group of even order.
    Therefore, there is a unique quadratic character of
    $C_p(A)$, that we will denote by $\chi_2$.
    \begin{prop}\label{pHES:1}
        For any character $\chi\neq \chi_2$, $\dim\calH_\chi=1$.
        In the symmetric case, the character $\chi_2$ does not
        appear in the decomposition, and in the nonsymmetric case $\dim\calH_{\chi_2}=2$.
    \end{prop}
    \begin{proof}
        Consider the projection operator
        \[\calP_\chi=\frac{1}{|C_p(A)|}\sum_{B\in
        C_p(A)}\chi^{-1}(B)U_p(B).\]
        The dimension of the corresponding eigenspace is then given
        by its trace
        \begin{equation}
        \label{eDIM}\dim(\calH_{\chi})=\Tr(\calP_\chi)=\frac{1}{|C_p(A)|}\sum_{B\in
        C_p(A)}\chi^{-1}(B)\Tr(U_p(B)).
        \end{equation}

        From corollary \ref{cTR:1} we have that,
        \[\abs{Tr(U_p(B))}=\sqrt{|\ker(B-I)|}.\]
        For any $B\in
        C_p(A)$, all eigenvalues are Galois conjugates and their inverses, hence $1$ is
        an eigenvalue of $B$ if and only if $B=I$. Therefore,
        for all $I\neq B\in C_p(A),\quad
        \abs{Tr(U_p(B))}=1$
        (and obviously $\Tr(U_p(I))=p^{d}=q$).

        In the symmetric case, $C_p(A)$ is isomorphic to the norm one elements in $\FF{q^2}/\FF{q}$
        and hence of order $q+1$.
        We can thus bound
        \[\dim\calH_{\chi}\leq \frac{2q}{q+1}<2,\]
        but since the dimension is an integer, $\dim (\calH_{\chi})\leq
        1$. Finally, there are $q+1$ characters and $\dim\calH_p=q$, so
        $q$ characters appear with multiplicity one.
        For now, denote the character that does not appear by $\tilde{\chi}_0$.

        In the nonsymmetric case, $|C_p(A)|=q-1$ and the
        corresponding bound is
        \[\dim\calH_{\chi}\leq \frac{2q-2}{q-1}=2.\]
        However, this inequality is actually an equality only if there
        is no cancellation in the sum (\ref{eDIM}), that is,
        \[\forall I\neq B\in C_p(A),\; \chi(B)=\Tr(U_p(B)).\]
        Such an equality can hold for at most one character.
        Thus, for any other character there is a strict
        inequality and $\dim \calH_\chi\leq 1$.
        Now, from dimension consideration we can deduce that there is a character
        with multiplicity $2$ (denoted again by $\tilde{\chi}_0$), and that all the other characters appear with
        multiplicity one.

        We now show that in both cases $\tilde{\chi}_0$ is the quadratic character.
        Notice, that for a cyclic group of even order the product of all the characters is the quadratic character.
        Therefore, for any $B\in C_p(A)$ the determinant of $U_p(B)$ is $\chi_2(B)\tilde{\chi}_0(B)^{-1}$ in the symmetric case,
        and $\chi_2(B)\tilde{\chi}_0(B)$ in the nonsymmetric. But since
        $\Sp(2d,\FF{p})$ has no nontrivial characters, then
        $\forall B\in\Sp(2d,\FF{p}),\;\det(U_p(B))=1$ and $\tilde{\chi}_0=\chi_2$.
    \end{proof}

    Since for $B\in C_p(A)-\{1\}$ the sum over all the characters vanish, the trace of $U_p(B)$ is $-\chi_2(B)$ in the symmetric case and $\chi_2(B)$ in the nonsymmetric.
    Consequently, we can find the constant in formula
    \ref{eUNA:1}.
    \begin{cor}\label{cUNA:1}
    For any $B\in C_p(A)$,
    \[U_p(B)=\pm\frac{\chi_2(B)}{q}\sum_{\vec{n}\in\FF{p}^d}\T{p}(\vec{n})\T{p}(-\vec{n}B),\]
    where the minus sign is for the symmetric case and the plus
    sign for the nonsymmetric.
    \end{cor}

\subsubsection{Explicit formulas and exponential sums}\label{sFORMULA}
    We now show that the matrix elements of
    elementary operators can be written explicitly as exponential sums.
    In \cite{GH} Gurevich and Hadani observed that matrix elements of
    elementary observables could be expressed as $\Tr(\T{p}(\vec{n})\calP_{\chi})$
    (where $\calP_{\chi})$ is the projection operator to the
    corresponding eigenspace). Using this observation, together with the formula for the propagator (corollary
    \ref{cUNA:1}), we obtain explicit formulas for the matrix
    elements.

    Denote by $e_q(x)=e_p(\Tr_{\FF{q}/\FF{p}}(x))$
    the corresponding additive character of $\FF{q}$. For
    notational convenience, we will denote by $\calC\cong C_p(A)$,
    the group of norm one elements
    in $\FF{q^2}/\FF{q}$ in the symmetric case, and the multiplicative group of $\FF{q}$, in the nonsymmetric.
    \begin{defn}\label{dEXP}
    For any character $\chi$ of
    $\calC$, and any element $\nu\in\FF{q}$, define the exponential sum:
    \[E_q(\nu,\chi)=\frac{1}{|\calC|}\sum_{1\neq x\in\calC}
    e_{q}(\nu\kappa\frac{x+1}{x-1} )\chi\chi_2(x)\]
    where $\kappa=(2\omega(\vec{v},\vec{v}^*))^{-1}$ (
    note that in the symmetric case, indeed
    $\kappa\frac{x+1}{x-1}\in\FF{q}$ so this is well defined).
    \end{defn}

    \begin{prop}\label{pFORM}
    Let $0\neq\vec{n}\in \FF{p}^{2d}$ and $\T{p}(\vec{n})$ the corresponding
    elementary operator. Let $\calQ(\vec{n})=\omega(\vec{n},\vec{v})\omega(\vec{n},\vec{v}^*)\in\FF{q}$,
    as in corollary \ref{cORBITS}.
    Let $\psi$ be a joint eigenfunction, with
    corresponding character $\chi$.
    Then, when $\chi\neq \chi_2$ is not the quadratic
    character (relevant only in the nonsymmetric case),
    \[\langle \T{p}(\vec{n})\psi,\psi\rangle=\pm E_q(\calQ(\vec{n}),\chi),\]
    where the minus sign is for symmetric case and plus for
    nonsymmetric.
    \end{prop}
    \begin{proof}
    Since the joint eigenspaces are one
    dimensional, an alternative way to write the matrix element
    is:
    \[\langle \T{p}(\vec{n})\psi,\psi\rangle=\Tr(\T{p}(\vec{n})\calP_{\chi}),\]
    where
    $\calP_{\chi}=\frac{1}{|C_p(A)|}\sum_{C_p(A)}\chi^{-1}(B)U_p(B)$, is
    the projection operator to $\calH_{\chi}$ \cite{GH}. Plugging in the formula for $U_p(B)$ (corollary \ref{cUNA:1}) gives,
    \[\langle \T{p}(\vec{n})\psi,\psi\rangle=\frac{\pm 1}{q|C_p(A)|}
    \sum_{C_p(A)}\chi^{-1}\chi_2(B)\sum_{\FF{p}^{2d}}\Tr(\T{p}(\vec{n})\T{p}(\vec{m})\T{p}(-\vec{m}B)).\]
    (where the minus sign is for the symmetric case).
    Notice that when $\vec{n}=\vec{m}(B-I)$,
    \[\Tr(\T{p}(\vec{n})\T{p}(\vec{m})\T{p}(-\vec{m}B))=qe_p(\frac{p+1}{2}\omega(\vec{n},\vec{m})),\]
    and that otherwise the trace vanishes.
    Therefore, when $B=I$ we get no contribution from the inner sum, and otherwise
    the only contribution is from $\vec{m}=\vec{n}(B-I)^{-1}$. Consequently
    \[\langle \T{p}(\vec{n})\psi,\psi\rangle=\frac{\pm 1}{|C_p(A)|}\sum_{C_p(A)\setminus\{I\}}\chi^{-1}\chi_2(B)
    e_p(\frac{p+1}{2}\omega(\vec{n},\vec{n}(B-I)^{-1})).\]

    We now use the identification with finite fields described in section \ref{sFIELDS}.
    Replace the sum over
    the elements in the centralizer with a sum over the elements in $\calC$, and for the symplectic form
    use formula (\ref{eSFORM}).
    Consequently, the formula for the matrix elements now takes the form
     \[\langle
    \T{p}(\vec{n})\psi,\psi\rangle=\frac{\pm 1}{|\calC|}\sum_{\calC-\{1\}}
    e_q(\calQ(\vec{n})\kappa\frac{1+\beta}{1-\beta})\chi\chi_2(\beta^{-1}).\]
    Changing summation variable $x=\beta^{-1}$
    concludes the proof.
\end{proof}
    We note that in both the symmetric and nonsymmetric cases for $\chi\neq\chi_2$, the Riemann Hypothesis for
    curves over finite fields imply the bound
    $|E_q(\nu,\chi)|\leq\frac{2}{\sqrt{q}}+O(\frac{1}{q})$ (see e.g. \cite[chapter 6]{Winnie} or \cite{KRR}). We can thus
    deduce:
\begin{cor}\label{cBOUND}
    For any $0\neq\vec{n}\in\FF{p}^{2d}$ and any $\psi\in\calH_\chi$ with $\chi\neq\chi_2$
    \[|\langle\T{p}(\vec{n})\psi,\psi\rangle|\leq
    \frac{2}{\sqrt{q}}+O(\frac{1}{q}).\]
\end{cor}

\begin{rem}
    Note that for $d=1$, this gives an alternative proof of the Kurlberg-Rudnick rate conjecture
    originally proved by Gurevich and Hadani \cite{GH}.
\end{rem}

    For the quadratic character (in the nonsymmetric case), $E_q(\calQ(\vec{n}),\chi_2)$
    is no longer a formula for the corresponding matrix
    element, but rather for
    \[\Tr(\T{p}(\vec{n})|_{\calH_{\chi_2}})=\langle \T{p}(\vec{n})\psi_0,\psi_0\rangle+\langle
    \T{p}(\vec{n})\psi_1,\psi_1\rangle,\]
    where $\set{\psi_0,\psi_1}$ is
    an orthonormal basis for $\calH_{\chi_2}$. Nevertheless, in this case we can find formulas for the
    eigenfunctions and use them to bound the individual matrix elements.
    \begin{lem}\label{lTRIV}
        In the nonsymmetric case, there is a normalized eigenfunction $\psi_0\in\calH_{\chi_2}$, such
        that \[\langle
        \T{p}(\vec{n})\psi_0,\psi_0\rangle=\left\lbrace\begin{array}{cc}
        0 & \omega(\vec{n},\vec{v})\neq 0\\
        1 & \omega(\vec{n},\vec{v})=0\end{array}\right.\]
        Furthermore, if $\calQ(\vec{n})\neq 0$ then for any normalized $\psi\in
        \calH_{\chi_2}$.
        \[|\langle \T{p}(\vec{n})\psi,\psi\rangle|\leq\frac{2}{\sqrt{q}}\]
    \end{lem}
    \begin{proof}
        We use a similar construction to the eigenfunctions constructed
        by Degli Esposti, Graffi and Isola  for two dimensional cat maps for splitting primes \cite{DGI}.

        In the nonsymmetric case, there is a decomposition
        $\FF{p}^{2d}=E\oplus E^*$ into two invariant Lagrangian subspaces.
        Therefore, there is
        $M\in\Sp(2d,\FF{p})$ such that for any $B\in C_p(A)$,
        $M^{-1}BM=\begin{pmatrix} \tilde{B}^t & 0 \\  0 & \tilde{B}^{-1} \\ \end{pmatrix}$.
        Consequently (by formula \ref{eUN:2}), the functions
        $\psi_0=\sqrt{q}U_p(M)\delta_0$ and
        $\psi_1=\sqrt{\frac{q}{q-1}}U_p(M)(1-\delta_0)$,
        are two orthonormal joint eigenfunctions of $U_p(B),\;B\in C_p(A)$ with the same
        eigenvalues, and hence a basis for $\calH_{\chi_2}$.

        Denote by $T_{i,j}=\langle
        \T{p}(\vec{n})\psi_i,\psi_j\rangle$.
        If we denote $\vec{m}=\vec{n}M$, then (by the intertwining equation)
        \[T_{0,0}=\langle \T{p}(\vec{n})\psi_0,\psi_0\rangle=q\langle
        \T{p}(\vec{m})\delta_0,\delta_0\rangle.\]
        By lemma \ref{lFIELD:1}, the projection of $\vec{n}$
        to the Lagrangian subspace $E$ vanishes (i.e., $\vec{m}=(0,\vec{m}_2)$), if and only if
        $\omega(\vec{n},\vec{v}^*)=0$.
        Now calculate directly,
        \[T_{0,0}=\sum_{\vec{x}}e_p(\frac{1}{2}\vec{m}_1\cdot\vec{m}_2)
        e_p(\vec{m}_2\cdot\vec{x})\delta_0(\vec{x}+\vec{m}_1)\delta_0(\vec{x}).\]
        Therefore, indeed
        $T_{0,0}=0$ if $\vec{m}_1\neq 0$ and $T_{0,0}=1$ if
        $\vec{m}_1= 0$.

        When $\calQ(\vec{n})\neq 0$, the projections to both Lagrangian subspaces do not vanish.
        By a similar computation, one can show that $T_{1,0}$ and $T_{0,1}$ are bounded by
        $\frac{1}{\sqrt{q-1}}$, and that $T_{1,1}$ is bounded by $\frac{2}{q-1}$.
        Therefore, since any normalized $\psi\in\calH_{\chi_2}$ is of
        the form $\psi=a_0\psi_0+a_1\psi_1$, with
        $|a_0|^2+|a_1|^2=1$, we have
        \[|\langle \T{p}(\vec{n})\psi,\psi\rangle|\leq\sum_{i,j=0}^1 |a_ia_jT_{i,j}|\leq \frac{2}{\sqrt{q}}.\]
    \end{proof}

    \subsubsection{Moments}
    Let $A\in\Sp(2d,\FF{p})$ be a matrix with one irreducible
    symplectic orbit (symmetric or nonsymmetric), and
    fix $\vec{n}\in\FF{p}^{2d}$ with $\calQ(\vec{n})\neq 0$.  Let $\{\psi_i\}$ be
    an orthonormal basis of joint eigenfunctions of $C_p(A)$.
    The different matrix elements $\langle
    \T{p}(\vec{n})\psi_i,\psi_i\rangle$ fluctuate around their average
    \[\frac{1}{q}\sum_i\langle \T{p}(\vec{n})\psi_i,\psi_i\rangle=\frac{1}{q}\Tr(\T{p}(\vec{n}))=0.\]
    \begin{rem}\label{rNONSYM}
    In the nonsymmetric case,
    for $0\neq\vec{n}\in\FF{p}^{2d}$ such that $\calQ(\vec{n})=0$,
    proposition \ref{pFORM} imply that for
    all characters $\chi\neq\chi_2$ the corresponding matrix elements are identical (and equal $-\frac{1}{p-1}$),
    so that the fluctuations are trivial.
    \end{rem}

    In \cite{KR1} Kurlberg and Rudnick gave a conjecture regarding
    the limiting distribution of these fluctuations (for $d=1$).
    Considering that in the formula for the matrix elements (proposition \ref{pFORM}),
    the dimension $d$ only determines the ground field $\FF{q}=\FF{p^d}$, we can reformulate their conjecture
    to predict the fluctuations of the corresponding exponential sums (formulated here as conjecture \ref{cLIMIT:2}).
    We now calculate (asymptotically) the second and fourth moments
    and show agreement with this conjecture.

    \begin{prop}\label{pMOMENT2}
    Let $\vec{n},\vec{m}\in\FF{p}^{2d}$ with $\calQ(\vec{n}),\calQ(\vec{m})\neq 0$.
    Then the mixed second moment, satisfies
        \[\frac{1}{q}\sum_i \langle \T{p}(\vec{n})\psi_i,\psi_i\rangle\overline{\langle
        \T{p}(\vec{m})\psi,\psi\rangle}=\left\lbrace\begin{array}{cc}
        \frac{1}{q}+O(\frac{1}{q^2}) &
        \calQ(\vec{n})=\calQ(\vec{m})\\
        O(\frac{1}{q^2}) & \calQ(\vec{n})\neq\calQ(\vec{m})\
        \end{array}\right.\]
    \end{prop}
    \begin{proof}
    First, we can replace the sum over eigenfunction to a sum over
    characters and the matrix element by corresponding exponential
    sums. By lemma \ref{lTRIV}, the error that comes from the quadratic character, is bounded by $O(\frac{1}{q^2})$ (recall $\calQ(\vec{n}),\calQ(\vec{m})\neq 0$).
    Now, since the sum over the characters $\chi(x)$ vanish unless
    $x=1$,
    \[\frac{1}{q}\sum_{\chi}E_q(\calQ(\vec{n}),\chi)\overline{E_q(\calQ(\vec{m}),\chi)}=
    \frac{1}{q|\calC|}\sum_{x\neq 1}e_q((\calQ(\vec{n})-\calQ(\vec{m}))\kappa\frac{x+1}{x-1}).\]
    If $\calQ(\vec{n})=\calQ(\vec{m})$ we indeed get
    $\frac{|\calC|-1}{q|\calC|}=\frac{1}{q}+O(\frac{1}{q^2})$.
    Otherwise, note that the map $x\mapsto \frac{x+1}{x-1}$ is
    injective, hence the sum is over $q-2$ distinct
    points in $\FF{q}$ (or $q$ in the symmetric case), and is therefore bounded by $O(\frac{1}{q^2})$.
    \end{proof}

    \begin{prop}\label{pFOURTH:1} For $\vec{n}\in\FF{p}^{2d}$ with $\calQ(\vec{n})\neq 0$, the fourth moment satisfies
    \[\frac{1}{q}\sum_i|\langle
    \T{p}(\vec{n})\psi_i,\psi_i\rangle|^4=\frac{2}{q^2}+O(\frac{1}{q^{5/2}}).\]
    \end{prop}
    \begin{proof}
    We follow the same lines as in the proof of proposition \ref{pFOURTH}.
    Consider the averaged operator
    \[D=\frac{1}{|\calC|}\sum_{B\in C_p(A)}\T{p}(\vec{n}B).\]
    Then, for any eigenfunction $\psi_i$, the diagonal matrix elements are the same $\langle D\psi_i,\psi_i\rangle=\langle
    \T{p}(\vec{n})\psi_i,\psi_i\rangle$, and
    for any two eigenfunctions $\psi_i,\psi_j$,
    corresponding to different characters the corresponding off diagonal terms vanish $\langle D\psi_i,\psi_j\rangle=0$.
    Consequently,
     \[\frac{1}{q}\sum_i|\langle
    \T{p}(\vec{n})\psi_i,\psi_i\rangle|^4=\frac{1}{q}\Tr((DD^*)^2)+O(\frac{1}{q^3}),\]
    where the error comes from the eigenfunctions corresponding to
    the quadratic character.

    We can calculate $\Tr((DD^*)^2)$ differently, by writing it as a product of
    4 sums and then taking trace, recalling that
    \[\Tr(\T{p}(\vec{n})\T{p}(\vec{m}))=\left\lbrace\begin{array}{cc}
    q & \vec{n}+\vec{m}= 0\\
    0  & \vec{n}+\vec{m}\neq 0\
    \end{array}\right.\]
    Define the set
    $X=\set{B_1,\ldots ,B_4\in C_p(A)|\vec{n}(B_1-B_2+B_3-B_4)=0}$,
    then this calculation gives
    \[\frac{1}{q}\Tr((DD^*)^2)=\frac{1}{|C_p(A)|^4}\sum_{X}e_p(\frac{1}{2}(\omega(\vec{n}B_2,\vec{n}B_1)+\omega(\vec{n}B_4,\vec{n}B_3)))\]
    Rewrite this expression using the identification with finite fields.
    The set $X$ transforms to
    \[X=\set{\beta_1,\ldots ,\beta_4\in \calC|
    \begin{array}{c}\nu(\beta_1-\beta_2+\beta_3-\beta_4)=0\\
    \nu^*(\beta_1^{-1}-\beta_2^{-1}+\beta_3^{-1}-\beta_4^{-1})=0\end{array}}\]
    and
    \[\frac{1}{q}\Tr((DD^*)^2)=\frac{1}{|\calC|^4}\sum_{X}e_q(\nu\nu^*\kappa(\beta_2\beta_1^{-1}-\beta_2^{-1}\beta_1+\beta_4\beta_3^{-1}-\beta_4^{-1}\beta_3)),\]
    where $\nu=\omega(\vec{n},\vec{v}^*)$,
    $\nu^*=\omega(\vec{n},\vec{v})$ and $\kappa=(2\omega(\vec{v},\vec{v}^*))^{-1}$ as in section \ref{sFIELDS}.

    Since we assumed $\calQ(\vec{n})=\nu\nu^*\neq 0$, the set
    $X$ is actually
    \[X=\set{\beta_1,\ldots ,\beta_4\in \calC|
    \begin{array}{c}\beta_1-\beta_2=\beta_4-\beta_3\\
    \beta_1^{-1}-\beta_2^{-1}=\beta_4^{-1}-\beta_3^{-1}\end{array}}.\]

    Now make a change of variables:
    \[x=\beta_2\beta_1^{-1},\;y=\beta_4\beta_3^{-1},\;z=\beta_3\beta_1^{-1},\;w=\beta_1,\]
    or equivalently $\beta_1=w,\beta_2=xw,\beta_3=zw$ and $\beta_4=yzw$.
    In these variables we get
    \[\frac{1}{q}\Tr((DD^*)^2)=\frac{1}{|\calC|^4}\sum_{Y}e_q(\calQ(\vec{n})\kappa(x-x^{-1}+y-y^{-1})),\]
    where the set
    \[Y=\set{x,y,z,w\in\calC\left|\begin{array}{c} (1-x)=z(y-1)\\yz(x-1)=x(1-y)\end{array}\right.}.\]
    The set $Y$ can be rewritten as
    \[Y=\set{x,y,z,w\in\calC\left|\begin{array}{cc} x=y=1 & \mbox{or} \\
     x=z=y^{-1} & \mbox{or}\\ x=y \mbox{ and } z=-1 & \end{array}\right.}.\]
    Indeed, if $x=1$, then from the second equation $y=1$ and $z$ is arbitrary.
    Otherwise, replace $y-1=z^{-1}(1-x)$ in the second equation to get
    $yz^2(x-1)=x(x-1)$, implying that $x=yz^2$. Plug this back to
    the first equation to get
    $1-yz^2=z(y-1)$, that is equivalent to
    $(yz-1)(1+z)=0$. Hence, either $z=-1$ or $yz=1$ which imply (by the first equation) $x=y$, or $x=z$ respectively.

    Therefore,
    \[\frac{1}{q}\Tr((DD^*)^2)=\frac{1}{|\calC|^3}\bigg(\sum_{z\in\calC}1+\sum_{x\in\calC}1+\sum_{x\in\calC}e_q(2\calQ(n)\kappa\frac{x^2-1}{x})-3\bigg).\]

    Both in the symmetric and nonsymmetric cases
    , $\calC$ is an irreducible algebraic curve of genus $1$,
    defined over the field
    $\FF{q}$, and the function $\frac{x^2-1}{x}$ has two
    simple poles at $0,\infty$. Hence, by \cite[theorem 5]{Bo}
        \[|\sum_{x\in\calC}e_q(2\calQ(n)\kappa\frac{x^2-1}{x})|\leq 2\sqrt{q},\]
    (in fact, in the nonsymmetric case $\calC=\FF{q}^*$ and this is a Kloosterman sum).
    This estimate implies,
     \[\frac{1}{q}\Tr((DD^*)^2)=\frac{2}{q^2}+O(\frac{1}{q^{5/2}}),\]
    concluding the proof.
    \end{proof}

\subsection{Formulas for matrix elements}
    Let $A\in\Sp_\theta(2d,\bbZ)$, be a matrix with distinct eigenvalues,
    and let $p>\Delta(P_A)$ be a sufficiently large prime.
    We showed that the quantization of the centralizer $C_p(A)$, is equivalent to a
    tensor product of the quantizations of $C_p(A_\bthet)\subset
    \Sp(2d_\bthet,\FF{p})$. To each irreducible element, we
    showed that the joint eigenfunctions are essentially unique,
    and found explicit formulas for the matrix elements.
    We now describe the Hecke eigenfunctions and corresponding matrix elements
    in the general case.

    Since $C_p(A)\cong \prod_\bthet C_p(A_\bthet)$, we can identify any character of $C_p(A)$ as a product
    $\chi=\prod_{\bthet}\chi_\bthet$, where $\chi_\bthet$ are characters of $C_p(A_\bthet)$.
    Denote by $\calH^{\bthet}_{\chi_{\bthet}}\subseteq L^2(\FF{p}^{d_{\bthet}})$, the joint eigenspace of all the
    operators $U_p^{(d_{\bthet})}(B_{\bthet})$, $B_{\bthet}\in C_p(A_\bthet)$ (with eigenvalues $\chi_\bthet$).
    Then, the map $\calU$ from proposition \ref{pTENZ}, maps the eigenspace $\calH_\chi$ isomorphically on to the space
    $\bigotimes\calH^{\bthet}_{\chi_{\bthet}}$. Furthermore,
    from proposition \ref{pHES:1} we know that these eigenspaces
    are essentially one dimensional. We can thus deduce:

    \begin{prop}\label{pHES:2}
    Let $\chi=\prod_{\bthet}\chi_\bthet$ be a character of $C_p(A)$.
    \begin{itemize}
            \item If $\forall {\bthet},\;\chi_{\bthet}$ is not the
            quadratic character, then $\dim\calH_\chi=1$.

            \item If $\chi_{\bthet}$ is the quadratic
            character for some symmetric orbit $\bthet$, then $\dim \calH_\chi=0$.
            \item Otherwise,  $\dim\calH_\chi=2^k$, where $k$ is
            the number of (nonsymmetric) orbits $\bthet$ for which
            $\chi_{\bthet}$ is the quadratic
            character.
            \item A basis for this space is given by
            $\set{\psi^{\eta}_\chi|\eta\in (\Z/2\Z)^k}$,
            \[\psi^{\eta}_\chi=\calU^{-1}\big(\bigotimes_{\chi_{\bthet}\neq  \chi_2}\psi_{\chi_{\bthet}}^{\bthet}\otimes
            \bigotimes_{\chi_{\bthet}= \chi_2}\psi_{\eta_\bthet}^{\bthet}\big),\]
            where $\set{\psi_0^\bthet,\psi_1^\bthet}$ is a basis for
            $\calH^\bthet_{\chi_2}$.
    \end{itemize}
    \end{prop}
    Note that the number of characters for which the quadratic character appears in the decomposition
    is bounded by $O(p^{d-1})$.
    Hence, the set $J_p\subseteq\{\psi_1,\ldots, \psi_{p^d}\}$ of Hecke eigenfunctions for which
    the quadratic character does not appear in the decomposition
    is of density one (i.e., $\lim_{p\to\infty}\frac{\sharp
    J_p}{p^d}=1$). For these eigenfunctions we can express the matrix elements as a product of
    exponential sums.

    For $\vec{n}\in\bbZ^{2d}$, and any
    symplectic Frobenius orbit $\bthet\in\Lambda_p/\pm G_p$, let
    $\nu_\bthet=\calQ_\thet(\vec{n}_\bthet)=\omega(\vec{n}_\bthet,\vec{v}_\bthet)\omega(\vec{n}_\bthet,\vec{v}_\bthet^*)$ as in proposition \ref{pFORM}
    (where $\vec{v}_\bthet,\vec{v}_\bthet^*$ are eigenvectors of $A_\bthet$ and $\vec{n}_\bthet$ is the projection of $\vec{n}\pmod{p}$ to $E_\bthet$).
    Let $\psi$ be a Hecke eigenfunction with corresponding
    character $\chi=\prod \chi_\bthet$. Define
    \[E^\bthet(\vec{n}_\bthet,\chi_\bthet)=\left\lbrace\begin{array}{cc}
    -E_{q_\bthet}(\nu_\bthet,\chi_\bthet) & \vec{n}_\bthet\neq 0,\;\thet=\thet^*\\
    E_{q_\bthet}(\nu_\bthet,\chi_\bthet) & \vec{n}_\bthet\neq 0,\;\thet\neq\thet^*\\
    1 & \vec{n}_\bthet= 0\
    \end{array}\right.,\]
    where $E_{q_\bthet}(\nu_\bthet,\chi_\bthet)$ are the exponential sums defined in \ref{dEXP}
    and $q_\bthet=p^{d_\bthet}$.

    If $\chi_\bthet\neq\chi_2$ is not the quadratic character for any orbit, then
    $\psi$ is uniquely determined and
    $E^\bthet(\vec{n}_\bthet,\chi_\bthet)=\langle T^{(d_\bthet)}(\vec{n}_\bthet)\psi^\bthet_{\chi_\bthet},\psi^\bthet_{\chi_\bthet}\rangle$.
    Consequently, the corresponding matrix element is a product of exponential
    sums,
     \begin{equation}\label{eMATEL:1}\langle
     \T{p}(\vec{n})\psi,\psi\rangle=\prod_{\Lambda_p/\pm G_p}
     E^\bthet(\vec{n}_\bthet,\chi_\bthet).
    \end{equation}

    For characters $\chi$, such that the quadratic character appears in
    the decomposition, the corresponding eigenfunction is no
    longer unique. However, any $\psi\in\calH_\chi$ is of the form
    $\psi=\sum_{\eta} a_\eta \psi^\eta_\chi$, where $\psi^\eta_\chi$ are defined in proposition \ref{pHES:2} and $\sum
    |a_\eta|^2=1$.
    Consequently, the corresponding matrix element is of the form
    \begin{equation}\label{eMATEL:2}\langle
    \T{p}(\vec{n})\psi,\psi\rangle=F(\vec{n},\psi)\prod_{\bthet\not\in
    W_\chi}E^\bthet(\vec{n}_\bthet,\chi_\bthet),\end{equation}
    where $W_\chi$ is the set
    of nonsymmetric orbits $\bthet$ for which $\chi_\bthet$ is the
    quadratic character and
    \[F(\vec{n},\psi)=\sum_{\eta,\eta'} a_\eta
    a_{\eta'}\prod_{\thet\in W_\chi} \langle
    \T{p}^{(d_\bthet)}(\vec{n}_\bthet)
    \psi^\bthet_{\eta_\thet},\psi^\bthet_{\eta_\thet'}\rangle.\]

\section{Super Scars}\label{sSCARS}
    This section is devoted to the proof of theorem \ref{tSCARS:1}. To
    any rational isotropic subspace $E_0\subset \Q^{2d}$ that is
    invariant under the action of $A$ ( i.e., $\vec{n}\in E_0 \Rightarrow \vec{n}A\in E_0$), we
    assign a corresponding submanifold of the torus  $X_0\subseteq \bbT^{2d}$, of
    dimension $\dim X_0=2d-\dim E_0$ that is invariant under the induced dynamics
    (i.e., $\vec{x}=(\begin{array}{c}\vec{p}\\\vec{q}\end{array})\in X_0\Rightarrow A\vec{x}\in X_0$).
    We then construct, for each
    prime $N=p$, a corresponding Hecke eigenfunction $\psi=\psi^{(p)}$ such
    that the distribution on the torus given by
    $f\mapsto \langle \Op_p(f)\psi,\psi\rangle$, weekly converges
    to Lebesgue measure on $X_0$.

\subsection{Invariant manifolds}
    Let $A\in\Sp_\theta(2d,\bbZ)$ be a matrix with distinct eigenvalues. To
    any invariant isotropic rational subspace $E_0\subset\Q^{2d}$, define the lattice $Z_0=E_0\cap \bbZ^{2d}$
    and assign a
    closed subgroup of the torus $X_{E_0}\subseteq\bbT^{2d}$ defined
    \[X_{E_0}=\set{\vec{x}\in\bbT^{2d}|\vec{n}\cdot\vec{x}=0\pmod{\bbZ},\;\forall\;\vec{n}\in
    Z_0}.\]
    The group
    $X_{E_0}\cong\bbT^{2d-d_0}$ is a submanifold with codimension $d_0=\dim E_0$, and is invariant under the action of
    $A$. In general the submanifold
    $X_{E_0}$ is co-isotropic, nevertheless, when $E_0$ is a Lagrangian
    subspace, $X_{E_0}$ is also Lagrangian.

    \begin{lem}
    Let $E_0$ be an invariant rational isotropic subspace. Then there is
    $\vec{x}_0\in \bbT^{2d}$ such that
    \[\vec{n}\cdot\vec{x}_0=\frac{\vec{n}_1\cdot\vec{n}_2}{2}\pmod\bbZ,\;\forall
    \vec{n}\in Z_0.\]
    \end{lem}
    \begin{proof}
    It is sufficient to show that there is $\vec{x}\in\R^{2d}$
    such that
    \begin{equation}\label{eSCAR}
    \vec{n}\cdot\vec{x}\equiv\vec{n}_1\cdot\vec{n}_2\pmod{2},
    \end{equation}
    for all $\vec{n}\in Z_0$
    (then $\vec{x}_0$ is the class of $\frac{1}{2}\vec{x}$ modulo
    $\bbZ$).
    Notice, that if (\ref{eSCAR}) is satisfied for
    $\vec{n},\vec{m}\in Z_0$, then it is also satisfied for
    $\vec{n}+\vec{m}$. Indeed,
    for any $\vec{n},\vec{m}\in Z_0$, because $E_0$ is isotropic we have $\vec{n}_1\cdot
    \vec{m}_2=\vec{m}_1\cdot \vec{n}_2$, hence
    \[(\vec{n}_1+\vec{m}_1)\cdot (\vec{n}_2+\vec{m}_2)\equiv\vec{n}_1\cdot
    \vec{n}_2+\vec{m}_1\cdot\vec{m}_2\pmod{2}.\]
    Therefore, it is sufficient to check the condition for an
    integral basis of the lattice $Z_0$.

    Let $\{\vec{n}^{(i)}\}_{i=1}^{d_0}$ be an integral basis. The vectors $\vec{n}^{(i)}$ are linearly
    independent, hence the set of equations
    $\vec{n}^{(i)}\cdot\vec{x}=b_i$ has a solution for any
    $(b_1,\ldots,b_{d_0})\in\R^{d_0}$ and in particular for $b_i=\vec{n}^{(i)}_1\cdot\vec{n}^{(i)}_2$.
    \end{proof}
    We can now define the manifold $X_0$ to be the coset $X_0=\vec{x}_0+X_{E_0}$, that is,
    \[X_0=\set{\vec{x}\in\bbT^{2d}\big|\vec{n}\cdot\vec{x}=\frac{\vec{n}_1\cdot\vec{n}_2}{2}\pmod{\bbZ},\quad\forall\vec{n}\in Z_0}.\]
    The condition that $A\in\Sp_\theta(2d,\bbZ)$ is quantizable, implies that $X_0$ is still invariant under the induced dynamics.

    \subsection{Rational orbits and Frobenius orbits} \label{sFROBRAT}
    For the proof of theorem \ref{tSCARS:1}, we would like to use
    the properties of the Hecke eigenfunctions and matrix elements
    described in section \ref{sHECKP}. However, since all the
    results in section \ref{sHECKP} were described in terms of the finite
    field $\FF{p}$, we first need to establish the correspondence between
    invariant rational subspaces for $A$ and invariant subspaces for $A$
    modulo $p$.

    Let $A\in \Sp_\theta(2d,\bbZ)$ with distinct eigenvalues. Then for any
    prime $p>\Delta(P_A)$,
    we can think of $A$ also as an element of
    $\Sp(2d,\FF{p})$ with distinct eigenvalues. Denote by
    $\Lambda_\Q$, the set of complex eigenvalues of $A$, and by
    $\Lambda_p$ the set of eigenvalues of $A$ (modulo $p$) in  $\bar{\mathbb{F}}_{\!p}$ (the
    algebraic closure of $\FF{p}$). Let
    $\Q^{2d}=\bigoplus_{\lambda_\Q/G_\Q}E_\theta$, and
    $\FF{p}^{2d}=\bigoplus_{\Lambda_p/G_p}E_\thet$ be the
    decompositions into irreducible invariant subspaces.

    To each rational orbit $\theta\in \Lambda_{\Q}/G_{\Q}$, denote by $P_\theta=\irr_{\Q}(\theta)$
    the minimal polynomial for some $\lambda_\theta\in\theta$ (this is independent of representative).
    We say that a Frobenius orbit, $\thet\in\Lambda_p/G_p$, lies under $\theta$ (denoted by $\thet|\theta$) if
    $\irr_{\FF{p}}(\thet)$ divides $P_\theta$ modulo $p$.
    We denote by $\theta^*$ the orbit of $\lambda_\theta^{-1}$ and note that
    $\thet|\theta\Leftrightarrow \thet^*|\theta^*$, in particular if $\theta$ is nonsymmetric (i.e., $\theta\neq \theta^*$) then so is any Frobenius orbit $\thet$ that lies under $\theta$.

    For every rational orbit $\theta\in\Lambda_\Q/G_\Q$, fix an eigenvalue $\lambda_\theta$.
    For every Frobenius orbit $\thet\in \Lambda_p/ G_p$ lying under $\theta$, fix a representative
    $\lambda_\thet$. For any such choice, there is a corresponding ring homomorphism
    \[\pi_{\lambda_\theta,\lambda_\thet}:\bbZ[\lambda_\theta]\rightarrow \FF{p}(\lambda_\thet),\]
    sending $\lambda_\theta$ to $\lambda_\thet$.

    \begin{lem}\label{lSURJ}
    Let $\calD_K\subseteq\calO_K$ be a subring of the integral ring of a number field $K/\Q$, let
    $\FF{q}$ be
    a finite field of characteristic $p$, and let
    $\pi:\calD_K\rightarrow \FF{q}$ be any ring homomorphism.
    Then, for any $\alpha\in\calO_K$ such that
    $\calN_{K/\Q}(\alpha)\neq 0\pmod{p}$, the image $\pi(\alpha)\neq 0$ as
    well.
    \end{lem}
    \begin{proof}
    Let $f=\irr_\Q(\alpha)$, then $f$ is a unit integral
    polynomial such that $f(\alpha)=0$. Consequently, if we take
    $\bar{f}\in \FF{q}[t]$ (by reduction of $f$ modulo $p$), then
    $\bar{f}(\pi(\alpha))=0$ as well. On the other hand we have
    that $f(0)=\pm\calN_{K/\Q}(\alpha)\neq 0\pmod{p}$, hence $\bar{f}(0)\neq 0$ and
    in particular $\pi(\alpha)\neq 0$.
    \end{proof}

    For any rational orbit $\theta\in \Lambda_\Q/G_\Q$, take eigenvectors $\vec{v}_\theta,\vec{v}_\theta^*$
    with coefficients in $\bbZ[\lambda_\theta]$ and
    eigenvalues $\lambda_\theta,{\lambda_\theta}^{\!\!\!-1}$ respectively.
    For $\vec{n}\in\bbZ^{2d}$ define
    \[N_\theta(\vec{n})=\calN_{\Q(\lambda_\theta)/\Q}(\omega(\vec{n},\vec{v}_\theta^*)).\]

    \begin{lem}\label{lPOJRF}
    For any  element $\vec{n}\in\bbZ^{2d}$ and any orbit $\theta\in \Lambda_\Q/G_\Q$.
    \begin{itemize}
    \item If the projection of $\vec{n}$ to $E_\theta$ vanishes, then for any $\thet|\theta$
    the projection of $\vec{n}\pmod{p}$ to
    $E_\thet$ also vanishes.
    \item If $p>N_\theta(\vec{n})$ and the projection of $\vec{n}$ to $E_\theta$ does not vanish,
    then for any $\thet|\theta$, the projection to $E_\thet$ does'nt vanish as well.
    \end{itemize}
    \end{lem}
    \begin{proof}
    For any Frobenius orbit $\thet|\theta$, let
    $\vec{v}_\thet^*=\pi_{\lambda_{\theta},\lambda_\thet}(\vec{v}_\theta^*)$.
    The vectors $\vec{v}_\thet^*$, are then eigenvectors with eigenvalues
    $\lambda_\thet^{-1}$, and
    \[\omega(\vec{n},\vec{v}_\thet^*)=\pi_{\lambda_{\theta},\lambda_\thet}(\omega(\vec{n},\vec{v}_\theta^*)).\]
    By corollary \ref{cPROJ:1} the projection of $\vec{n}$ to
    $E_\theta$ vanishes if and only if
    $\omega(\vec{n},\vec{v}_\theta^*)=0$ and the projection of $\vec{n}\pmod{p}$ to
    $E_\thet$ vanishes if and only if
    $\omega(\vec{n},\vec{v}_\thet^*)=\pi_{\lambda_{\theta},\lambda_\thet}(\omega(\vec{n},\vec{v}_\theta^*))=0$.
    The first part is now immediate,
    and the second part follows from lemma \ref{lSURJ}.
    \end{proof}

    \subsection{Construction of eigenfunctions}
    Let $Q^{2d}=\bigoplus_{\Lambda_\Q/G_\Q} E_\theta$ be the
    unique decomposition into irreducible (rational) invariant subspaces. Then, any
    invariant isotropic subspace $E_0$ is a direct sum
    \[E_0=\bigoplus_{\theta\in\Theta} E_\theta,\]
    where $\Theta\subseteq \Lambda_\Q/G_\Q$,
    is a subset containing nonsymmetric orbits such that $\theta\in\Theta\Rightarrow \theta^*\not\in\Theta$.

    Fix a large prime $p\geq \Delta(P_A)$, and recall the
    reduction to irreducible orbits described in section
    \ref{sHECKP} and the formulas for the eigenfunctions given in proposition \ref{pHES:1}.
    We will now construct a Hecke eigenfunction by prescribing the
    characters $\chi_\bthet$ and eigenstates $\psi_\bthet$ for each symplectic Frobenius
    orbit $\bthet\in\Lambda_p/\pm G_p$.

    We first determine the characters.
    For any symmetric orbit $\bthet$ fix an arbitrary character
    $\chi_\btheta\neq \chi_2$. For any nonsymmetric orbit $\bthet$, there is a unique nonsymmetric
    rational orbit $\btheta$ such that $\bthet|\btheta$. If $\btheta=\theta\cup\theta^*$ with
    $\theta,\theta^*\not\in\Theta$ then we take $\chi_\bthet\neq\chi_2$ to be
    any character except the quadratic, and otherwise we take $\chi_\bthet=\chi_2$ to be the quadratic one.

    Now for the eigenfunctions, when $\chi_\bthet\neq\chi_2$ the eigenspace $\calH^\bthet_{\chi_\bthet}$
    is one dimensional and $\psi^\bthet_{\chi_{\bthet}}$ is determined.
    Otherwise, there is $\theta\in\Theta$ such that $\btheta=\theta\cup\theta^*$.
    Let $\thet|\theta$ be the Frobenius orbit under
    $\theta$ and let $\vec{v}_\thet$ be an eigenvector for
    $A_\bthet$ with eigenvalue $\lambda_\thet\in\thet$. We then take $\psi^\bthet_0\in \calH^\bthet_{\chi_2}$,
    to be the eigenfunction (constructed in lemma \ref{lTRIV})
    satisfying
    \[\langle
    \T{p}(\vec{n}_\bthet)\psi^\bthet_0,\psi^\bthet_0\rangle=\left\lbrace\begin{array}{cc}
    1 & \omega(\vec{n}_\bthet,\vec{v}_\thet)=0\\
    0 &  \mbox{otherwise.}\
    \end{array}\right.\]
    To conclude, we take the character $\chi=\prod \chi_\bthet$ and eigenfunction
    \[\psi=\psi_\chi=\calU^{-1}\big(\bigotimes_{\chi_\bthet\neq\chi_2} \psi^\bthet_{\chi_\bthet}\otimes \bigotimes_{\chi_\bthet=\chi_2} \psi^\bthet_{0}\big)\]
     as in proposition \ref{pHES:2}.

    \begin{prop}\label{pSCARS:1}
    \[ |\langle
    \T{p}(\vec{n})\psi,\psi\rangle|=\left\lbrace\begin{array}{cc}
    1 & \vec{n}\in E_0\\
    O(p^{-1/4}) & \vec{n}\not\in E_0\
    \end{array}\right.,\]
    \end{prop}
    \begin{proof}
    The matrix elements corresponding to $\psi$ are of the form.
    \[\langle \T{p}(\vec{n})\psi,\psi\rangle=\prod_{\Lambda_p/\pm G_p} \langle
    \T{p}(\vec{n}_\bthet)\psi_\thet,\psi_\thet\rangle.\]

    First for $\vec{n}\in E_0$. For any rational orbit
    $\theta\in\Lambda_\Q/G_\Q$ such that $\theta,\theta^*\not\in\Theta$
    and any $\thet|\theta$, by lemma \ref{lPOJRF},
    $\vec{n}_\bthet=0$ and $\langle
    \T{p}(\vec{n}_\bthet)\psi_\thet,\psi_\thet\rangle=1$.
    On the other hand, for $\theta\in\Theta$, the projection
    of $\vec{n}$ to $E_{\theta^*}$ vanishes. Since
    $\thet|\theta\Rightarrow \thet^*|\theta^*$, again by lemma
    \ref{lPOJRF}, the projection to $E_{\thet^*}$ vanishes implying
    $\omega(\vec{n}_\bthet,\vec{v}_\bthet)=0$. Therefore, by
    construction again $\langle \T{p}(\vec{n}_\bthet)\psi_\thet,\psi_\thet\rangle=1$.
    This covers all symplectic Frobenius orbits in the product, hence
    $\langle \T{p}(\vec{n})\psi,\psi\rangle=1$.

    Now for $\vec{n}\not\in E_0$. There is some $\theta\not\in\Theta$ such that the projection of
    $\vec{n}$ to $E_\theta$ does not vanish. Then, by the second
    part of lemma \ref{lPOJRF} (we can assume $p$ is sufficiently large)
    for any $\thet|\theta$, the projection $\vec{n}_\thet\neq 0$.
    There are two possibilities, either $\theta^*\in \Theta$ or
    $\theta^*\not\in\Theta$.
    If $\theta^*\in \Theta$, then $\vec{n}_{\thet^*}\neq 0$ implying that $\omega(n_\bthet,\vec{v}_\thet)\neq 0$
    so $\langle
    \T{p}(\vec{n}_\bthet)\psi_\thet,\psi_\thet\rangle=0$ by our construction.
    Otherwise, the corresponding character is not
    the quadratic character, and by corollary \ref{cBOUND} we have
    $|\langle
    \T{p}(\vec{n}_\bthet)\psi_\thet,\psi_\thet\rangle|=O(p^{-d_\bthet/2})$.
    Therefore, the whole product satisfies
    \[|\langle \T{p}(\vec{n})\psi,\psi\rangle|\leq O(\prod_{\thet|\theta}p^{-d_\bthet/2})=O(p^{-d_\theta/2}).\]
    \end{proof}

    The eigenfunctions constructed above, satisfy $\langle\T{p}(\vec{n})\psi,\psi\rangle=1$
    for all $\vec{n}\in Z_0$.
    This implies, that $\psi$ is also a
    joint eigenfunction of the operators $\T{p}(\vec{n})$ with trivial eigenvalue
    \footnote{I thank St\'{e}phane Nonnenmacher for pointing
    that out.}.
    This property can be used in order make an
    alternative construction of these eigenfunctions.
    Given the isotropic invariant subspace $E_0$, the operators
    $\T{p}(\vec{n}),\;\vec{n}\in Z_0$ all commute (because it is isotropic)
    and one can consider the decomposition into joint eigenspaces.
    The joint eigenspace corresponding to the trivial eigenvalue
    is not empty, and is invariant under the action of all Hecke
    operators (because the space $E_0$ is invariant). Therefore
    there is a basis for this space composed of Hecke
    eigenfunctions each satisfying $\langle
    \T{p}(\vec{n})\psi,\psi\rangle=1$ for any $\vec{n}\in Z_0$.
    Furthermore, if $\vec{m}\in E_0^*$ (the symplectic complement of $E_0$)
    then there is $\vec{n}\in Z_0$ such that
    $\omega(\vec{n},\vec{m})\neq 0$. Consequently, for a
    sufficiently large $p$, $\T{p}(\vec{m})\psi$ is an eigenfunction of $\T{p}(\vec{n})$
    with eigenvalue $\neq 1$ and so $\langle\T{p}(\vec{m})\psi,\psi\rangle=0$.
    If we assume in addition that the space $E_0$ is a maximal
    isotropic invariant subspace, then any $\vec{n}\in\bbZ^{2d}$
    is either in $E_0\cup E_0^*$ or that is does not belong to
    any invariant isotropic subspace, in which case we have the
    estimate $\langle\T{p}(\vec{n})\psi,\psi\rangle=O(p^{-\frac{1}{2}})$.
    We thus see that any Hecke eigenfunction constructed in this manner, also satisfies proposition \ref{pSCARS:1}.

    \subsection{Proof of theorem \ref{tSCARS:1}}
    We now turn to prove theorem \ref{tSCARS:1}, that is we prove
    the following proposition.
    \begin{prop}\label{pSCARS:2}
    As $p\rightarrow \infty$ through primes, the distribution on the torus
    given by
    \[f\mapsto \langle \Op_p(f)\psi,\psi\rangle,\]
    (where $\psi$ are the Hecke eigenfunctions constructed above)
    converge to Lebesgue measure on $X_0$.
    \end{prop}
    \begin{proof}
    It is sufficient to show convergence for the test functions $e_{\vec{n}}(\vec{x})=\exp(2\pi
    i\vec{n}\cdot\vec{x}),\;\vec{n}\in\bbZ^{2d}$. For these
    functions,
    \[\int_{\bbT^{2d}}e_{\vec{n}}(\vec{x})d\mu_{X_0}(\vec{x})=\left\lbrace\begin{array}{cc}
    (-1)^{\vec{n}_1\cdot\vec{n}_2}& \vec{n}\in Z_0\\
    0 & \mbox{ otherwise}\
    \end{array}\right.\]
    where $\mu_{X_0}$ is Lebesgue measure on $X_0$.
    For $N=p$ a large (and in particular odd) prime the
    corresponding operator
    $\Op_p(e_{\vec{n}})=(-1)^{\vec{n}_1\cdot\vec{n}_2}\T{p}(\vec{n})$.
    Therefore, it is sufficient to show that as $p\rightarrow
    \infty$
    \[\langle \T{p}(\vec{n})\psi,\psi\rangle\rightarrow
    \left\lbrace\begin{array}{cc}
    1 & \vec{n}\in Z_0\\
    0 & \mbox{ otherwise}\\
    \end{array}\right.,\]
    and this follows from proposition \ref{pSCARS:1}.
   \end{proof}

    \section{Quantum Variance}\label{sVAR}
    In the following section, we assume that $A\in\Sp_\theta(2d,\bbZ)$ has
    no invariant isotropic rational subspaces, and compute the
    quantum variance when Planck's constant is the inverse of a large prime $N=p$.
    First we introduce a quadratic form
    $Q:\bbZ^{2d}\rightarrow \calD$, that characterizes the Hecke
    orbits of an element $\vec{n}\in\bbZ^{2d}$ (in the sense of
    proposition \ref{pnm:1}). We then define modified Fourier
    coefficients, grouping together coefficients belonging to the
    same Hecke orbits. Finally, we use the structure of the Hecke
    eigenfunctions described in section \ref{sHECKP} and the
    relations between the rational orbits and Frobenius orbits
    described in section \ref{sFROBRAT} to calculate
    the quantum variance proving theorem \ref{tVAR:1}.
    \subsection{A quadratic form}\label{sVAR:1}
        Let $A\in\Sp_\theta(2d,\bbZ)$ with $2d$ distinct eigenvalues.
        Recall the notation of section \ref{sHECKE:2}.
        Let $\Lambda_\Q/G_\Q$ denote the orbits of the Galois
        group $G_\Q$ on the set of eigenvalues $\Lambda_\Q$.
        Let $\Q^{2d}=\bigoplus E_\theta$ be the decomposition in
        to irreducible invariant subspaces. Further assume,
        that there are no invariant rational isotropic subspaces,
        implying that all orbits are symmetric $\theta=\theta^*=\btheta$ (hence all $E_\theta$ are symplectic).
        Recall the map $\iota^*:\bbZ^{2d}\rightarrow\calD$ (sending $\vec{n}\mapsto \omega(\vec{n},\vec{v})$) and the
        norm map $\calN:\calD\rightarrow\calD$ (sending
        $\beta\mapsto \beta\beta^*$) and define the quadratic form
        \[\begin{array}{c}
        Q:\bbZ^{2d}\rightarrow \calD\\
        \vec{n}\mapsto \calN(\iota^*(\vec{n}))\
        \end{array}\]
        The projection of $Q(\vec{n})$ to each component is given
        by
        $$Q_\theta(\vec{n})=\calN_{K_\theta/F_\theta}(\omega(\vec{n},\vec{v}_\theta)),$$
        where, $\vec{v}_\theta$
        is a left eigenvectors with eigenvalue $\lambda_\theta$, $K_\theta=\Q(\lambda_\theta)$, and
        $F_\theta=\Q(\lambda_\theta+\lambda_\theta^{-1})$.

        \begin{prop}\label{pnm:1}
        Let $\vec{n},\vec{m}\in\bbZ^{2d}$. Then
        $Q(\vec{n})=Q(\vec{m})$ if and only if for all sufficiently large
        primes, the classes of $\vec{n}$ and $\vec{m}$ modulo $p$ are in the same $C_p(A)$ orbit.
        \end{prop}
        \begin{proof}
            We now use the relations between rational orbits and
            Frobenius orbits described in section \ref{sFROBRAT},
            to relate corollary \ref{cORBITS} to the rational arithmetics.
            First assume that $Q(\vec{n})=Q(\vec{m})=\nu$. Let
            $N_0(\nu)=\max_\theta(\calN_{F_\theta/\Q}(\nu_\theta))$.
            We show that for any prime $p>N_0(\nu)$ there is $B\in C_p(A)$ such that $\vec{n}B=\vec{m}\pmod{p}$.
            It is sufficient to show that for any Frobenius orbit $\bthet\in\Lambda_p/\pm G_p$
            there is $B_\bthet\in C_p(A_\bthet)$ such
            that $\vec{n}_\bthet B_\bthet=\vec{m}_\bthet$.
            For $\theta\in \Lambda_\Q/G_\Q$ such that $Q_\theta(\vec{n})\neq
            0$,
            \[\calN_{F_\theta/\Q}(Q_\theta(\vec{n}))=\calN_{F_\theta/\Q}(Q_\theta(\vec{m}))\neq
            0\pmod{p}.\]
            Notice that $\vec{v}_\bthet=\pi_{\lambda_\theta,\lambda_\thet}(\vec{v}_\theta)$
            and
            $\vec{v}_\bthet^*=\pi_{\lambda_\theta,\lambda_\thet}(\vec{v}_\theta^*)$
            are eigenvectors for $A\pmod{p}$ with eigenvalues
            $\lambda_\thet$ and $\lambda_\thet^{-1}$ respectively.
            Consequently, by lemma \ref{lSURJ}, for any $\thet|\theta$,
            \[\calQ_\bthet(\vec{n}_\bthet)=\omega(\vec{n}_\bthet,\vec{v}_\bthet)\omega(\vec{n}_\bthet,\vec{v}_\bthet^*)=\pi_{\lambda_\theta,\lambda_\thet}(Q_\theta(\vec{n}))\neq 0,\]
            and by corollary \ref{cORBITS}, there is $B_\bthet\in C_p(A_\bthet)$
            such that $\vec{n}_\bthet B_\bthet=\vec{m}_\bthet$.
            On the other hand, if
            $Q_\theta(\vec{n})=Q_\theta(\vec{m})=0$, then by lemma \ref{lPOJRF} for any
            $\thet|\theta$, $\vec{n}_\thet=\vec{m}_\thet=0$. Since $\theta=\theta^*$ is symmetric then
            $\vec{n}_{\thet^*}=\vec{m}_{\thet^*}=0$ as well, hence $\vec{n}_\bthet=\vec{m}_\bthet=0$ and we can
            take any element of $C_p(A_\bthet)$.

            For the other direction, assume
            $Q(\vec{n})\neq Q(\vec{m})$. Then there is at least one orbit
            $\theta$ such that $Q_\theta(\vec{n})\neq
            Q_\theta(m)$. Consequently, for any prime
            $p>\calN_{F_\theta/\Q}(Q_\theta(\vec{n})-
            Q_\theta(m))$ and for any $\thet|\theta$ we have that
            $\calQ_\bthet(\vec{n}_\bthet)\neq\calQ_\bthet(\vec{n}_\bthet)$.
            Therefore $\vec{m}_\bthet$ and $\vec{n}_\bthet$ are not
            in the same $\calC_p(A_\bthet)$ orbit implying that
            $\vec{n},\vec{m}\pmod{p}$ are not in the same $\calC_p(A)$ orbit.
        \end{proof}
        \begin{cor}\label{cnm:1}
            Let $\vec{n},\vec{m}\in\bbZ^{2d}$ such that
            $Q(\vec{n})=Q(\vec{m})=\nu$. For any prime $p>N_0(\nu)$,
            and any Hecke
            eigenfunction $\psi\in \calH_p$,
            \[\langle \T{p}(\vec{n})\psi,\psi\rangle=\langle \T{p}(\vec{m})\psi,\psi\rangle.\]
        \end{cor}

        \subsection{Rewriting of matrix elements}
        We now use the form $Q$ to define modified fourier
        coefficients and rewrite the matrix elements, incorporating the Hecke symmetries.
        \begin{defn}
        For $f\in C^\infty(\bbT^{2d})$ and $\nu\in \calD$, define modified Fourier
        coefficients,
        \[f^\sharp(\nu)=\sum_{Q(\vec{n})=\nu}(-1)^{\vec{n}_1\cdot\vec{n}_2}\hat{f}(\vec{n}).\]
        For $\nu\in\calD$, and any Hecke eigenfunction $\psi$,
        define
        \[V_\nu(\psi)=\langle \T{p}(\vec{n}) \psi,\psi\rangle,\]
        where $\vec{n}\in\bbZ^{2d}$ is any element such that $Q(\vec{n})=\nu$.
        \end{defn}

        For $\nu\in\calD$ define
        $N_0(\nu)=\max_\theta(\calN_{F_\theta/\Q}(\nu_\theta))$ (as
        in the proof of proposition \ref{pnm:1}). For any
        trigonometric polynomial $f$,
        let $N_0(f)=\max_{\hat{f}(\vec{n})\neq
        0}(N_0(Q(\vec{n})))$.
        \begin{prop}\label{pREW}
            For any trigonometric polynomial $f$, any prime $p>N_0(f)$, and any Hecke eigenfunction $\psi\in\calH_p$:
            \[\langle Op_p(f)\psi,\psi\rangle=\sum_{\nu}f^\sharp(\nu)V_{\nu}(\psi).\]
        \end{prop}
        \begin{proof}
        Apply corollary \ref{cnm:1}.

        \end{proof}
        \begin{rem}
        Notice that it is possible to have $f\neq 0$ such that all the coefficients
        $f^\sharp(\nu)=0$ vanish. For example
        fix some $\vec{n}\in\bbZ^{2d}$ and take
        $f(\vec{x})=e_{\vec{n}}(\vec{x})-e_{\vec{nA}}(\vec{x})\neq 0$.
        \end{rem}

        \subsection{Proof of theorem \ref{tVAR:1}}\label{sVAR:2}
        We now want to prove theorem \ref{tVAR:1}, that is to show
        that as $p\rightarrow\infty$,
            \[S_2^{(p)}(f)=\frac{V(f)}{p^{d_f}}+O(\frac{1}{p^{d_f+1}}),\]
        where $d_f=\min_{f^\sharp(\nu)\neq 0} d_\nu$, $d_\nu=\sum_{\nu_\theta\neq 0}\frac{|\theta|}{2}$ and $V(f)=\sum_{d_\nu=d_f} |f^\sharp(\nu)|^2$.

        First, we compute mixed moments of elementary operators
        \[S_2^{(p)}(\vec{n},\vec{m})=\frac{1}{p^d}\sum_i \langle \T{p}(\vec{n})\psi_i,\psi_i\rangle\overline{\langle \T{p}(\vec{m})\psi_i,\psi_i\rangle}.\]
        \begin{lem}\label{lMIXED}
        Let $0\neq \vec{n},\vec{m}\in\bbZ^{2d}$ with
        $Q(\vec{n})=\nu$, $Q(\vec{m})=\mu$ and assume $d_\nu\leq d_\mu$. Then, for $p>\max(N_0(\nu),N_0(\mu))$ the mixed
        second moment satisfy
        \[S_2^{(p)}(\vec{n},\vec{m})=\left\lbrace\begin{array}{cc}
        \frac{1}{p^{d_\nu}}+ O(\frac{1}{p^{d_\nu+1}}) & \nu=\mu \\
        O(\frac{1}{p^{d_\nu+1}}) & \nu\neq \mu \
        \end{array}\right.\]
    \end{lem}

    \begin{proof}
        First assume that the matrix elements for all Hecke eigenfunctions are in the form
        of (\ref{eMATEL:1}).
        Consequently, we can rewrite
        \[S_2^{(p)}(\vec{n},\vec{m})=\prod_{\bthet\in \Lambda_p/\pm G_p}
        \bigg(\frac{1}{p^{d_\bthet}}\sum_{\chi_\bthet}
        E^{(\bthet)}(\vec{n}_\bthet,\chi_\bthet)E^{(\bthet)}(\vec{m}_\bthet,\chi_\bthet)\bigg).\]
        Recall that we assumed that there are no nonsymmetric rational
        orbits so (by the proof of proposition \ref{pnm:1}) if $\nu_\theta\neq 0$ then $\forall \thet|\theta,\; \calQ_\bthet(\vec{n}_\btheta)\neq 0$ and if
        $\nu_\theta=0$ then $\forall \thet|\theta,\; \vec{n}_\btheta= 0$ (similarly for $\vec{m}$ and $\mu$).
        The result is now immediate from proposition
        \ref{pMOMENT2}.

        Now for a general Hecke basis. Any Hecke eigenfunctions for which the quadratic character does not appear
        in the decomposition, gives the same contribution to the sum as before (because such an eigenfunction is unique).
        It is thus sufficient to show that the contribution of all other eigenfunctions
        is bounded by $O(\frac{1}{p^{d_\nu+1}})$. The number of
        these eigenfunctions is bounded by $O(p^{d-1})$, so it is
        sufficient to show that each summand contributes at most
        $O(\frac{1}{p^{d_\nu}})$ and this is immediate from corollary \ref{cBOUND} and lemma \ref{lTRIV}.
    \end{proof}

     Theorem \ref{tVAR:1}, now follows from lemma \ref{lMIXED} and proposition \ref{pREW}.
     \begin{proof}
     We first prove in the case where $f$ is a trigonometric
     polynomial. Define $N_0(f)=\max_{\hat{f}(\vec{n})\neq
     0}\{N_0(Q(\vec{n}))\}$. Then for $p>N_0(f)$ (by proposition \ref{pREW}) we can rewrite,
     \[\langle \Op_p(f)\psi_i,\psi_i\rangle-\int fdx=\sum_{0\neq
     \nu\in\calD}f^\sharp(\nu)V_\nu(\psi_i).\]
     Consequently, (after changing the order of summation), the quantum variance takes the form
     \[S_2^{(p)}(f)=\sum_{0\neq
     \nu,\mu\in\calD}f^\sharp(\nu)\overline{f^\sharp(\mu)}\frac{1}{p^d}\sum_i V_\nu(\psi_i)\overline{V_\mu(\psi_i)}.\]
     The second term (by lemma \ref{lMIXED}) contributes $\frac{1}{p^{d_\nu}}+O(\frac{1}{p^{d_\nu+1}})$ when $\nu=\mu$
     and $O(\frac{1}{p^{d_\nu+1}})$ otherwise. Therefore, the
     leading term is indeed
     \[S_2^{(p)}(f)=\frac{1}{p^{d_f}}\sum_{d_\nu=d_f}|f^\sharp(\nu)|^2+O(\frac{1}{p^{d_f+1}}).\]

     Now for any smooth $f\in C^\infty(\bbT^{2d})$. Approximate $f$
     by trigonometric polynomials $f_R=\sum_{\norm{\vec{n}}\leq
     R}\hat{f}(\vec{n})e_{\vec{n}}$. Note that since
     $N_0(Q(\vec{n}))\ll \norm{\vec{n}}^{4d^2}$, then
     $N_0(f_R)\ll R^{4d^2}$. We can thus define $R=R(p)\sim
     p^{1/4d^2}$, so that $\norm{\vec{n}}\leq R$ implies $N_0(Q(\vec{n}))\leq p$. We can take $p$ sufficiently large, so that
     $d_f=d_{f_R}$, then from the first part
     \[S_2^{(p)}(f_R)=\frac{V(f_R)}{p^{d_f}}+O(\frac{1}{p^{d_f+1}}).\]
     On the other hand, we can bound the difference
     \[|S_2^{(p)}(f)-S_2^{(p)}(f_R)|\ll_f
     \sum_{\norm{n}>R}\hat{f}(\vec{n})\ll_{f,\delta}
     \frac{1}{R^\delta},\]
     for any power $R^\delta$. In particular
     $|S_2^{(p)}(f)-S_2^{(p)}(f_R)|\ll_f\frac{1}{p^{d+1}}$, and in the
     same way, we also have $|V(f)-V(f_R)|\ll_f\frac{1}{p^{d+1}}$.
     We thus get that the quantum variance for smooth $f\in C^\infty(\bbT^{2d})$ satisfies
     \[S_2^{(p)}(f)=\frac{V(f)}{p^{d_f}}+O_f(\frac{1}{p^{d_f+1}}).\]
     \end{proof}

\section{Limiting Distributions}\label{sDIST}
    Let $A\in\Sp_\theta(2d,\bbZ)$, with distinct eigenvalues and no invariant rational isotropic
    subspaces as in the previous section. Given a smooth observable $f$ and a large prime $p$,
    consider the normalized matrix elements in the Hecke basis,
    $$\calW_i(f,p)=p^{d_f/2}(\langle \Op_p(f)\psi_i,\psi_i\rangle-\int fdx).$$
    As $p\rightarrow\infty$ these points fluctuate around zero with variance tending to $V(f)$,
    and we can ask whether they converge to some limiting
    distribution.
    Throughout this section, we will assume the validity of the Kurlberg-Rudnick conjecture for the limiting distribution (formulated here as conjecture \ref{cLIMIT:2}),
    and deduce the limiting distributions for $\calW_i(f,p)$.

    First, for any trigonometric polynomial $f$, and Hecke eigenfunction $\psi_i$ for which
    the quadratic character does not appear the
    decomposition,
    by formula \ref{eMATEL:1} and proposition \ref{pREW} we
    have
    \begin{equation}\label{eMATEL:3}
    \calW_i(f,p)=\sum_{d_\nu=d_f} f^\sharp(\nu)\prod_{\nu_\theta\neq 0}\prod_{\thet|\theta}\sqrt{q_\bthet}E_{q_\bthet}(\nu_\bthet,\chi_\bthet)
    +O(\frac{1}{\sqrt{p}}),
    \end{equation}
    where $\chi=\prod \chi_\bthet$ is the character corresponding to
    $\psi_i$, $\nu_\bthet=\pi_{\lambda_\theta,\lambda_\thet}(\nu_\theta)$ and the error term comes from the elements with $d_\nu>d_f$.
    By approximating a smooth function $f$ by trigonometric polynomials $f_R$ (as in the proof of theorem \ref{tVAR:1})
    formula \ref{eMATEL:3} is also valid for smooth functions.
    Finally, recall that the subset $J_p\subseteq\set{\psi_1,\ldots,\psi_{p^d}}$
    of eigenfunctions $\psi_i$ for which formula \ref{eMATEL:3} is valid is of density
    1.
    Therefore, conjecture \ref{cLIMIT:2} for the limiting distributions of the exponential sums, imply the following
    limiting distributions for the matrix elements:
    \begin{conj}\label{conLIMIT:3}
    For any tuple $k=(k_\theta),\;1\leq k_\theta\leq
    d_\theta$ consider the set of primes $\mathbf{P}_{k}$
    for which under every rational (symmetric) orbit $\theta$ there are precisely $k_\theta$ symplectic Frobenius
    orbits $\bthet|\theta$.
    Then, as $p\rightarrow\infty$ through primes from $\mathbf{P}_{\mathbf{k}}$ there is a
    limiting distribution for $\calW_i(f,p)$, and it is that of
    the random variable
    \[X_f=\sum_{d_\nu=d_f}f^\sharp(\nu)\prod_{\nu_\theta\neq 0}X^\theta_{\nu_\theta},\]
    where the random variables $X^\theta_{\nu_\theta}$ are all
    independent random variables. Furthermore, each of the variables $X^\theta_{\nu_\theta}$ is a product of
    $k_\theta$ independent random variables with Sato-Tate distribution.
    \end{conj}

    In particular, if we restrict to elementary observables
    $e_{\vec{n}}=\exp(2\pi i \vec{n}\cdot\vec{x})$, we recover
    conjecture \ref{cnLIMIT:1}.

    We now give an algorithm for
    determining which of the sets $\mathbf{P}_k$ are infinite, that is,
    to determine for a given matrix $A\in\Sp_\theta(2d,\bbZ)$ which limiting distributions can actually
    occur.

    Denote by $P_A$ the
    characteristic polynomial for $A$, and assume that $P_A$ is irreducible over the
    rationals (if it is reducible, one can repeat this process for each irreducible factor).
    Let $\lambda$ be a root of $P_A$ and denote by
    $\tilde{P}_A=\irr_\Q(\lambda+\lambda^{-1})$ the minimal polynomial for $\lambda+\lambda^{-1}$. Then
    $\tilde{P}_A$ is an irreducible integral unit polynomial of degree $d$.
    Furthermore, the space $\FF{p}^{2d}=\bigoplus_{\bthet}E_\bthet$ decomposes into $k$ irreducible invariant symplectic
    subspaces,
    if and only if $\tilde{P}_A=\prod_{\bthet}\tilde{P}_\bthet$ is a
    product of $k$ irreducible polynomials over $\FF{p}$ (where $\tilde{P}_\bthet=\irr_{\FF{p}}(\lambda_\bthet+\lambda_\bthet^{-1})$).
    Therefore, the set $\mathbf{P}_k$ is precisely
    the set of primes for which the polynomial $\tilde{P}_A\pmod{p}$ is a
    product of $k$ irreducible polynomials. The density of these
    sets,
    $\frac{1}{\pi(X)}\#\set{p\leq X|p\in\mathbf{P}_k}$, can be calculated by the Chebotarev theorem after
    calculating the Galois groups for $\tilde{P}_A$.
    To do this, consider the Galois group as a subgroup
    of the symmetric group $S_d$ (via its action on the roots of $\tilde{P}_A$).
    Recall that any element of $S_d$ can be uniquely presented as a product of disjoint cycles.
    The Chebotarev theorem says that the density of the set $\mathbf{P}_k$ is
    the relative number of elements in the Galois group that are
    a product of $k$ cycles.
    Furthermore, if there are no elements that are
    a product of $k$ cycles then $\mathbf{P}_k$ contains at most
    finitely many primes. For a precise statement and some
    background on the Chebotarev theorem see \cite[theorem 6.3.1]{FJ}.
    We demonstrate this calculation for a few simple examples.

    Our first example, is a $4$-dimensional symplectic matrix
    $A\in\Sp(4,\bbZ)$ for which $P_A$ is irreducible (i.e., no invariant rational subspaces).
    In this case the polynomial $\tilde{P}_A$ is a quadratic
    irreducible polynomial. In fact if
    $P_A(t)=t^4-at^3+bt^2-at+1$, then $\tilde{P}_A=t^2-at+b-2$.
    Consequently, the condition that $p\in\mathbf{P}_2$ is equivalent to
    the condition that the quadratic polynomial $t^2-at+b-2$
    has roots in $\FF{p}$, which is equivalent to the integer $c=a^2-4(b-2)$ being a square
    modulo $p$. Therefore the sets $\mathbf{P}_1,\mathbf{P}_2$ are both unions of
    arithmetic progressions, and each have density $1/2$.

    Our next examples are for matrices $A\in\Sp(6,\bbZ)$ with an irreducible characteristic
    polynomial (i.e., the polynomial $\tilde{P}_A$ is an irreducible
    polynomial of degree 3). In this case, we can no longer describe the sets $\mathbf{P}_k$ as arithmetic
    progressions. However, the classification of the Galois group for
    degree 3 polynomials is still relatively easy, and we can give the corresponding densities in each case.
    There are only two possible cases,
    either the splitting field for $\tilde{P}_A$ is a degree $6$ extension in which case the Galois group is isomorphic to the symmetric
    group $S_3$, or that the splitting field is of degree $3$ and the Galois group is cyclic of order
    $3$. We will now consider each case separately.

    In the symmetric group $S_3$ there are a total of $6$ elements,
    $2$ of them ($(1,2,3)$ and $(1,3,2)$) are composed of one cycle,
    $3$ of them ($(1,2)(3),(1,3)(2)$ and $(2,3)(1)$) are composed of
    two cycles
    and $1$ element (the identity) is composed of
    three cycles.
    Consequently, if the Galois group for $\tilde{P}_A$ is $S_3$ then by the ~Chebotarev theorem the densities of the sets $\mathbf{P}_1,\mathbf{P}_2$ and $\mathbf{P}_3$, are $2/6,3/6$ and $1/6$
    respectively.

    The cyclic group has $3$ elements, $2$ of them ($(1,2,3)$ and $(1,3,2)$) are
    composed of one cycle, and $1$ is composed of $3$ cycles
    (there are no elements composed of $2$ cycles).
    Therefore, when the Galois group for $\tilde{P}_A$ is cyclic
    the ~Chebotarev theorem implies that the density of $\mathbf{P}_1,\mathbf{P}_2$ and $\mathbf{P}_3$ are $2/3,0$ and $1/3$ respectively.
    Furthermore, $\mathbf{P}_2$ contains at most finitely many primes
    and the corresponding limiting distribution is not obtained.

\appendix
\section{Galois Orbits and Invariant Subspaces}\label{sORBITS}
    Let $E$ be a $2d$ dimensional vector space, defined over a
    perfect field $F$ (we will consider only the cases where $F$ is a
    number field or a finite field). Let $\omega:E\times
    E\rightarrow F$ be a symplectic form, and let $A\in \Sp(E,\omega)$ be
    a symplectic linear map with distinct eigenvalues acting on $E$ from the left.
    Denote by $\Lambda_F$, the set of
    eigenvalues of $A$ (in the algebraic closure of $F$). Let $G_F$, be the
    absolute Galois group and denote by $\Lambda_F/G_F$ the orbits of
    the eigenvalues under the action of $G_F$ (in fact it is sufficient to consider $\Gal(P_A/F)$,
    the Galois group of the splitting field of the characteristic polynomial $P_A$).

    Since the matrix $A$ is symplectic, if $\lambda\in\Lambda_F$ is an
    eigenvalue, then $\lambda^{-1}\in\Lambda_F$ as well. To each orbit
    $\theta\in \Lambda_F/G_F$ there is a unique orbit $\theta^*$ such
    that $\lambda\in\theta\Leftrightarrow \lambda^{-1}\in\theta^*$.
    If $\theta=\theta^*$ we say that the orbit is symmetric,
    otherwise we say that the orbit is nonsymmetric.
    \begin{lem}\label{lDECOMP}
    There is a unique decomposition into irreducible invariant
    subspaces:
    $E=\bigoplus_{\Lambda_F/G_F} E_\theta.$
    \begin{itemize}
    \item To each orbit $\theta\in\Lambda_F/G_F$, there is a
    corresponding subspace (denoted by $E_\theta$), such that the
    eigenvalues of the restriction $A_{|E_\theta}$ are the eigenvalues
    $\lambda\in\theta$. In particular $\dim
    E_\theta=|\theta|$.
    \item For any two orbits $\theta,\theta'$, unless
    $\theta'=\theta^*$, $E_\theta$ and $E_{\theta'}$ are
    orthogonal with respect to the symplectic form.
    \end{itemize}
    \end{lem}
    \begin{proof}
    Take representatives $\lambda_\theta\in\theta$ with eigenvectors $\vec{v}_\theta$.
    The space
    \[E_\theta=\set{\Tr_{F(\lambda_\theta)/F}(t\vec{v}_\theta)|t\in
    F(\lambda_\theta)},\]
    is a subspace of $E$ invariant under $A$, and the
    eigenvalues of the restriction of $A$ to $E_\theta$ are
    $\lambda\in\theta$. Furthermore, $E_\theta$ and $E_{\theta'}$
    are orthogonal, unless there is $\sigma\in G_F$ such that
    $\omega(\sigma(\vec{v}_\theta),\vec{v}_{\theta'})\neq 0$, and this
    happens only when $\theta'=\theta^*$. It remains to show that
    this is the only decomposition. Indeed, if $\tilde E$ is an
    invariant irreducible subspace,
    then there is an eigenvector $\vec{v}_\theta\in \tilde{E}\otimes \bar{F}$,
    and since $\tilde{E}$ is defined over $F$
    then all the Galois conjugates $\sigma(\vec{v}_\theta)$ are in this space as well.
    Therefore, the space $E_\theta\subseteq \tilde{E}$ and since we assumed
    $\tilde{E}$ is irreducible then $\tilde{E}=E_\theta$.
    \end{proof}

    \begin{defn}
    To each orbit $\theta\in\Lambda_F/G_F$ we define a symplectic
    orbit $\bar\theta=\theta\cup\theta^*$. Correspondingly, to
    each symplectic orbit, we assign the symplectic subspace
    $E_{\bar\theta}=E_\theta+E_{\theta^*}$. Then for symmetric orbits
    $E_{\bar\theta}=E_\theta$,
    and for nonsymmetric orbits $E_{\bar\theta}=E_\theta\oplus E_{\theta^*}$.
    \end{defn}
    Denote by $\Lambda_F/\pm G_F$ the set of symplectic orbits. Then
    \[E=\bigoplus_{\Lambda_F/\pm G_F}E_{\bar\theta},\]
    is a decomposition to a direct sum of orthogonal symplectic subspaces.

    \begin{lem}\label{lFIELD:1}
    Let $\lambda_\theta\in\theta$ with corresponding eigenvector
    $\vec{v}_\theta$ (with coefficients in $F(\lambda_\theta)$). Let
    $\vec{v}_\theta^*$, be an
    eigenvector with eigenvalue $\lambda_\theta^{-1}$.
    Then the map
    \[\begin{array}{ccl}
    F(\lambda_\theta) & \rightarrow & E_\theta\\
    t & \mapsto & \Tr_{F(\lambda_\theta)/F}(t\vec{v}_\theta)\end{array},\]
    is linear isomorphism, with an inverse map given by
    \[\begin{array}{ccc}
    E_\theta & \rightarrow & F(\lambda_\theta)\\
    \vec{n}& \mapsto &
    \frac{\omega(\vec{n},\vec{v}_\theta^*)}{\omega(\vec{v}_\theta,\vec{v}_\theta^*)}\end{array}.\]
    \end{lem}
    \begin{proof}
        The Galois conjugates of
        $\vec{v}_\theta$ are all eigenvectors with distinct eigenvalues in $\theta$.
        Therefore, they are linearly independent and the map $t \mapsto\Tr_{F(\lambda)/F}(t\vec{v}_\theta)$ is
        injective. On the other hand, $F(\lambda_\theta)/F$ is a
        vector space of dimension $[F(\lambda_\theta):F]=|\theta|$, hence it is isomorphic to $E_\theta$.

        Now let $\vec{n}\in E_\theta$, from the first part there is a
        decomposition
        \[\vec{n}=\Tr_{F(\lambda)/F}(t\vec{v}_\theta)=\sum_{\sigma\in\Mor_F(F(\lambda),\bar{F})}
        \sigma(t\vec{v}_\theta).\]
        Note that for any morphism, $\sigma\in \Mor_F(F(\lambda),\bar{F})$,
        the symplectic form
        \[\omega(\sigma(\vec{v}_\theta),\vec{v}_\theta^*)=\omega(\sigma(\vec{v}_\theta A),\vec{v}_\theta^*A)=
        \sigma(\lambda)\lambda^{-1}\omega(\sigma(\vec{v}_\theta),\vec{v}_\theta^*).\]
        Therefore, for any nontrivial morphism $\sigma$ we have
        $\omega(\sigma(\vec{v}_\theta),\vec{v}_\theta^*)=0$, and indeed
        $\omega(\vec{n},\vec{v}_\theta^*)=t\omega(\vec{v}_\theta,\vec{v}_\theta^*)$.
    \end{proof}

    \begin{cor}\label{cPROJ:1} For any element $\vec{n}\in E$,
    the projection of $\vec{n}$ to $E_\theta$ vanishes if and only
    if $\omega(\vec{n},\vec{v}_\theta^*)=0$, where
    $\vec{v}_\theta^*$ is any eigenvector with eigenvalue in
    $\theta^*$.
    \end{cor}

    To each symplectic orbit $\btheta\in \Lambda_F/\pm G_F$, fix a representative $\lambda_\btheta$
    and let
    $\vec{v}_\btheta,\vec{v}_\btheta^*$, be eigenvectors for
    $\lambda_\btheta,\lambda_\btheta^{-1}$. In the symmetric case,
    where $\lambda_\btheta^{-1}=\tau(\lambda_\btheta)$ are Galois
    conjugates, we take $\vec{v}_\btheta^*=\tau(\vec{v}_\btheta)$ to
    be Galois conjugates as well. To each symplectic orbit we also assign a
    field extension, $F_{\bar\theta}=F(\lambda_\btheta+\lambda_\btheta^{-1})$
    (note that for $\btheta$ nonsymmetric $F(\lambda_\btheta)=F_{\bar\theta}$
    and for $\btheta$ symmetric $[F(\lambda_\btheta):F_{\bar\theta})]=2$).

    \begin{lem}\label{lFIELD:2}
    Let $\vec{n},\vec{m}\in E$, and denote by $\vec{n}_\btheta,\vec{m}_\btheta$ their projection to $E_\btheta$.
    Then, the symplectic form
    \[\omega(\vec{n}_\btheta,\vec{m}_\btheta)=\Tr_{F_{\bar\theta}/F}(\kappa(\mu\nu^*-\nu\mu^*)),\]
    where
    $\nu=\omega(\vec{n},\vec{v}_\btheta^*),\nu^*=\omega(\vec{n},\vec{v}_\btheta)$,
    $\mu=\omega(\vec{m},\vec{v}_\btheta^*),\mu^*=\omega(\vec{m},\vec{v}_\btheta)$,
    and $\kappa=\omega(\vec{v}_\btheta,\vec{v}_\btheta^*)^{-1}$.
    \end{lem}
    \begin{proof}
    We prove first in the symmetric case.
    By lemma \ref{lFIELD:1},
    \[\vec{n}_\btheta=\Tr_{F(\lambda_\btheta)/F}(\kappa\nu\vec{v}_\btheta)=\sum_\sigma \sigma(\kappa\nu\vec{v}_\btheta),\]
    where the sum is over $\sigma\in\Mor_F(F(\lambda_\btheta),\bar{F})$.
    Therefore
    \[\begin{array}{cl}
    \omega(\vec{n}_\btheta,\vec{m}_\btheta) &=\omega(\sum_\sigma \sigma(\kappa\nu\vec{v}_\btheta),\sum_{\sigma'} \sigma'(\kappa\mu\vec{v}_\btheta))\\
    & =\sum_{\sigma,\sigma'}\sigma(\kappa\nu)\sigma'(\kappa\mu)\omega(\sigma(\vec{v}_\btheta),\sigma'(\vec{v}_\btheta))\\
    & =\Tr_{F(\lambda_\btheta)/F}[\sum_\sigma
    \kappa\nu\sigma(\kappa\mu)\omega(\vec{v}_\btheta,\sigma(\vec{v}_\btheta))].\
    \end{array}\]
    Now notice that $\omega(\vec{v}_\btheta,\sigma(\vec{v}_\btheta))\neq 0
    \Leftrightarrow \sigma=\tau$, in which case
    $\omega(\vec{v}_\btheta,\tau(\vec{v}_\btheta))=\omega(\vec{v}_\btheta,\vec{v}_\btheta^*)=\kappa^{-1}=-\tau(\kappa^{-1})$ and $\tau(\nu)=\nu^*$. Therefore
    \[\omega(\vec{n}_\btheta,\vec{m}_\btheta)=\Tr_{F(\lambda_\btheta)/F}(-\kappa\nu\mu^*)=\Tr_{F_\btheta/F}(\kappa(\mu\nu^*-\mu^*\nu)).\]

    In the nonsymmetric case,
    \[\vec{n}_\btheta=\Tr_{F_\btheta/F}(\kappa\nu\vec{v}_\btheta)+\Tr_{F_\btheta/F}(-\kappa\nu^*\vec{v}_\btheta^*).\]
    Here, $\omega(\vec{v}_\btheta,\sigma(\vec{v}_\btheta))=0$ for all automorphisms,
    and $\omega(\vec{v}_\btheta,\sigma(\vec{v}_\btheta)^*)\neq 0$
    only if $\sigma$ is the
    trivial automorphism. Hence, in this case as well
    \[\omega(\vec{n}_\btheta,\vec{m}_\btheta)=\Tr_{F_\btheta/F}(\kappa(\nu^*\mu-\nu\mu^*)).\]
     \end{proof}

\section{Counting elements in quotient rings}\label{sCOUNT}
    Let $F$ be a number field, and $K/F$ a quadratic Galois
    extension. Denote by $\calO_F,\calO_K$ the
    corresponding integral rings. For any ideal $a\subseteq\calO_F$,
    consider the map $\calN_a:(\calO_K/a\calO_K)^*\rightarrow
    (\calO_F/a)^*$ induced by the norm map $\calN_{K/F}$, and let
    $\calC(a)=\ker(\calN_a)$ denote its kernel.

    To each ideal $a\subseteq\calO_F$ define:
    \[\begin{array}{l}
     S_1(a)=\sum_{\beta\in\calC(a)}\sqrt{\#\set{\nu\in\calO_K/a\calO_K|\nu(\beta-1)\equiv 0\pmod{aO_K}}},\\
    S_2(a)= \#\set{\beta_1,\beta_2\!\in\!
    \calC(a)|(1-\beta_1)(1-\beta_2)(\beta_1+\beta_2)\equiv 0\!\!\!\!\!\pmod{a\calO_K}}.\\
    \end{array}\]

    Eventually we will be interested in estimating these quantities for ideals of the form
    $N\calO_F$ where $N\in\mathbb{N}$ are large integers.
    By the Chinese reminder theorem, if $a,b\subseteq\calO_F$ are
    co-prime (i.e., $a+b=O_F$), then
    $\calO_F/ab\cong\calO_F/a\times\calO_F/b$ and
    $\calO_K/ab\calO_K\cong\calO_K/a\calO_K\times\calO_K/b\calO_K$.
    Consequently, $\calC(ab)\cong \calC(a)\times\calC(b)$ and the quantities $S_1,S_2$ are multiplicative (i.e., $S_i(ab)=
    S_i(a)S_i(b)$). Therefore, it suffices to calculate them
    for powers of prime ideals.

    \subsection{Prime ideals}
    In the following proposition we summarize some facts
    regarding factorization of ideals in extensions of number fields
    (for proofs and general background on the subject we refer to \cite{Cohn}).
    \begin{prop}
    Let $K/F$ be an extension of number fields, and $\calO_K,\;\calO_F$ the corresponding integral
    rings. Let $P\subseteq\calO_F$ be a prime ideal, then the ideal $P\calO_K$ decomposes into prime ideals of
    $\calO_K$, $P\calO_K=\prod_{i=1}^r\calP_i^{e_i}$ where the ideals $\calP_i$ are all the ideals lying above $P$
    (i.e., $\calP_i\cap\calO_F=P$). Furthermore:
    \begin{enumerate}
    \item The fields $\calO_K/\calP_i$ are all finite field extensions
    of $\calO_F/P$. The degree $[\calO_K/\calP_i:\calO_F/P]=f_i$ is called the inertia degree.
    If the inertia degree $f_i=1$, then $\forall k\in\bbN$
    the corresponding rings are
    isomorphic $\calO_K/\calP_i^k\cong \calO_F/P^k$.
    \item  The exponent $e_i$ is called the ramification index. When not
    all the ramification indices
    $e_i=1$, the ideal $P$ is said to be ramified in $\calO_K$.
    For any number field $F/\Q$, there are only a finite number of ramified ideals
    (all lying above prime factors of the discriminant).
    \item The ramification indices $e_i$ and the inertia degrees $f_i$, satisfy  $[K:F]=\sum_{i=1}^r e_i f_i$.
    \item If $K/F$ is a Galois extension then all prime ideals of $\calO_K$ lying above a prime ideal $P\subseteq\calO_F$ are Galois conjugates,
    the ramification indices and the inertia degrees are fixed
    $e_i=e,\;f_i=f$, and the former equation takes the form $[K:F]=r e f$.

    \end{enumerate}
    \end{prop}
    In particular, in our case $[K:F]=2$, hence for any fixed prime ideal $P\subseteq
    \calO_F$,
    there are only 3 possibilities:
    \begin{enumerate}
    \item $P\calO_K=\calP\bar{\calP}$ ($P$ splits),
    \item $P\calO_K=\calP$ ($P$ is inert),
    \item $P\calO_K=\calP^2$ ($P$ is ramified),
    \end{enumerate}
    where $x\mapsto\bar{x}$ denotes the nontrivial
    automorphism of $K/F$.

    In the following proposition we describe the norm map
    $\calN_{P^k}$ in each of these cases.
    \begin{prop}\label{pPRIME:1}
    Let $\calP\subseteq \calO_K$ and $P=\calP\cap\calO_F$ be prime
    ideals.
    \begin{enumerate}
    \item If $P\calO_K=\calP\bar{\calP}$ splits, then $\calO_K/P^k\calO_K\cong \calO_F/P^k\times \calO_F/P^k$ as
    rings. Under this isomorphism, the norm map $\calN_{P^k}$ induces the map
    \[\begin{array}{ccc}
    (\calO_F/P^k)^*\times(\calO_F/P^k)^* & \rightarrow & (\calO_F/P^k)^*\\
    (x,y) & \mapsto & xy.\
    \end{array}\]
    \item If $P\calO_K=\calP$ is inert then the norm map
    $\calN_{P^k}$ is onto.
    \item If $P\calO_K=\calP^2$ ramifies then the image of
    $\calN_{P^k}$ is a subgroup of $(\calO_F/P^k)^*$ with index
    $2$ if $P$ lies above an odd prime and index bounded by $2^{d+1}$ if it lies above $2$.
    \end{enumerate}
    \end{prop}
    \begin{proof} We prove for each case
    separately:

    \ttl{1} When $P$ splits, by the Chinese reminder theorem
    $\calO_K/P^k\calO_K\cong\calO_K/\calP^k\times\calO_K/\bar{\calP}^k$,
    and since the inertia degree $f=1$ we can identify
    $\calO_F/P^k\cong\calO_K/\calP^k\cong\calO_K/\bar\calP^k$.
    Under this identification the norm map
    $N_{P^k}$ sends $(x,y)\in(\calO_F/P^k)^*\times(\calO_F/P^k)^*$ to
    $xy\in(\calO_F/P^k)^*$.

   \ttl{2} When $P$ is inert we prove by induction on $k$. For $k=1$, the inertia degree
    $[\calO_K/\calP:\calO_F/P]=2$ and the nontrivial
    automorphism of $K/F$ induces the nontrivial automorphism of
    $(\calO_K/\calP)/(\calO_F/P)$. Consequently, the norm map
    $\calN_{P}$ is the field extension norm
    map, that is surjective for finite fields. For $k>1$ by
    induction, let $\alpha\in(\calO_F/P^k)^*$, and
    $\alpha_0\in\calO_F$ its representative. By induction $\exists
    \beta_0\in\calO_K$ such that
    $\calN_{K/F}(\beta_0)\equiv\alpha_0\pmod{P^{k-1}}$.
    Denote by $\eta=\calN_{K/F}(\beta_0)-\alpha_0\in P^{k-1}$ and
    let $x\in\calO_K$ be an element such that
    $\Tr_{K/F}(\bar\beta_0x)=-1\pmod{P}$ (such an element exists because the trace
    for extension of finite fields is onto). Now,
    $\calN_{K/F}(\beta_0+\eta x)-\alpha_0\in
    P^k$, hence for $\beta=[\beta_0+\eta x]\in\calO_K/\calP^k$ (the class of $\beta_0+\eta
    x$), the norm map $\calN_{P^k}(\beta)=\alpha$.

   \ttl{3} When $P$ is ramified and lies above an odd prime again by induction. For $k=1$, $P$ ramifies implies
    $[\calO_K/\calP:\calO_F/P]=1$. Consequently, the nontrivial
    automorphism of $K/F$ induces the trivial automorphism of
    $(\calO_K/\calP)/(\calO_F/P)$ and the induced map $\calN_{P}$ (after
    identifying $\calO_K/\calP\cong\calO_F/P$)
    is the squaring map $x\mapsto x^2$.
    When the ideal $P$ lies
    above an odd prime $p$, the multiplicative group $(\calO_F/P\calO_F)^*$
    is a cyclic group of an even order ($p^{f_P}-1$) and the image of the map
    $x\mapsto x^2$ has index 2.
    For $k>1$ by induction. Let $\alpha\in(\calO_F/P^k)^*$
    and $\alpha_0\in\calO_F$ its representative.
    Then $\exists\beta_0\in\calO_K$ such that $\eta=\xi \calN_{K/F}(\beta_0)-\alpha_0\in
    P^{k-1}$, where $\xi$ is a representative of one of the classes of $(\calO_F/P^{k-1})^*/\im(\calN_{P^{k-1}})$.
    The map induced by $\Tr_{K/F}$ on $\calO_K/\calP\cong\calO_F/P$ is simply multiplication by $2$ and hence
    onto. We can thus take
    $x\in\calO_K$ such that $\xi\Tr(\beta_0 x)=-1\pmod{P}$.
    Now $\xi\calN_{K/F}(\beta_0+x\eta)-\alpha_0\in P^k$, meaning
    $\alpha$ is in one of the two classes as well.

    When $P$ lies above $2$, let $h$ denote the largest integer such that
    $2\in P^h$. For any $\alpha\in \calO_F$ we have
    that $\alpha^2\equiv 1\pmod{P^k}$ implies $\alpha\equiv
    \pm 1\pmod{P^{k-h}}$. Consequently, the kernel of squaring map
    has order bounded by $2|\calO_F/P^h|\leq 2|\calO_F/2\calO_F|\leq 2^{d+1}$.
    \end{proof}

    \subsection{Counting elements}
    \begin{prop}\label{pN1}
    The number of norm one elements satisfy
    \[\big(\frac{N}{\log N}\big)^d\ll |\calC(N\calO_F)|\ll
    (N\log N)^d.\]
    \end{prop}

    We first compute $|\calC(P^k)|$ for $P\subseteq \calO_F$ a prime
    ideal.
    \begin{lem}\label{lN1}
    Let $P\in\calO_F$ be a prime
    ideal lying above a rational prime $p\in\bbZ$. Then, if $p$
    is odd
    \[|\calC(P^k)|=|\calO_F/P^k|\cdot\left\lbrace\begin{array}{cc}
    (1-\frac{1}{p^{f_P}}) & P \mbox{ splits}\\
    (1+\frac{1}{p^{f_P}}) & P \mbox{ is inert}\\
    2 & P \mbox{ is ramified} \
    \end{array}\right.\]
    where $f_P=[\calO_F/P:\bbZ/p\bbZ]$ is the inertia degree.
    If $P$ lies above $2$, we can bound
    \[|\calC(P^k)|\leq 2^{d+1}|\calO_F/P^k|.\]
    \end{lem}
    \begin{proof}
    We compute $|\calC(P^k)|$, in each case separately.

    \ttl{1} When $P$ splits, by proposition \ref{pPRIME:1} we can
    identify the group of norm one elements
    \[\calC(P^k)\cong\set{(x,y)\in{(\calO_F/P^k)^*}^2|xy=1\pmod{P^k}}\cong
    (\calO_F/P^k)^*.\]
    Therefore,
    $|\calC(P^k)|=|(\calO_F/P^k)^*|=|\calO_F/P^{k}|(1-\frac{1}{|\calO_F/P|})$,
    and recall that $\calO_F/P$ is the finite field with $p^{f_P}$ elements.

    \ttl{2} When $P$ is inert the map
    $\calN_{P^k}:(\calO_K/\calP^k)^*\rightarrow (\calO_F/P^k)^*$
    is onto. Therefore,
    \[|\calC(P^k)|=|\ker(\calN_{P^K})|=\frac{|(\calO_K/\calP^k)^*|}{|(\calO_F/P^k)^*|}=\frac{|(\calO_K/\calP^k)||(1-\frac{1}{p^{f_{\calP}}})|}{|(\calO_F/P^k)||(1-\frac{1}{p^{f_{P}}})|}.\]
    Now, the inertia degree $[\calO_K/\calP:\calO_F/P]=2$, which implies
    $f_{\calP}=2f_P$ and $|\calO_F/\calP|=|\calO_F/P|^2$.

    \ttl{3} For $P$ ramified and odd, the image of $\calN_{P^k}$ is of index
    2 in $(\calO_F/P^k)^*$.
    Therefore,
     \[|\calC(P^k)|=2\frac{|(\calO_K/\calP^{2k})^*|}{|(\calO_F/P^k)^*|}=2\frac{|(\calO_K/\calP^{2k})||(1-\frac{1}{p^{f_{\calP}}})|}{|(\calO_F/P^k)||(1-\frac{1}{p^{f_{P}}})|}.\]
    In this case the inertia degree $[\calO_K/\calP:\calO_F/P]=1$,
    so that $f_{\calP}=f_P$ and $|\calO_K/\calP|=|\calO_F/P|$.
    When $P$ lies above $2$ the image is of index bounded by $2^{d+1}$, which implies the bound on $|\calC(P^k)|$.
    \end{proof}
    We now give the proof of proposition \ref{pN1} for composite $N$.
    \begin{proof}
    Let $N\calO_F=\prod P_i^{k_i}$ be the decomposition to prime
    ideals. By the Chinese reminder theorem,
    \[|\calC(N\calO_F)|=\prod_{i=1}^{r}|\calC(P_i^{k_i})|.\]
    Using lemma \ref{lN1} for each component:
    For all prime ideals $P_i$ there is a common term of
    $|\calO_F/P_i^{k_i}|$, that contributes precisely
    \[\prod
    |\calO_F/P_i^{k_i}|=|\prod(\calO_F/P_i^{k_i})|=|\calO_F/N\calO_F|=N^d.\]
    The additional contribution from the inert primes is bounded from below by 1 and from above by
    \[\prod_i (1+\frac{1}{p^{f_{P_i}}})\leq \prod_{p|N}(1+\frac{1}{p})^d\ll(\log N)^d,\]
    (since for every prime $p|N$ there are at most $d$ ideal
    primes that lie above it). Similarly, the contribution from the split
    primes is bounded from above by $1$ and from below by
    \[\prod_i (1-\frac{1}{p^{f_{P_i}}})\geq\prod_{p|N}(1-\frac{1}{p})^d\gg\big(\frac{1}{\log N}\big)^d.\]
    Finally, the contribution from the even and ramified primes, is bounded by some constant (recall that there are a bounded number of
    ramified primes).
    \end{proof}

    Given a prime ideal $P\subset \calO_F$, with ramification index $e\in\{1,2\}$
    and any $1\leq l\leq ek$ consider the
    congruence subgroup
    \[\calC^{(l)}(P^k)=\set{\beta\in\calC(P^k)|\beta\equiv
    1\pmod{\calP^l}},\]
    where $\calP\subset\calO_K$ is a prime ideal above $P$
    (note that it is indeed well defined and does not depend on $\calP$).

    \begin{lem}\label{lClk}
    If $P$ lies above an odd prime then
    \[|\calC^{(l)}(P^k)|=|\calO_F/P|^{k-\lfloor\frac{l}{e}\rfloor}.\]
    Otherwise,
    \[|\calO_F/P|^{k-\lfloor\frac{l}{e}\rfloor}\leq|\calC^{(l)}(P^k)|\leq 2^{d+1}|\calO_F/P|^{k-\lfloor\frac{l}{e}\rfloor}.\]
    \end{lem}
    \begin{proof}
    We prove it separately for $P$ split inert or ramified.

    \ttl{1} When $P$ splits we can identify
    \[C(P^k)\cong\set{(x,x^{-1})\in(\calO_F/P^k)^*\times(\calO_F/P^k)^*}\cong(\calO_F/P^k)^*.\]
    Denote by $(1+P^l)/(1+P^k)$ the kernel of the natural
    projection $(\calO_F/P^k)^*\rightarrow
    (\calO_F/P^l)^*$.
    Then, under this identification $C^{(l)}(P^k)\cong (1+P^l)/(1+P^k)$,
    and hence of order
    \[|C^{(l)}(P^k)|=|(1+P^l)/(1+P^k)|=|\calO_F/P|^{k-l}.\]
    \ttl{2} For $P$ inert, denote by $\calN_{P^k}^{(l)}$ the restriction of the norm map
    to $(1+\calP^l)/(1+\calP^k)$ (then $\calC^{(l)}(P^k)=\ker(\calN_{P^k}^{(l)})$).
    We now show that for $P$ odd $\calN_{P^k}^{(l)}$ is
    onto $(1+P^l)/(1+P^k)$, whereas if $P$ lies above $2$, it's image has index bounded by $2^{d+1}$
    (this would conclude the proof for the inert case).
    First, the image of $\calN_{P^k}^{(l)}$ is indeed a subgroup of $(1+P^l)/(1+P^k)$
    (because if $\beta=1\pmod{\calP^k}$ then $\calN_{K/F}(\beta)=1\pmod
    {P^k}$).
    Next, note that the image of $\calN_{P^k}^{(l)}$ contains all the squares in $(1+P^l)/(1+P^k)$.
    Now, for odd prime, $|(1+P^l)/(1+P^k)|=|\calO_F/P|^{k-l}$ is a power of $p$ and hence odd.
    Consequently, the map $x\mapsto x^2$ is an automorphism of $(1+P^l)/(1+P^k)$,
    and $\calN_{P^k}^{(l)}$ is
    onto. When $P$ lies above $2$ the map $x\mapsto x^2$ has
    kernel bounded by $2|\calO_F/P|^h$ (as in the proof of lemma
    \ref{pPRIME:1}). Consequently, the image of the squaring map (and hence also the image of
    $\calN_{P^k}^{(l)}$) has index bounded by $2|\calO_F/P|^h\leq 2^{d+1}$.

    \ttl{3} For $P$ ramified as in the previous case we can restrict the norm map to the
    group $(1+\calP^l)/(1+\calP^{2k})$. Here, (again by the squaring argument) the restricted
    map $\calN_{P^k}^{(l)}$ is onto $(1+P^{\lceil\frac{l}{2}\rceil})/(1+P^k)$ for $P$ odd and has image of index bounded by
    $2^{d+1}$ if $2\in P$. Consequently, in this case for $P$ odd,
    \[|\calC^{(l)}(P^k)|=|\calO_F/P|^{k-\lfloor\frac{l}{2}\rfloor},\]
    while for even prime ideals,
     \[|\calO_F/P|^{k-\lfloor\frac{l}{2}\rfloor}\leq|\calC^{(l)}(P^k)|\leq 2^{d+1}|\calO_F/P|^{k-\lfloor\frac{l}{2}\rfloor}.\]
    \end{proof}

    \begin{prop}\label{pS1}
    \[S_1(N\calO_F)\ll_\epsilon N^{d+\epsilon}.\]
    \end{prop}
    Again we start by computing $S_1(P^k)$ for powers of prime ideals.
    \begin{lem}\label{lS1}
    Let $P\in\calO_F$ be a prime ideal.
    \\
    If $P$ lies above an odd prime then
    \[S_1(P^k)\leq|\calO_F/P^k|\cdot\left\lbrace\begin{array}{cc}
    (k+1) & P \mbox{ is inert or splits}\\
    (k+1)\sqrt{|\calO_F/P|} & P \mbox{ is ramified}\
    \end{array}\right.\]
    If $P$ lies above $2$, then
    \[S_1(P^k)\leq 2^{d+2}|\calO_F/P^k|\cdot\left\lbrace\begin{array}{cc}
    (k+1) & P \mbox{ is inert or splits}\\
    (k+1)\sqrt{|\calO_F/P|} & P \mbox{ is ramified}\
    \end{array}\right.\]
    \end{lem}
    \begin{proof}
    Let $e\in\{1,2\}$ be the ramification index of $P$ in $\calO_K$.
    The group $\calC(P^k)$ decomposes into a disjoint union
    $\bigcup_{l=0}^{ek} \calC^{(l)}(P^k)\setminus
    \calC^{(l+1)}(P^k)$. We can thus rewrite
    \[S_1(P^k)=\sum_{l=0}^{ek}\sum_{\calC^{(l)}(P^k)\setminus\calC^{(l+1)}(P^k)}\sqrt{\#\set{\nu\in
    \calO_K/P^kO_K|\nu(\beta-1)=0}}.\]
    For fixed $l$ and any $\beta\in \calC^{(l)}(P^k)\setminus
    \calC^{(l+1)}(P^k)$, we have $\beta-1\in \calP^l\setminus\calP^{l+1}$.
    Therefore, the number of elements
    $\nu\in\calO_K/P^k\calO_K$ satisfying $\nu(\beta-1)=0$
    is precisely $|\calO_F/P|^{2l/e}$ independent of $\beta$.
    We
    can thus take it out of the sum to get
    \[ S_1(P^k)=
    \sum_{l=0}^{ek}
    (|\calC^{(l)}(P^k)|-|\calC^{(l+1)}(P^k)|)|\calO_F/P|^{l/e}.\]
    The result now follows directly from lemma \ref{lClk}.
    \end{proof}

We now give the proof of proposition \ref{pS1} for composite $N$.
\begin{proof}
Decompose $N\calO_F=\prod_{i=1}^r P_i^{k_i}$ into prime ideals.
For each prime ideal apply lemma \ref{lS1} to get the bound
\[S_1(N\calO_F)=\prod_{i=1}^r S_1(P_i^{k_i})\ll |\calO_F/N\calO_F|\prod_{i=1}^r (k_i+1),\]
where the implied constant comes from the contribution of the
ramified and even prime ideals. The first term
$|\calO_F/N\calO_F|=N^d$ and the second term can be bounded by
$\prod_{i=1}^r (k_i+1)\ll_\epsilon N^{\epsilon}$ completing the
proof.
\end{proof}

\begin{prop}\label{pS2}
\[S_2(N\calO_F)\ll_\epsilon N^{d+\epsilon}.\]
\end{prop}
As before, we start by a computation for powers of prime ideals.
\begin{lem} \label{lS2}
    Let $P\in\calO_F$ be a prime
    ideal. If $P$ lies above an odd prime, then
    \[S_2(P^k) \leq |\calO_F/P^k|\left\lbrace
    \begin{array}{cc}
      6(k+1) & \text{ P is inert or splits} \\
      6(k+1)|\calO_F/P| & \text{ P is ramified} \\
    \end{array}\right.\]
    If $P$ is even, then
    \[S_2(P^k) \leq 2^{4d}6(k+1)|\calO_F/P^k|.\]
\end{lem}
\begin{proof}
First note that when $P$ splits, the equation
\[(1-\beta_1)(1-\beta_2)(\beta_1+\beta_2)\equiv 0\pmod{P^k\calO_K},\;\beta_i\in\calC(P^k),\]
is invariant under Galois conjugation. Thus, it is equivalent to
the equation
\[(1-\beta_1)(1-\beta_2)(\beta_1+\beta_2)\equiv 0\pmod{\calP^k},\;\beta_i\in\calC(P^k),\]
where $\calP$ is a prime ideal above $P$. Therefore, in any case
$S_2(P^k)$ is the number of solutions to
\begin{equation}\label{eS2:1}
(1-\beta_1)(1-\beta_2)(\beta_1+\beta_2)\equiv
0\pmod{\calP^{ek}},\quad \beta_i\in \calC(P^k).
\end{equation}

When $P$ lies above an odd prime, then $2\notin \calP$ and
$\beta_1\equiv\beta_2\equiv
1\pmod{\calP}\Rightarrow\beta_1+\beta_2\equiv 2\not\equiv
0\pmod{\calP}$. Therefore, the number of solutions to
(\ref{eS2:1}) is bounded by $3$ times the number of solutions to
\begin{equation}\label{eS2:2}
(1-\beta_1)(1-\beta_2)\equiv 0\pmod{\calP^{ek}},\quad \beta_i\in
\calC(P^k).
\end{equation}
Since any solution $\beta_1,\beta_2$ of (\ref{eS2:2}), satisfies
$\beta_1\in \calC^{l}(P^k)\setminus \calC^{l+1}(P^k) ,\;\beta_2\in
\calC^{(ek-l)}(P^k)$ for some $0\leq l\leq ek$, the number of
solutions is bounded by
\[S_2(P^k)\leq 3\sum_{l=0}^{ek} (|\calC^{l}(P^k)|-|\calC^{l+1}(P^k)|)|\calC^{ek-l}(P^k)|,\]
and the result follows from lemma \ref{lClk}.

When $2\in P$ denote by $h$ the largest integer such that $2\in
P^{h}$ ( so that, $\calP^{eh}|2\calO_K)$. Now, if
$\beta_1\equiv\beta_2\equiv 1\pmod{\calP^{eh+1}}$ then
$\beta_1+\beta_2\neq 0\pmod{\calP^{eh+1}}$. Therefore, as in the
case of the odd prime, the number of solutions to (\ref{eS2:1}) is
bounded by 3 times the number of solutions to
\begin{equation}\label{eS2:3}
(1-\beta_1)(1-\beta_2)\equiv 0\pmod{\calP^{ek-eh}},\quad
\beta_i\in \calC(P^k).
\end{equation}
Now, any such solution satisfies
$\beta_1\in\calC^{(l)}(P^k)\setminus \calC^{(l+1)}(P^k)$ and
$\beta_2\in\calC^{(ek-eh-l)}(P^k)$ for some $0\leq l\leq ek-eh$,
hence
\[S_2(P^k)\leq 3\sum_{l=0}^{ek-eh} (|\calC^{l}(P^k)|-|\calC^{(l+1)}(P^k)|)|\calC^{ek-eh-l}(P^k)|,\]
and the result follows from lemma \ref{lClk}.
\end{proof}

Now for the general case.
\begin{proof}
Decompose $N\calO_K=\prod_{i=1}^t P_i^{k_i}$, and apply lemma
\ref{lS2} for each component
\[S_2(N\calO_K)=\prod_{i=1}^r S_2(P_i^{k_i})\ll \prod_{i=1}^r|\calO_F/P_i^{k_i}|6(k_i+1)=N^d\prod_{i=1}^r6(k_i+1),\]
where the implied constant comes from the even and ramified
ideals. The estimate $\prod_{i=1}^r6(k_i+1)\ll_\epsilon
N^{\epsilon}$ concludes the proof.
\end{proof}


\begin{thebibliography}{1}

%*********************************************************************
%********************************************************************
 \bibitem{BD1} F.~Bonechi, and S.~De Bi\`{e}vre {\em Controlling strong
scarring for quantized ergodic toral automorphisms}, Duke Math. J.
{\bf 117(3) } (2003), 571--587.


\bibitem{BD} A.~Bouzouina, and S.~De Bi\`{e}vre {\em Equipartition of the eigenfunctions of quantized ergodic
maps on the torus}, Commun. Math. Phys.  {\bf 178 } (1996),
83--105.

\bibitem{BoS}
E.~Bogomolny, and C.~Schmit {\em Superscars} preprint 2004,
arXiv:nlin--CD/0402017

\bibitem{Bo}
E.~Bombieri. {\em On exponential sums in finite fields}, Amer. J.
Math. {\bf 88} (1966), 71--105.

\bibitem{Cohn}  H.~Cohn {\em A classical invitation to algebraic numbers and
class fields}, Springer, New York 1978.

\bibitem{DD} S.~De Bi\`{e}vre, and M. Degli Esposti {\em Egorov
theorems and equidistribution of eigenfunctions for sawtooth and
Baker maps}, Ann. Inst. Poincar\'e {\bf 69} (1998), 1--30.

\bibitem{DEG} M.\ Degli~Esposti and S.\ Graffi {\em ``Mathematical
aspects of quantum maps''} in M.\ Degli~Esposti and S.\ Graffi,
editors {\it The mathematical aspects of quantum maps}, volume 618
of Lecture Notes in Physics, Springer, 2003, pp. 49--90.

\bibitem{DGI} M.~Degli Esposti, S.~Graffi and S.~Isola {\em Classical
limit of the quantized hyperbolic toral automorphisms}, Comm. Math
Phys. {\bf 167} (1995), 471--507.

\bibitem{EFK1}
B.~Eckhardt, S.~Fishman, J.~Keating, O.~Agam, J.~Main, and
K.~M\"{u}ller,
  \emph{Approach to ergodicity in quantum wave functions.}, Phys. Rev. E
  \textbf{52(6)} (1995), 5893--5903.

\bibitem{FP}
M.~Feingold and A.~Peres, \emph{Distribution of matrix elements of
chaotic
  systems}, Phys. Rev. A \textbf{34(1)} (1986), 591--595.

\bibitem{FND}
F.~Faure, S.~Nonnenmacher and S.~De Bi\`{e}vre, \emph{Scarred
eigenstates for quantum
  cat maps of minimal periods}, comm. Math. Phys.  \textbf{239(3)} (2003), 449--492.

\bibitem{FJ}
  M. D.~Fried and M.~Jarden, \emph{Field Arithmetic}, Second Edition, revised and enlarged by
Moshe Jarden, Ergebnisse der Mathematik (3) 11, Springer,
Heidelberg, 2004.


\bibitem{Ge}
P.~G\'{e}rardin, \emph{Weil representations associated to finite
fields}, J.Algebra \textbf{46} (1977), 54--101.

\bibitem{G} S.~Gurevich, \emph{Weil Representation, Deligne Sheaf, and Proof of the Kurlberg-Rudnick Conjecture.}
\newblock {PhD Thesis Tel-Aviv University} (2005).

\bibitem{GH} S.~Gurevich   and  R.~Hadani , \emph{Proof of the
Kurlberg-Rudnick Rate Conjecture.}, preprint 2004,
arXiv:math--ph/0404074.

\bibitem{GH2} S.~Gurevich   and  R.~Hadani , \emph{The Higher-Dimensional Rudnick-Kurlberg Conjecture
.} , preprint 2004, arXiv:math--ph/0409031 .


\bibitem{HB}
J.H.~Hanny and M.V.~Berry, \emph{Quantization of linear maps on a
torus-Fresnel diffraction by
  a periodic grating}, Phys.D  \textbf{1} (1980), 267--290.

\bibitem{Hua} L.K.~Hua, and I.~Reiner,  \emph{On the generators of the symplectic
modular group}, Trans. Amer. Math. Soc. \textbf{65}, (1949),
415--426.

\bibitem{Iw}
 H.~Iwaniec, \emph{Spectral Methods of Automorphic
Forms }, Graduate Studies in Mathematics \textbf{53}, American
Mathematical Society, Rhode Islans, 2002.


\bibitem{K1}
D.~Kelmer \emph{On the quantum variance of matrix elements for the
cat map on the 4-dimensional torus}, IMRN {\bf 36} (2005),
2223--2236.

\bibitem{Knabe}
S.~Knabe {\em On the quantisation of Arnold's cat}, J.Phys. A:
Math. Gen. {\bf 23} (1990), 2013--2025.


\bibitem{KR2}
P.~Kurlberg and Z.~Rudnick, \emph{Hecke theory and
equidistribution for the
  quantization of linear maps of the torus}, Duke math. J. \textbf{103(1)}
  (2000), 47--77.

\bibitem{KR1}
P.~Kurlberg and Z.~Rudnick, \emph{On the distribution of matrix
elements for the quantum cat map}, Ann. of Math. \textbf{161}
(2005), 1-19.

\bibitem{KRR}
P.~Kurlberg, L.~Rosenzweig, and Z.~Rudnick, \emph{Matrix elements
for the quantum cat map: fluctuations in short windows}, In
preperation

\bibitem{Mo}
 C.~M{\oe}glin, M.-F.~Vign{\'e}ras, and J.-L.~Waldspurger, \emph{Correspondances de {H}owe sur un corps
{$p$}-adique},
              Lecture Notes in Mathematics \textbf{1291}, Springer-Verlag, 1987.

\bibitem{Ne}
M. Neuhauser {\em An explicit construction of the metaplectic
representation over a finite field}, J. Lie Theory \textbf{12(1)}
(2002), 15--30.

\bibitem{No}
S.~Nonnenmacher Private comunication (2003).

\bibitem{RSO}
A. M. F. Rivas,  M. Saraceno and  A. M. Ozorio de Almeida {\em
Quantization of multidimensional cat maps}, Nonlinearity
\textbf{13(2)} (2000), 341--376.

\bibitem{RS} Z. Rudnick and P. Sarnak {\em The behaviour of
eigenstates of arithmetic hyperbolic manifolds}, Comm. Math Phys.
{\bf 161} (1994), 195--213.


\bibitem{S}
A.I \v{S}hnirel'man, \emph{Ergodic properties of eigenfunctions},
Usp. Mat. Nauk
  \textbf{29} (1974), 181--182.


\bibitem{Winnie}
W. C. Winnie Li, \emph{Number theory with applications}. Series on
University Mathematics, 7. World Scientific Publishing Co., Inc.,
River Edge, NJ, 1996.

\bibitem{Weil}
A. Weil, \emph{On some exponential sums.}, Proc. Nat. Acad. Sci.
U.S.A. \textbf{34}, (1948) 204-–207.

\bibitem{Z}
S.~Zelditch, \emph{Uniform distribution of eigenfunctions on
compact hyperbolic surfaces}, Duke Math. J. \textbf{55(4)} (1987),
919--941.


\end{thebibliography}
\end{document}